\documentclass[final,5p,times,twocolumn,numbers,sort&compress]{elsarticle}

\usepackage{natbib}

 \usepackage{graphicx}

\usepackage{subfigure}
\usepackage{color}

\usepackage{latexsym,bm}

\usepackage{amssymb}
\usepackage{amsmath}

\usepackage[linesnumbered,ruled]{algorithm2e}
\usepackage{algorithmic}

\usepackage{cases}

\usepackage{threeparttable}
\usepackage{multirow}

\usepackage[colorlinks,linkcolor=red]{hyperref}

\newcommand{\abs}[1]{\left | #1 \right |}
\newcommand{\pref}[1]{(\ref{#1})}
\newcommand{\tabincell}[2]{\begin{tabular}{@{}#1@{}}#2\end{tabular}}
\allowdisplaybreaks
\begin{document}

\begin{frontmatter}

\title{Heterogeneous porous scaffold generation in trivariate B-spline solid with triply periodic minimal surface in the parametric domain}

\author[add1]{Chuanfeng Hu}
\author[add1,add2]{Hongwei Lin\corref{cor1}}
    \cortext[cor1]{Corresponding author: phone number: 86-571-87951860-8304, fax number: 86-571-88206681, email:
    hwlin@zju.edu.cn}
\address[add1]{School of Mathematics, Zhejiang University, Hangzhou, 310027, China}
\address[add2]{State Key Lab. of CAD\&CG, Zhejiang University, Hangzhou, 310027, China}

\date{}

\begin{abstract}
 A porous scaffold is a three-dimensional network structure composed of a
    large number of pores,
    and triply periodic minimal surfaces (TPMSs) are one of conventional tools for designing porous scaffolds.
 However, discontinuity, incompleteness,
    and high storage space requirements are the three main shortcomings of
    TPMSs for porous scaffold design.
 In this study,
    we developed an effective method for heterogeneous porous scaffold generation to overcome the abovementioned shortcomings of TPMSs.
 The input of the proposed method is a trivariate B-spline solid (TBSS)
    with a cubic parameter domain.
 The proposed method first constructs a threshold distribution field (TDF)
    in the cubic parameter domain,
    and then produces a continuous and complete TPMS within it.
 Moreover, by mapping the TPMS in the parametric domain to the TBSS,
    a continuous and complete porous scaffold is generated in the TBSS.
 In addition, if the TBSS does not satisfy engineering requirements,
    the TDF can be locally modified in the parameter domain,
    and the porous scaffold in the TBSS can be rebuilt.
 We also defined a new storage space-saving file format based on the TDF to store porous scaffolds.
 The experimental results presented in this paper demonstrate the effectiveness and
    efficiency of the method using a TBSS as well as the superior space-saving of the proposed storage format.
\end{abstract}

\end{frontmatter}


\section{Introduction}
\label{sec:introduction}
Porous structures are widely found in natural objects, such as trabecular
    bones, wood, and cork,
    which have many appealing properties,
    such as low weight and large internal surface area.
 In the field of tissue engineering,
    which aims to repair damaged tissues and organs,
    porous scaffolds play a critical role in the formation of new functional tissues for medical purposes, i.e.,
    they provide an optimum biological microenvironment for cell attachment, migration, nutrient delivery and product expression~\cite{Hutmacher2000Scaffolds}.
 To facilitate cell growth and diffusion of both cells and nutrients
    throughout the whole structure,
    high porosity, adequate pore size and connectivity are key requirements
    in the design of porous scaffolds~\cite{Starly2003Computer}.
 Therefore, it is important to be able to control the pore size
    and porosity when designing heterogenous tissue scaffolds.

 Recently, triply periodic minimal surfaces (TPMSs) have been widely employed in the design of porous scaffolds~\cite{Y2018Gyroid}.
 A TPMS is a type of minimal surface with periodicity in three independent
    directions in three-dimensional Euclidean space and is represented by an implicit equation~\cite{Schnering1991Nodal}.
 Generally speaking, porous scaffold design methods based on TPMS can be classified into two categories.
 In the first class of methods,
    a volume mesh model is embedded in an ambient TPMS,
    and the intersection of them is taken as the porous scaffold~\cite{Yoo2011Porous,Yoo2012Heterogeneous,Yang2014Effective,Feng2018Porous}.
 In the second class of methods,
    a regular TPMS unit is transformed into each hexahedron of a hexahedron mesh model,
    thus generating a porous scaffold~\cite{Yoo2011Computer,Chen2018Porous,Shi2018A}.
 However, the first class of methods can generate incomplete TPMS units near
    the boundary of a volume mesh model,
    and the second class of methods may cause discontinuities between two adjacent TPMS units.
 Moreover, in both method classes,
    porosity and pore size are difficult to control.

 In this study, we developed a method for generating heterogenous porous
    scaffolds in a trivariate B-spline solid (TBSS) with TPMS designed in the parametric domain of the TBSS.
 We also developed a porous scaffold storage format that saves significant storage space.
 Specifically, given a TBSS,
    a threshold distribution field (TDF) is first constructed in the cubic parameter domain of the TBSS.
 Based on the TDF,
    a TPMS is generated in the parameter domain.
 Finally, by mapping the TPMS in the parameter domain to the TBSS,
    a porous scaffold is produced in the TBSS.
 In addition, the TDF can be modified locally in the parameter domain to
    improve the engineering performance of the porous scaffold.
 All of the TPMS units generated in the porous scaffold is complete,
    and adjacent TPMS units are continuously stitched.
 To summarize, the main contributions of this study are as follows:
 \begin{itemize}
    \item A TBSS is employed to generate porous scaffolds,
        which ensures completeness of TPMS units,
        and continuity between adjacent TPMS units.
    \item Porosity is easy to control using the TDF defined in the parametric domain of the TBSS.
    \item A storage format for porous scaffolds is designed based on the TDF in the parametric domain,
        which saves significantly storage space.
 \end{itemize}

 The remainder of this paper is organized as follows.
 In Section~\ref{sec:related_work}, we review related work on
    porous scaffold design and TBSS generation methods.
 In Section~\ref{sec:theories},
    preliminaries on TBSS and TPMS are introduced.
 Moreover,
    the heterogeneous porous scaffold generation method using a TBSS and TDF in its parameter domain is presented in detail in Section~\ref{sec:method}.
 In Section~\ref{sec:implementations},
    some experimental examples are presented to demonstrate the effectiveness and efficiency of the developed method.
 Finally, Section~\ref{sec:conclusion} concludes the paper.

\subsection{Related work}
\label{sec:related_work}

 In this section, we review some related work on porous scaffold design and TBSS
    generation methods.

 \textbf{Porous scaffold design:}
 In recent years,
    TPMS has been of special interest to the porous scaffold design community owing to its excellent properties,
    and many scaffold design methods have been developed based on TPMS.
 Rajagopalan and Robb~\cite{Rajagopalan2006Schwarz} made the first attempt to design tissue scaffolds based on Schwarz's primitive minimal surface, which is a type of TPMS.
 Moreover, the other two typical TPMSs (Schwarz's diamond surface and
    Schoen's gyroid surface) are constructed by employing K3DSurf software to design tissue scaffolds~\cite{Melchels2010Mathematically},
    which achieve a gradient change of gyroid structure in terms of pore size by adding a linear equation in $z$ into the TPMS function.

 Yoo~\cite{Yoo2011Computer} developed a method for generating porous
    scaffolds in a hexahedral mesh model,
    where the coordinate interpolation algorithm and shape function method are employed to map TPMS units to hexahedron elements to generate tissue scaffolds.
 To reduce the time consume in trimming and re-meshing process of Boolean operations,
    a tissue scaffold design method based on a hybrid method of distance field and TPMS was further proposed in~\cite{Yoo2011Porous}.
 Moreover, to make the porosity easier to control in the design a heterogeneous porous scaffold,
    Yoo~\cite{Yoo2012Heterogeneous} introduced a method based on an implicit interpolation algorithm that uses the thin-plate radial basis function.
 Similar to the method of Yoo~\cite{Yoo2012Heterogeneous},
    Yang et al.~\cite{Yang2014Effective} introduced the sigmoid function and Gaussian radial basis function to design tissue scaffolds.
 However, the hexahedral mesh based porous scaffold generation methods cannot
    ensure continuity between adjacent TPMS units.

 Recently, in consideration of the increasing attention towards gradient porous scaffolds,
    Shi et al.~\cite{Shi2018A} utilized the TPMS and sigmoid function to generate functional gradient bionic porous scaffolds from Micro-CT data reconstruction.
 Feng et al.~\cite{Feng2018Porous} proposed a method to design porous
     scaffold based on solid T-splines and TPMS,
    and analyzed the parameter influences on the volume specific surface area and porosity.
 In addition, a heterogenous methodology for modeling porous scaffolds using a parameterized hexahedral mesh and TPMS was developed by Chen et al.~\cite{Chen2018Porous}.

 \textbf{TBSS generation:}
 TBSS modeling methods are developed mainly for producing three dimensional
    physical domain in isogeometric analysis~\cite{Hughes2005Isogeometric}.
 Specifically, to analyze arterial blood flow through isogeometric analysis,
    Zhang et al.~\cite{zhang2007patient} introduced a skeleton-based method of generating trivariate non-uniform rational basis spline (NURBS) solids.
 In~\cite{martin2009volumetric}, a tetrahedral mesh model is parameterized
    through discrete volumetric harmonic functions and a cylinder-like TBSS is generated.
 Aigner et al.~\cite{Aigner2009Swept} proposed a variational framework for
     generating NURBS parameterizations of swept volumes using the given boundary conditions and guiding curves.
 Optimization approaches have been developed for filling
    boundary-represented models to produce TBSSs with positive Jacobian values~\cite{Wang2014An}.
 Moreover, a discrete volume parameterization method for tetrahedral mesh
    models and an iterative fitting algorithm have been presented for TBSS generation~\cite{Lin2015Constructing}.

\section{Preliminaries }
\label{sec:theories}

 Preliminaries on TBSS and TPMS are introduced in this section.

\subsection{TBSS}
 \label{subsec:tbss}

 A B-spline curve of order $p+1$ is formed by several piecewise
    polynomial curves of degree $p$,
    and a B-spline curve has $C^p$ continuity at its breakpoints~\cite{Piegl1997The}.
 A knot vector $U=\{u_0,u_1,\ldots,u_{m+p+1}\}$ is defined by a set of breakpoints $u_0 \leq u_1 \leq \cdots \leq u_{m+p+1}$.
 The associated B-spline basis functions $N_{i,p}(u)$ of degree $p$ are
    defined as follows:
 \begin{equation}
  \label{eq:basisfunction}
  \begin{aligned}
      &N_{i,0} =
        \begin{cases}
            1, & for \quad u_i \leq u < u_{i+1}, \\
            0, & otherwise,
        \end{cases}\\
      &N_{i,p}(u)=\frac{u-u_i}{u_{i+p}-u_i}N_{i,p-1}(u)+\frac{u_{i+p+1}-u}{u_{i+p+1}-u_{i+1}}N_{i+1,p-1}(u)
  \end{aligned}
 \end{equation}

 A TBSS of degree $(p,q,r)$ is a tensor product volume defined as
 \begin{equation}
  \label{eq:BsplineSolid}
  P(u,v,w)=\sum_{i=0}^{m}\sum_{j=0}^{n}\sum_{k=0}^{l}N_{i,p}(u)N_{j,q}(v)N_{k,r}(w)P_{ijk}
 \end{equation}
 where $P_{ijk},\ i=0,1,\cdots,m,\ j=0,1,\cdots,n,\ k=0,1,\cdots,l$ are control points in the $u$, $v$ and $w$ directions and
 $$N_{i,p}(u),N_{j,q}(v),N_{k,r}(w)$$
 are the B-spline basis functions of degree $p$ in the $u$ direction, degree $q$ in the $v$ direction, and degree $r$ in the $w$ direction.

 In this study,
    the input to our porous scaffold generation algorithm is a TBSS
    that represents geometry at a macro-structural scale.
 The TBSS can be generated either by fitting the mesh vertices of a
    tetrahedral mesh model~\cite{Lin2015Constructing},
    or filling a closed triangular mesh model~\cite{Wang2014An}.

 \subsection{TPMS}
 \label{subsec:tpms}

 A TPMS is an implicit surface that is infinite and periodic in three independent directions.
 There are several ways to evaluate a TPMS,
    and the most frequently employed approach is to approximate the TPMS using a periodic nodal surface defined by a Fourier series~\cite{Gandy2001Nodal},
 \begin{equation}
  \label{eq:TPMS}
  \psi(r)=\sum_kA_kcos[2\pi (\bm{h}_k\cdot \bm{r})/\lambda_k - P_k]=C,
 \end{equation}
 where $\bm{r}$ is the location vector in Euclidean space,
    $A_k$ is amplitude,
    $\bm{h}_k$ is the $k^{th}$ lattice vector in reciprocal space,
    $\lambda_k$ is the wavelength of the period,
    $P_k$ is phase shift,
    and $C$ is a threshold constant.
 Please refer to~\cite{Yan2007Periodic} for more details on the abovementioned parameters.
 The nodal approximations of P, D, G, and I-WP types of TPMSs,
    which were presented in ~\cite{Schnering1991Nodal},
    are listed in Table~\ref{tbl:nodal}.
 The valid range of $C$ guarantees that the implicit surface is complete.

  \begin{table*}[!htb]
  \centering
  \caption{Nodal approximations of typical TPMS units.}
  \label{tbl:nodal}
  \begin{tabular}{ l|l|c }
   \hline
    TPMS & Nodal approximations & Valid range of $C$\\
   \hline
    Schwarz's P Surface   & $\psi_P(x,y,z)=cos(\omega_xx)+cos(\omega_yy)+cos(\omega_zz)=C$  & $[-0.8,0.8]$\\
    Schwarz's D Surface   & $\psi_D(x,y,z)=cos(\omega_xx)cos(\omega_yy)cos(\omega_zz)-sin(\omega_xx)sin(\omega_yy)sin(\omega_zz)=C$  & $[-0.6,0.6]$\\
    Schoen's G Surface    & $\psi_G(x,y,z)=sin(\omega_xx)cos(\omega_yy)+sin(\omega_yy)cos(\omega_zz)+sin(\omega_zz)cos(\omega_xx)=C$  & $[-0.8,0.8]$\\
    Schoen's I-WP Surface & \tabincell{l}{$\psi_{I-WP}(x,y,z)=2[cos(\omega_xx)cos(\omega_yy)+cos(\omega_yy)cos(\omega_zz)+cos(\omega_zz)cos(\omega_xx)]$ \\$-[cos(2\omega_xx)+cos(2\omega_yy)+cos(2\omega_zz)]=C$}  & $[-2.0,2.0]$\\
   \hline
  \end{tabular}
 \end{table*}

\begin{figure*}[!htb]
  \begin{center}
  \subfigure[]{
    \label{subfig:Psurface}
    \includegraphics[width=0.18\textwidth]{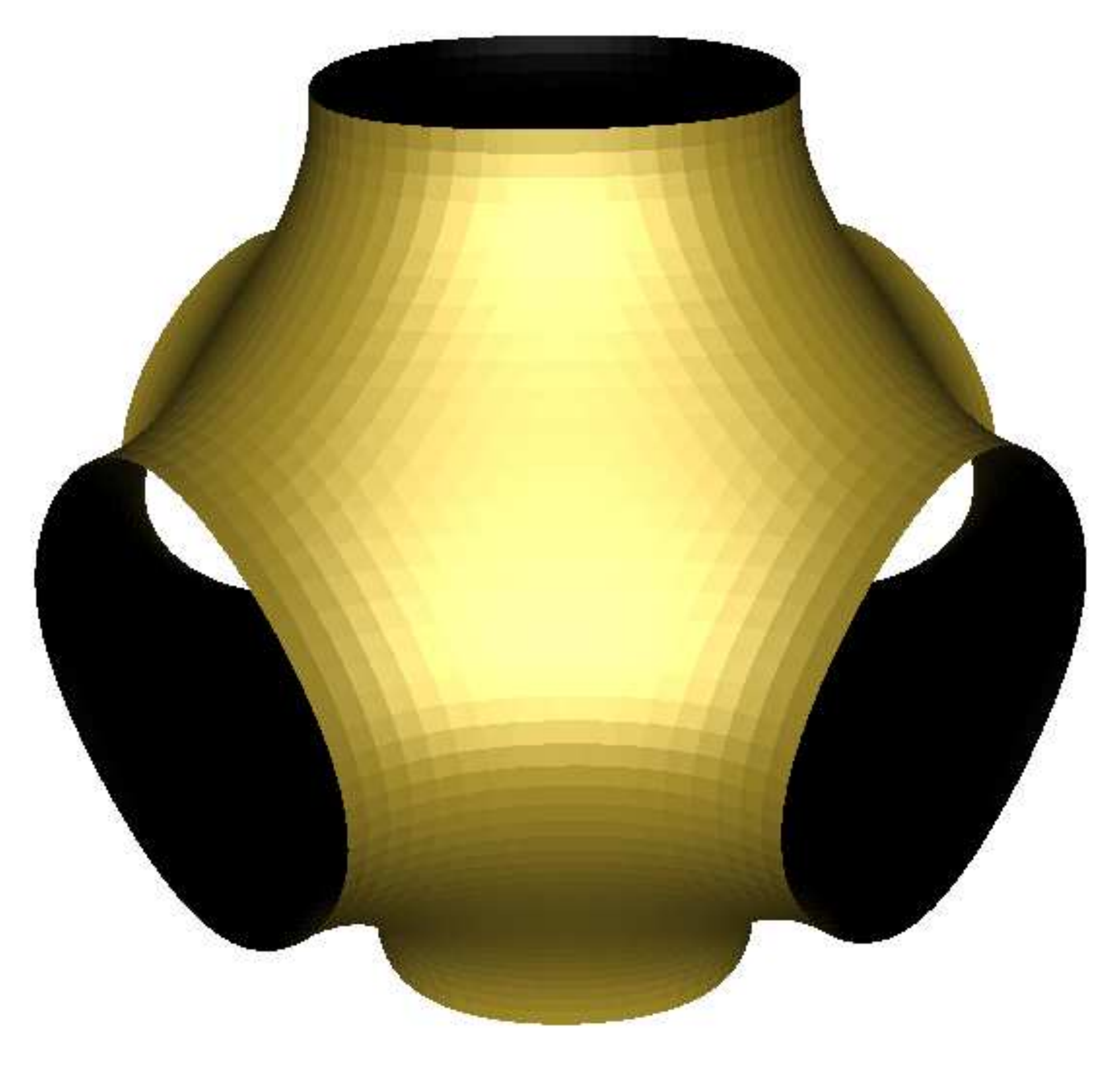}}
  \subfigure[]{
    \label{subfig:Dsurface}
    \includegraphics[width=0.18\textwidth]{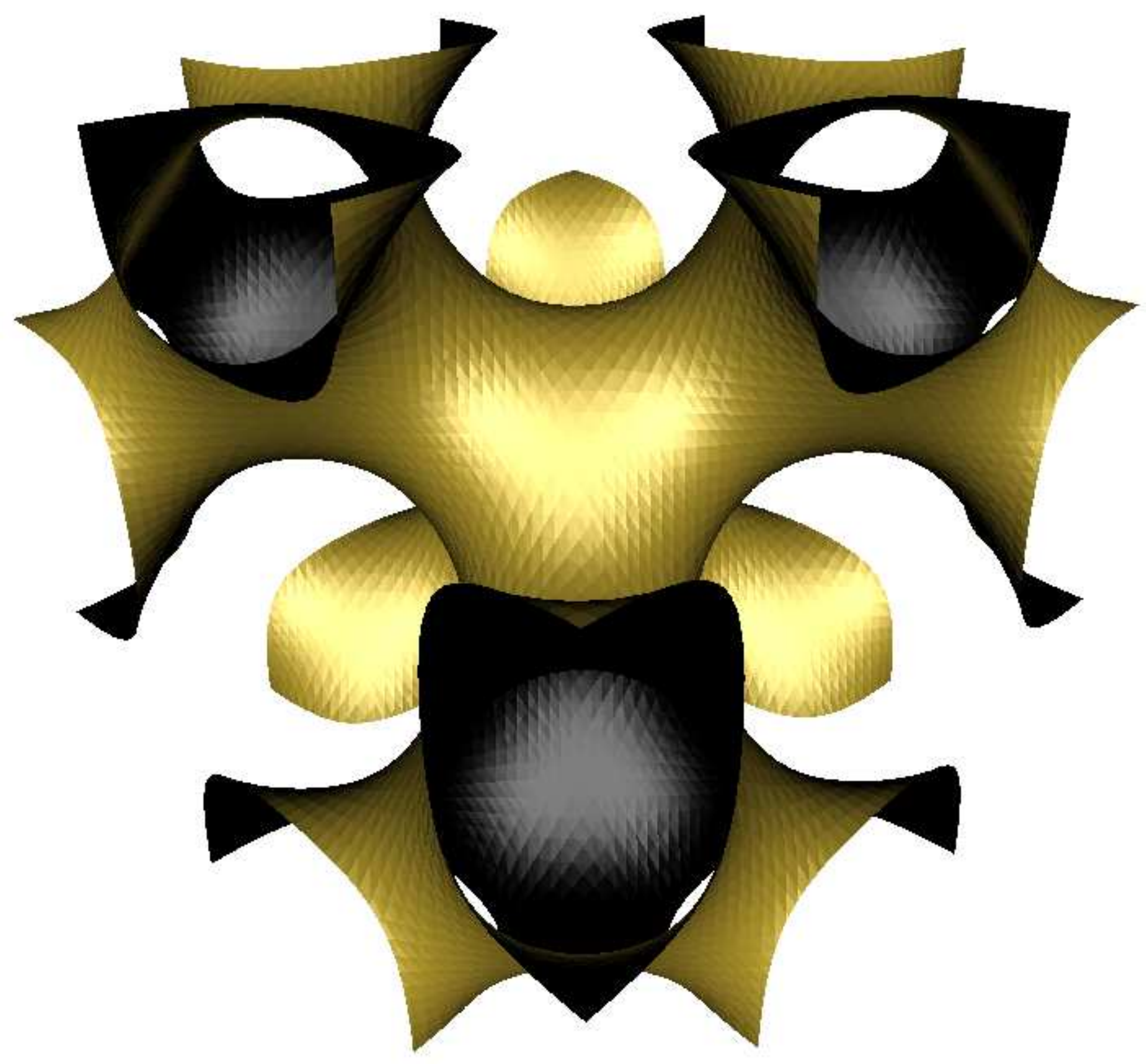}}
  \subfigure[]{
    \label{subfig:Gsurface}
    \includegraphics[width=0.18\textwidth]{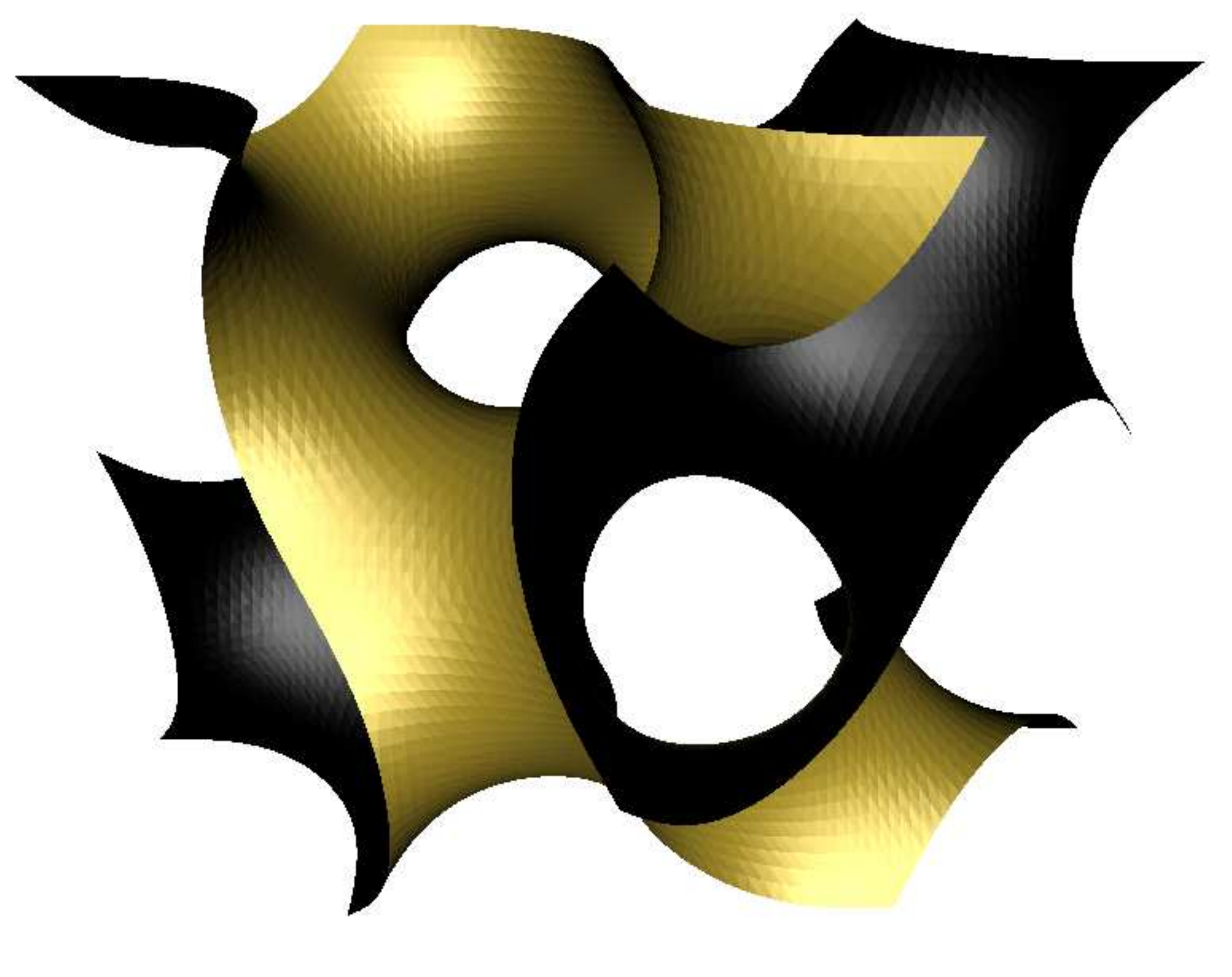}}
  \subfigure[]{
    \label{subfig:IWPsurface}
    \includegraphics[width=0.18\textwidth]{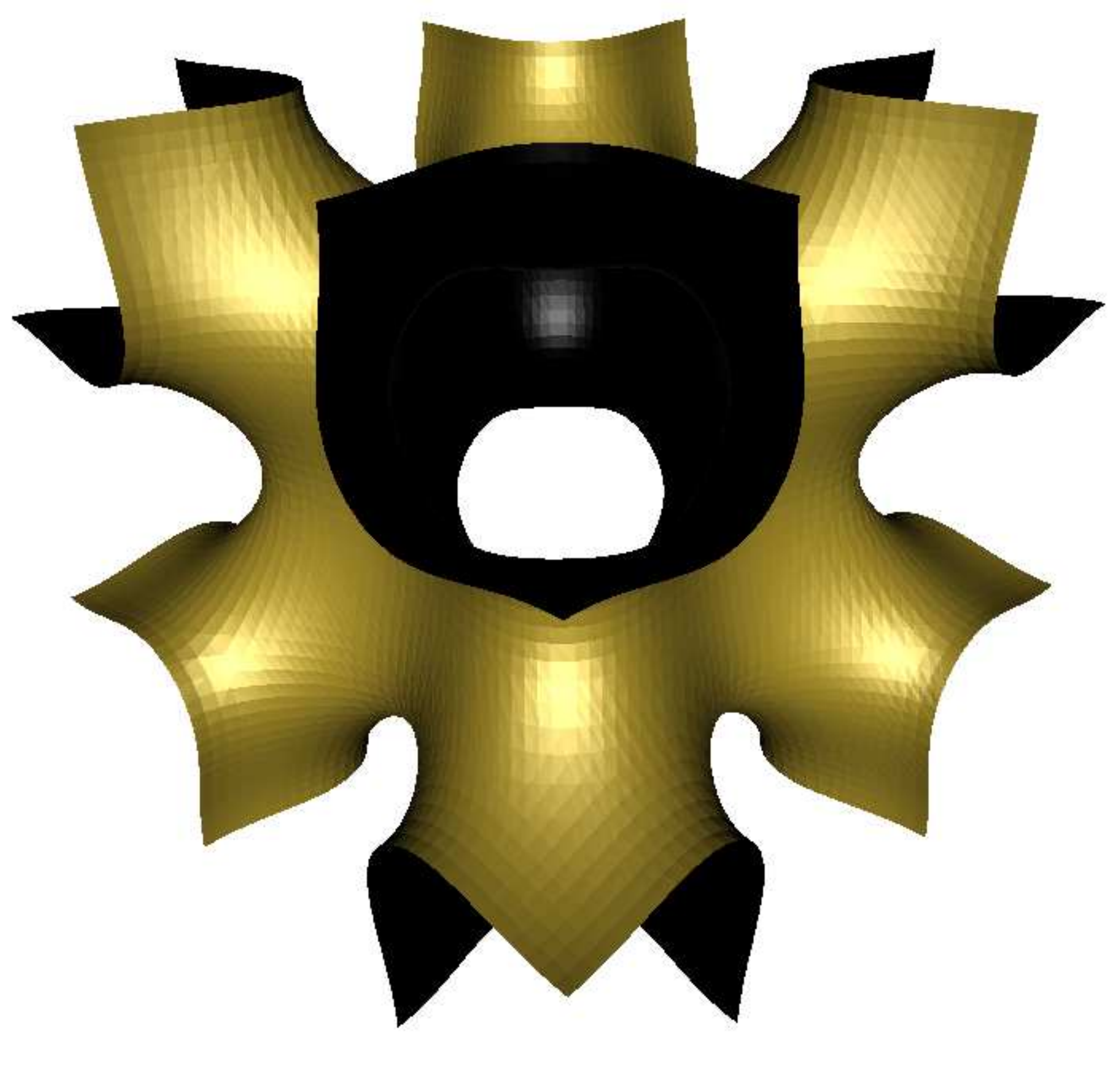}}
      \caption
      {
      \small
        Four types of TPMS units.
        (a) P-type. (b) D-type. (c) G-type. (d) I-WP-type.
      }
   \label{fig:TPMSunits}
  \end{center}
\end{figure*}

 In TPMS-based porous scaffold design methods,
    the threshold value $C$~\pref{eq:TPMS} controls the porosity,
    and the coefficients $\omega_x$, $\omega_y$, and $\omega_z$ (refer to Table~\ref{tbl:nodal}) which affect the period of the TPMS,
    are called \emph{period coefficients}.
 The effects of the two types of parameters in porous scaffold design
    have been discussed in detail in the literature~\cite{Feng2018Porous}.
 Additionally, in this study, the marching tetrahedra (MT) algorithm~\cite{Doi1991An} is
    employed to extract the TPMS (shown in Fig.~\ref{fig:TPMSunits}).

\section{Methodology of porous scaffold design}
\label{sec:method}

\begin{figure}[!htb]
  \centering
  \includegraphics[width=0.45\textwidth]{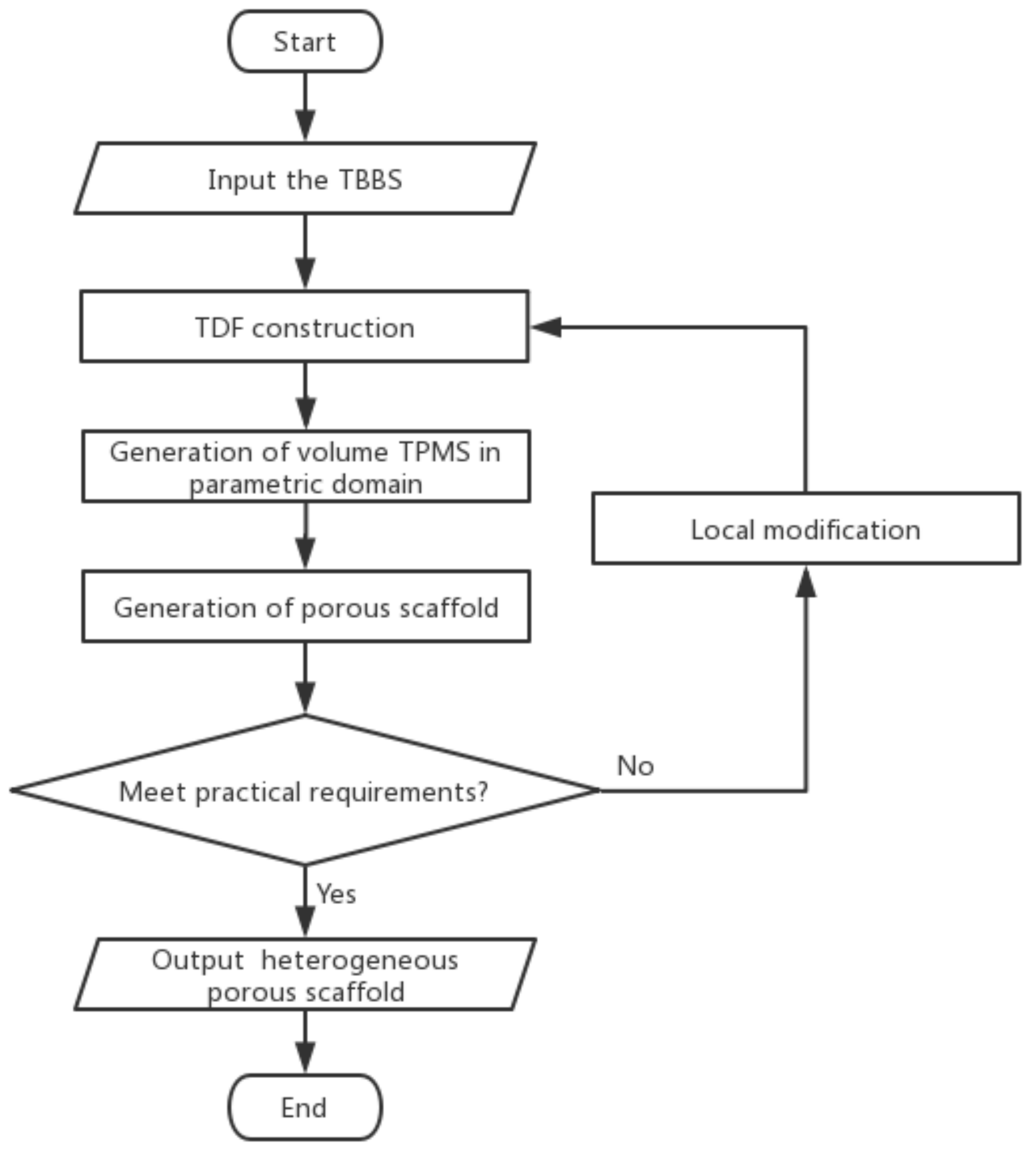}\\
  \caption{\small The procedure of the heterogenous porous scaffold generation method.}
  \label{fig:procedure}
\end{figure}

 The whole procedure of the heterogenous porous scaffold generation method
    based on TBSS and TPMS in its parametric domain is illustrated in Fig.\ref{fig:procedure}.
 Specifically,
    given a TBSS as input,
    we design a method for constructing the TDF in its cubic parameter domain.
 Based on the TDF, a TPMS is generated in the parameter domain using
    MT algorithm~\cite{Doi1991An}.
 Moreover, by mapping the TPMS in the parameter domain to the TBSS by
    the TBSS function~\pref{eq:BsplineSolid},
    a porous scaffold with compeleteness and continuity is produced in the TBSS.
 If the porous scaffold in the TBSS does not meet the engineering
    requirements,
    the TDF can be locally modified in the parameter domain,
    and the porous scaffold in the TBSS can be rebuilt.
 Finally, we develop a storage format for the porous scaffold based on the TDF
    in the parametric domain,
    which saves significant storage space.
 The details of the porous scaffold design method are elucidated in
    the following sections.

\subsection{Generation of TPMS and volume TPMS structures}
\label{subsec:tpms_generation}

 After the TDF is constructed in the parameter domain,
    and the period coefficients are assigned,
    the TPMS $\psi = C$ in the parameter domain can be defined such that it is polygonized into a triangular mesh.
 As a type of iso-surface, a TPMS can be polygonized by many algorithms
    such as iterative refinement~\cite{Rajagopalan2006Schwarz}, Delaunay triangulation~\cite{George1998Delaunay},
    marching cubes~\cite{Lorensen1987Marching} and MT~\cite{Doi1991An}.
 To avoid ambiguous connection problems~\cite{Newman2006A}
    and simplify the intersection types,
   the MT algorithm is adopted to polygonize the TPMS in the parametric domain of a TBSS.
 For this purpose, the cubic parameter domain is uniformly divided into
    a grid.
 In our implementation,
    to balance the accuracy and storage cost of the porous scaffold,
    the parametric domain is divided into a $100\times100\times100$ grid.

\begin{figure*}[!htb]
  \begin{center}
  \subfigure[]{
    \label{subfig:Ppore_units}
    \includegraphics[width=0.18\textwidth]{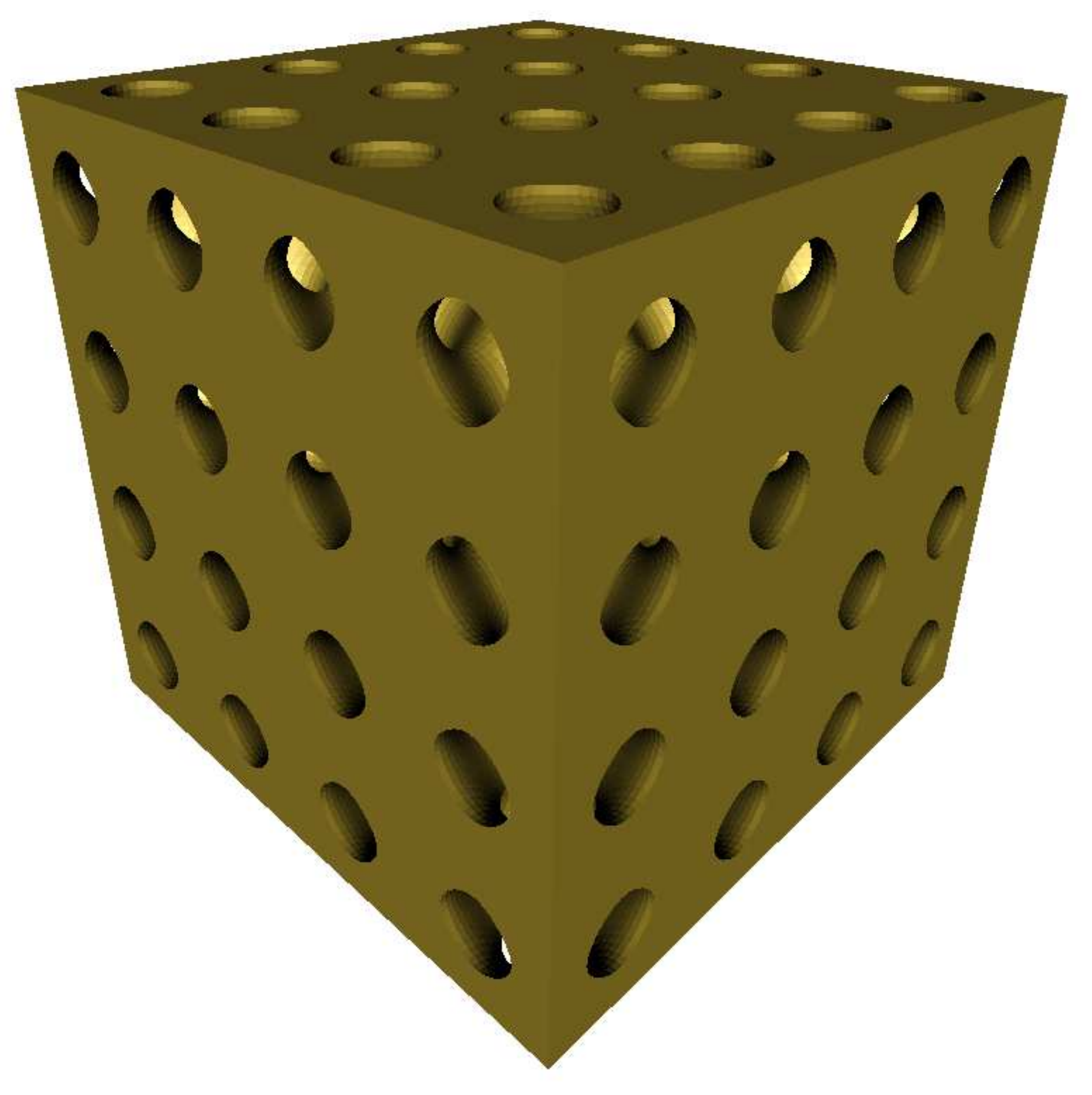}}
  \subfigure[]{
    \label{subfig:Dpore_units}
    \includegraphics[width=0.18\textwidth]{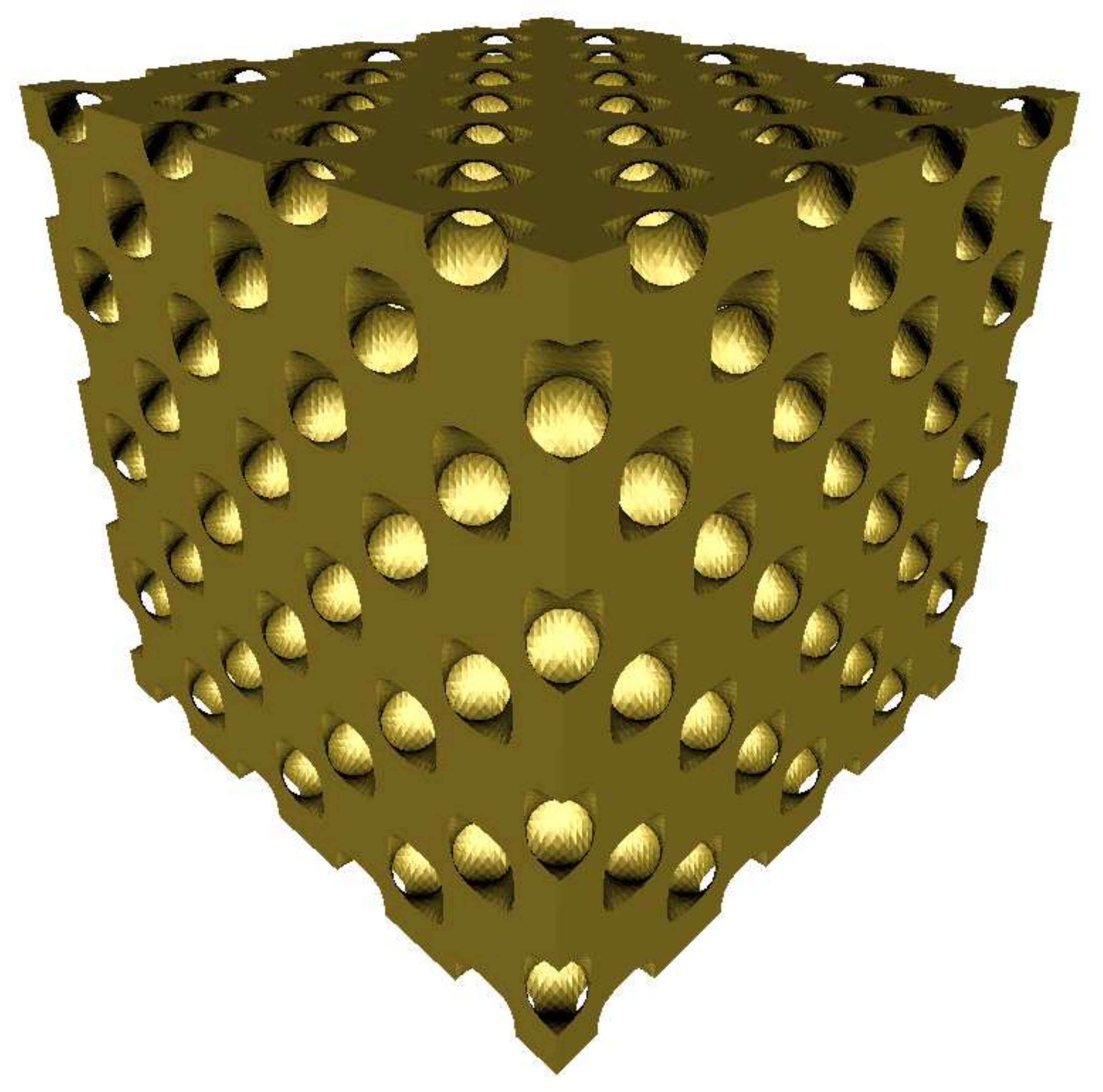}}
  \subfigure[]{
    \label{subfig:Gpore_units}
    \includegraphics[width=0.18\textwidth]{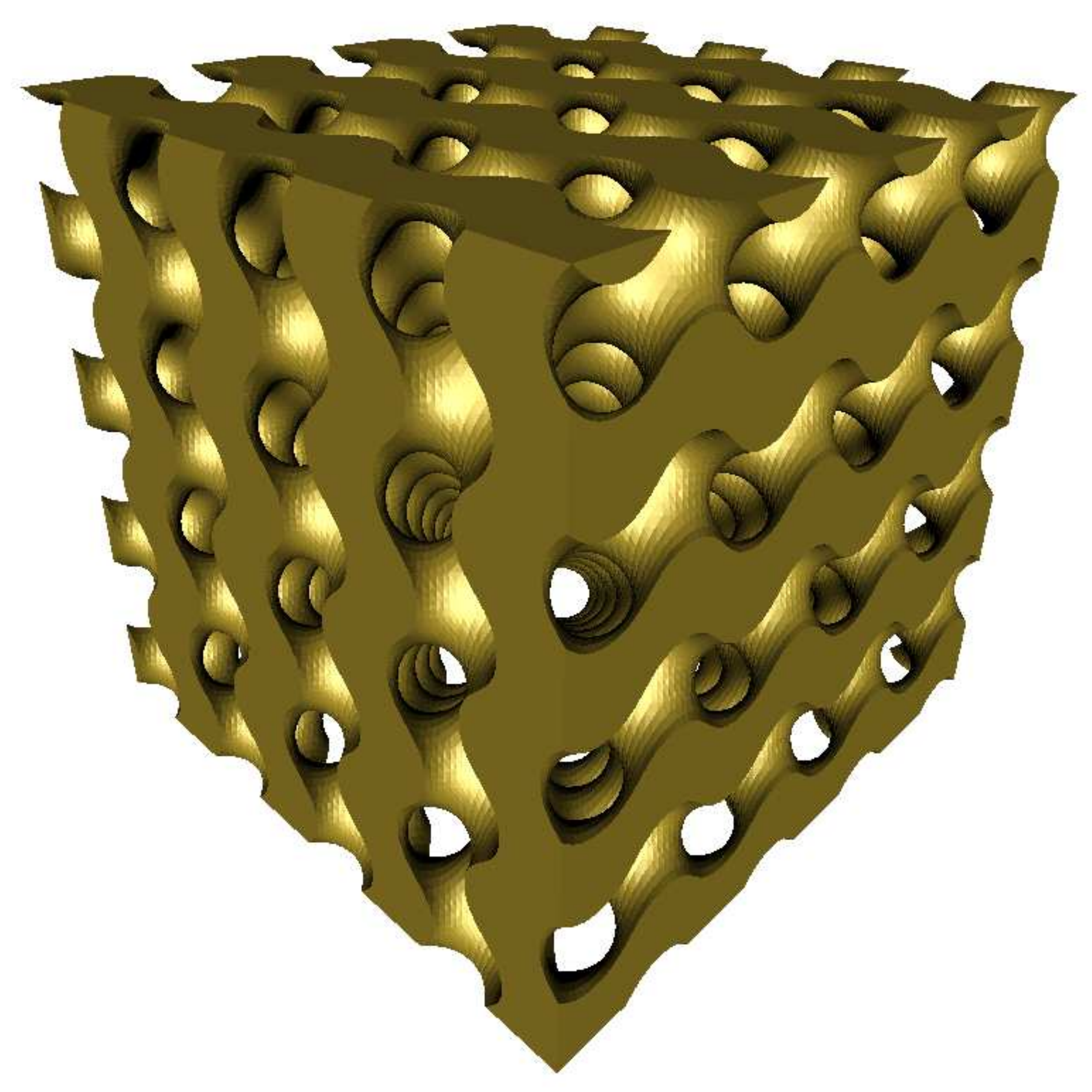}}
  \subfigure[]{
    \label{subfig:IWPpore_units}
    \includegraphics[width=0.18\textwidth]{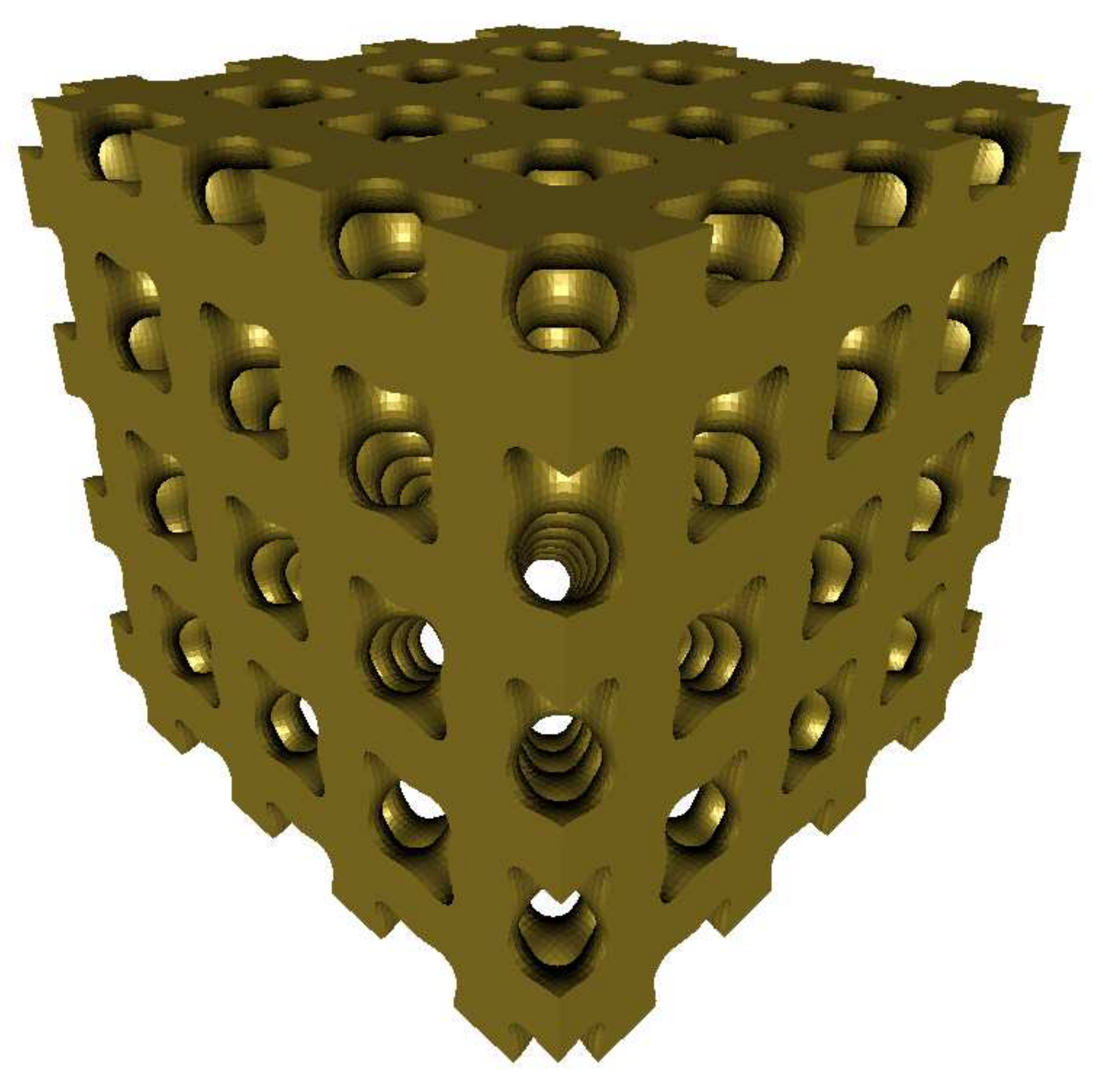}}
    \\
  \subfigure[]{
    \label{subfig:Prod_units}
    \includegraphics[width=0.18\textwidth]{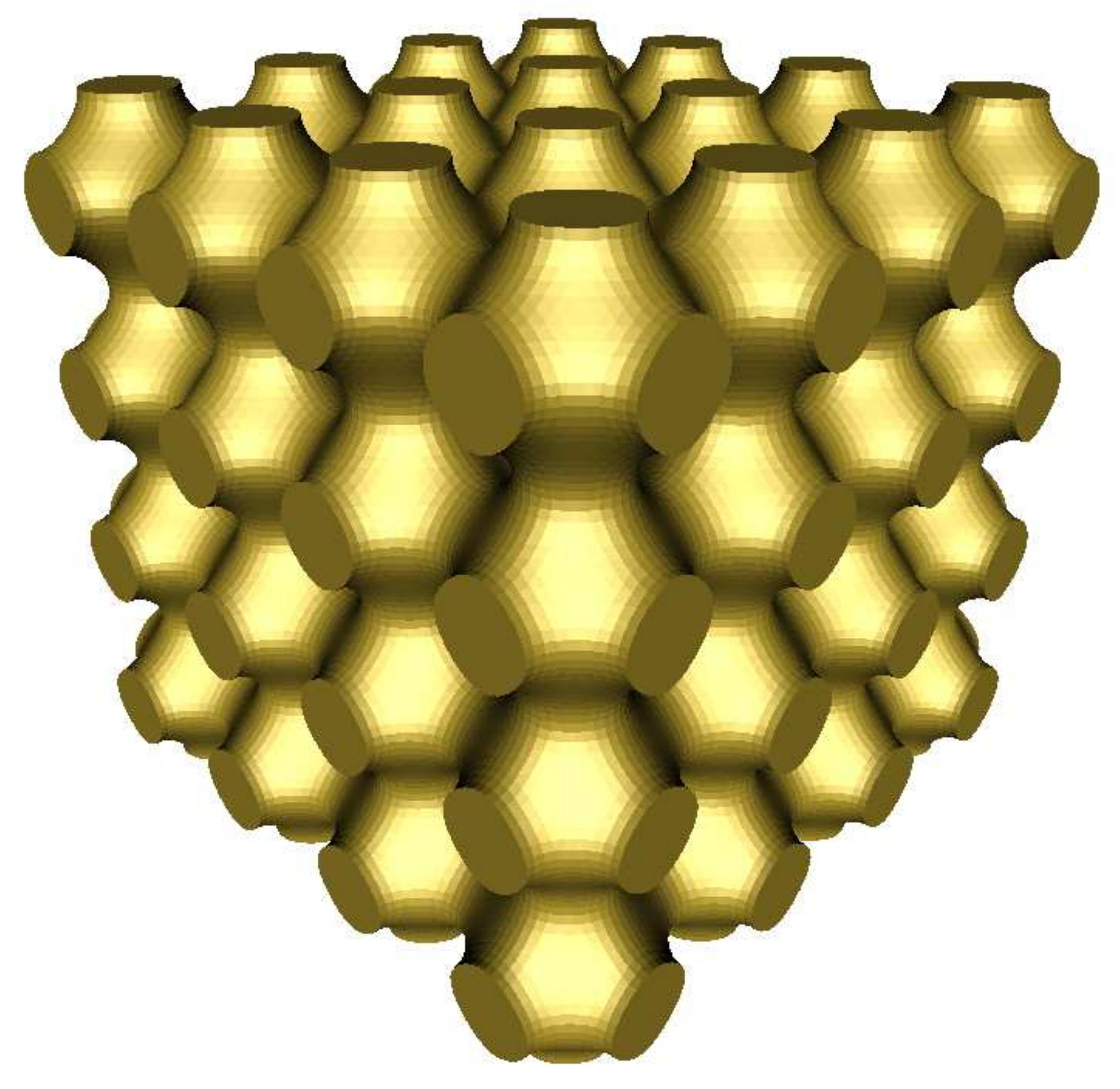}}
  \subfigure[]{
    \label{subfig:Drod_units}
    \includegraphics[width=0.18\textwidth]{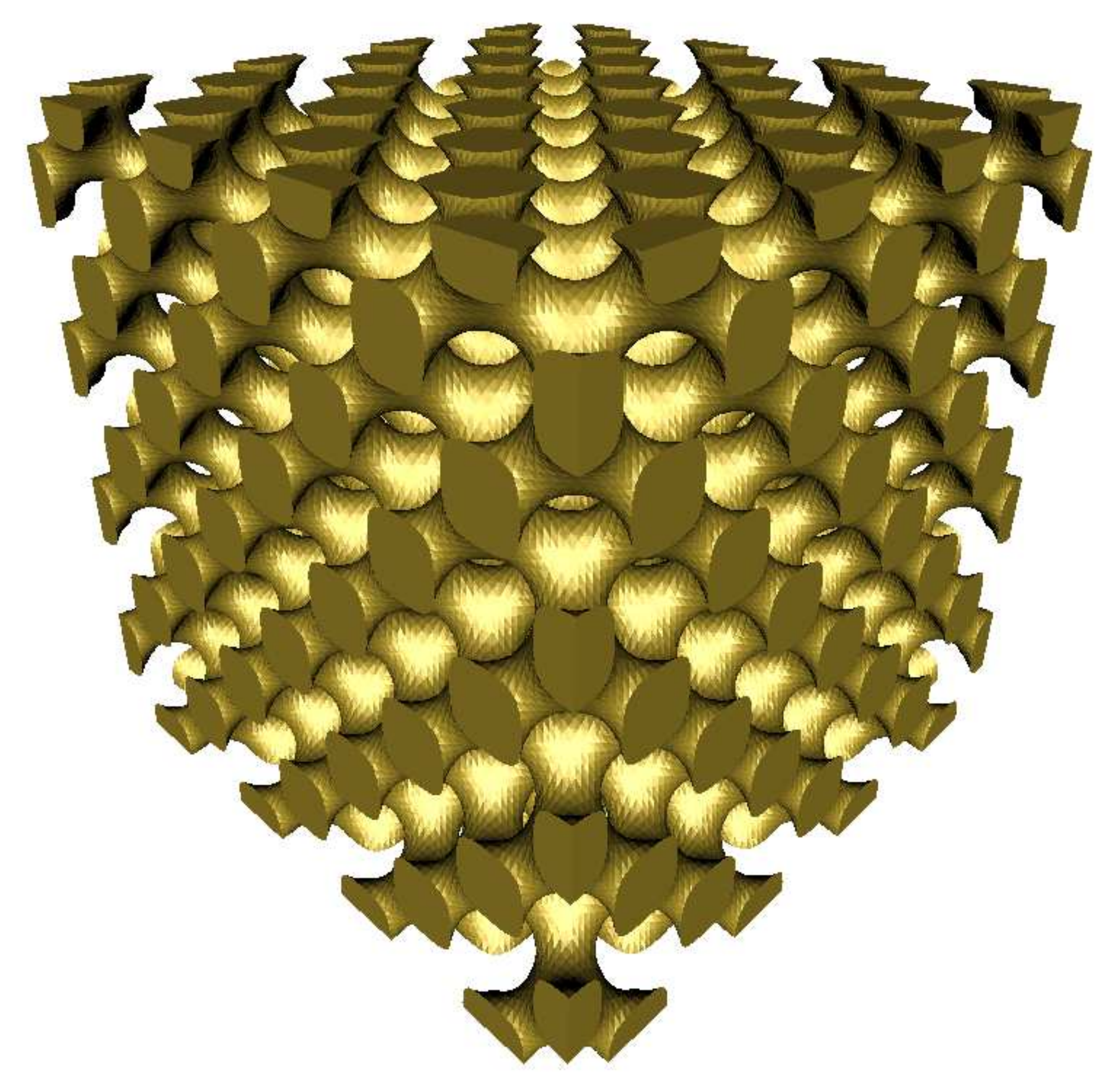}}
  \subfigure[]{
    \label{subfig:Grod_units}
    \includegraphics[width=0.18\textwidth]{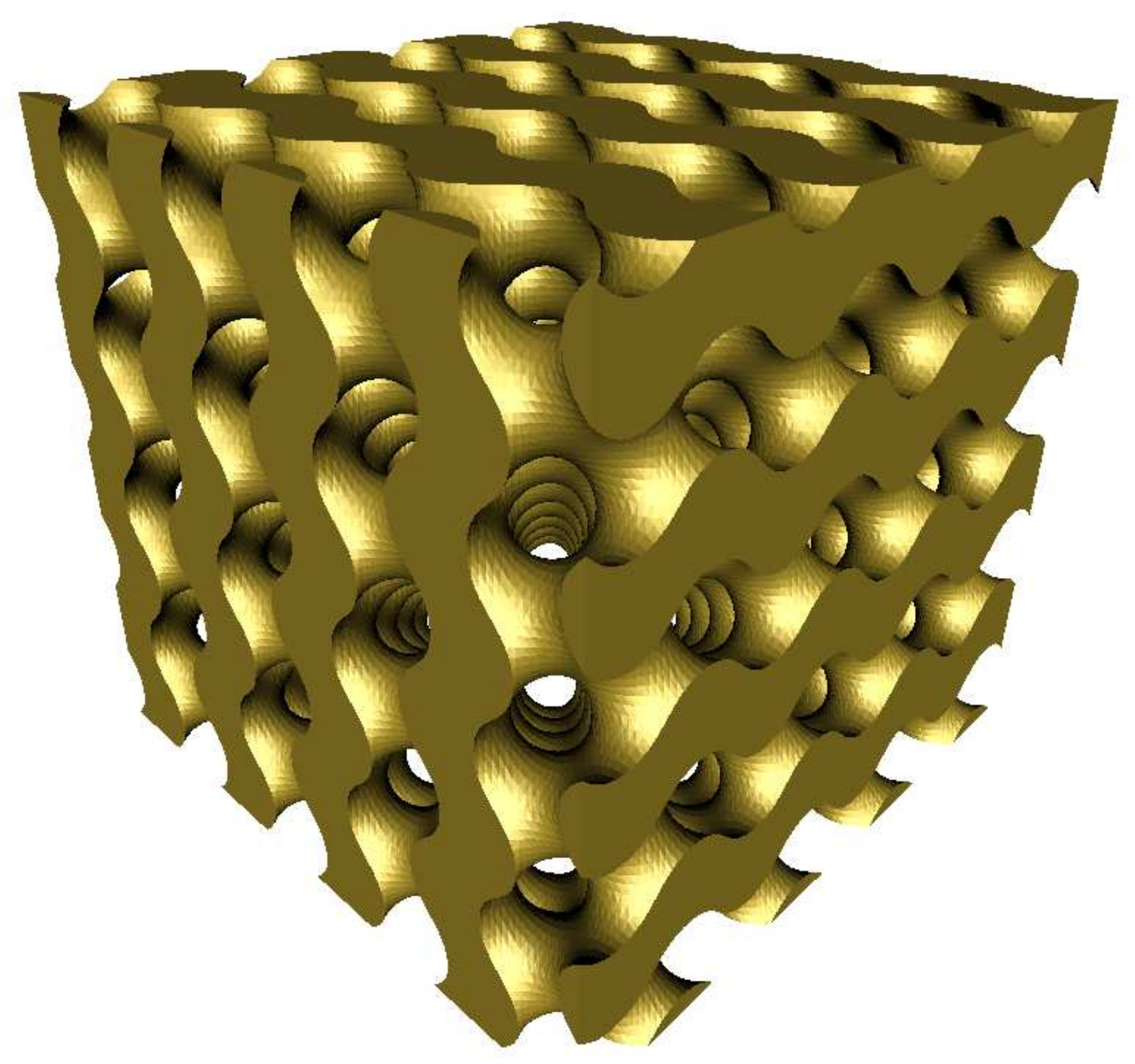}}
  \subfigure[]{
    \label{subfig:IWProd_units}
    \includegraphics[width=0.18\textwidth]{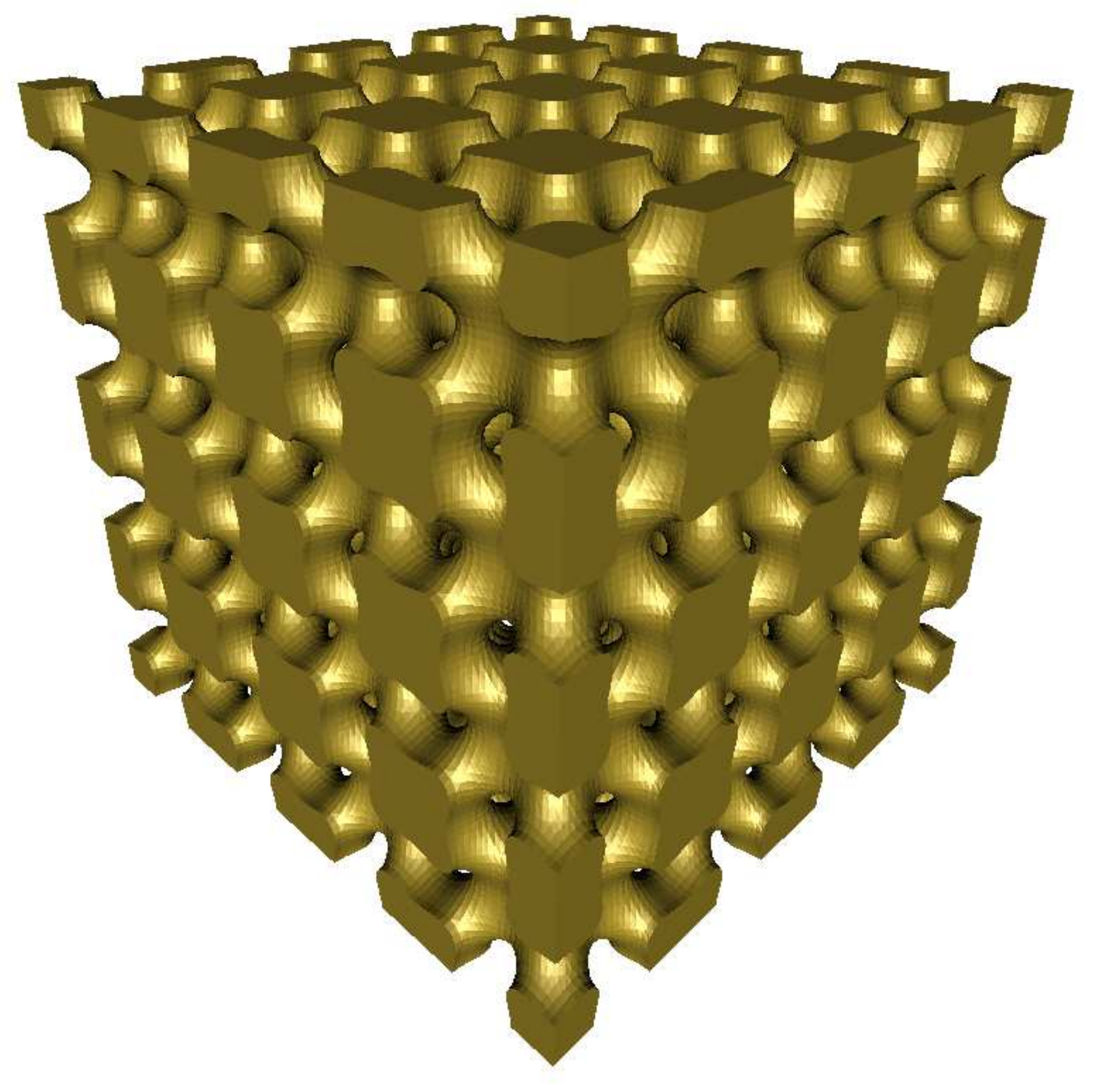}}
    \\
  \subfigure[]{
    \label{subfig:Psheet_units}
    \includegraphics[width=0.18\textwidth]{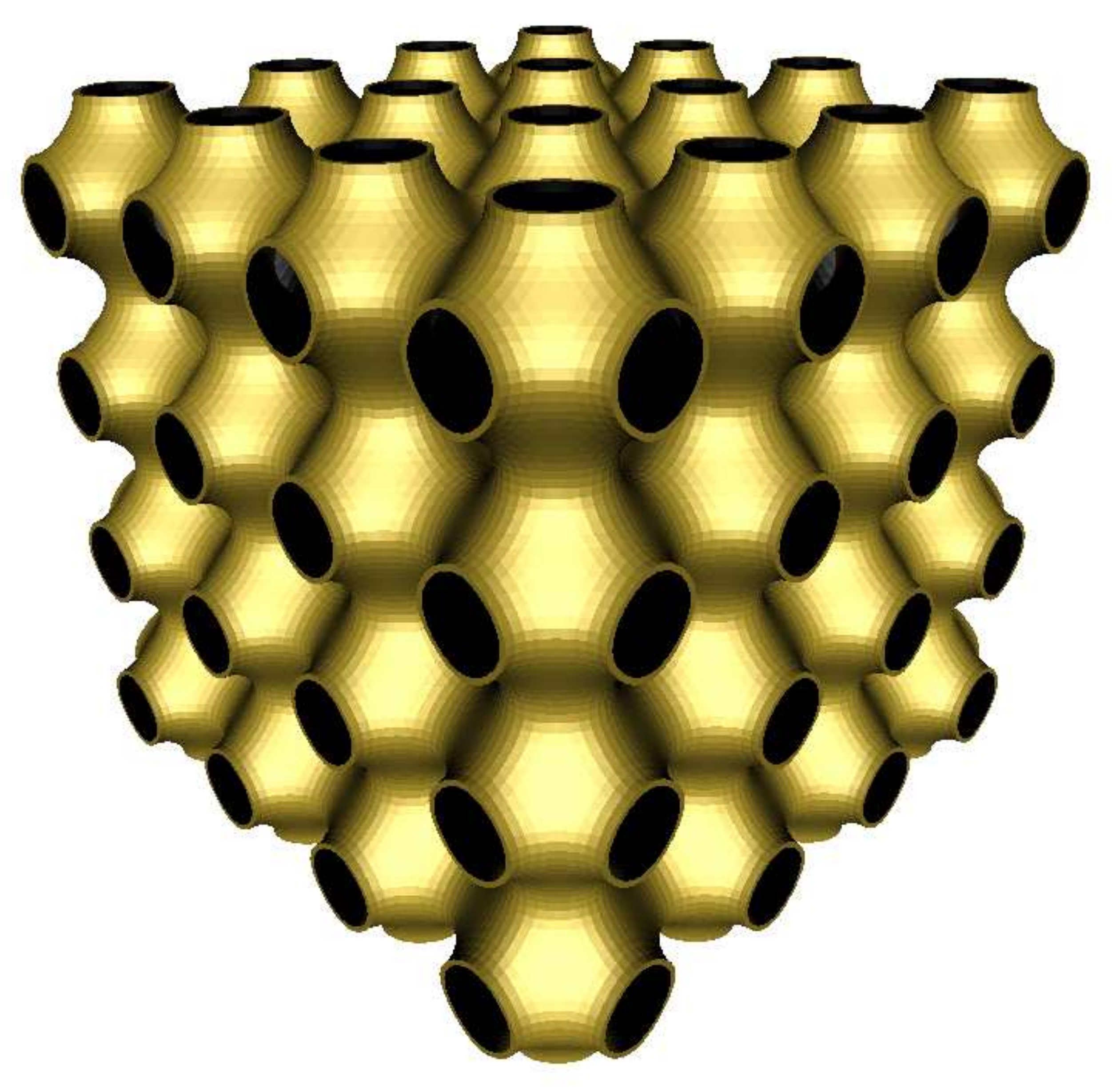}}
  \subfigure[]{
    \label{subfig:Dsheet_units}
    \includegraphics[width=0.18\textwidth]{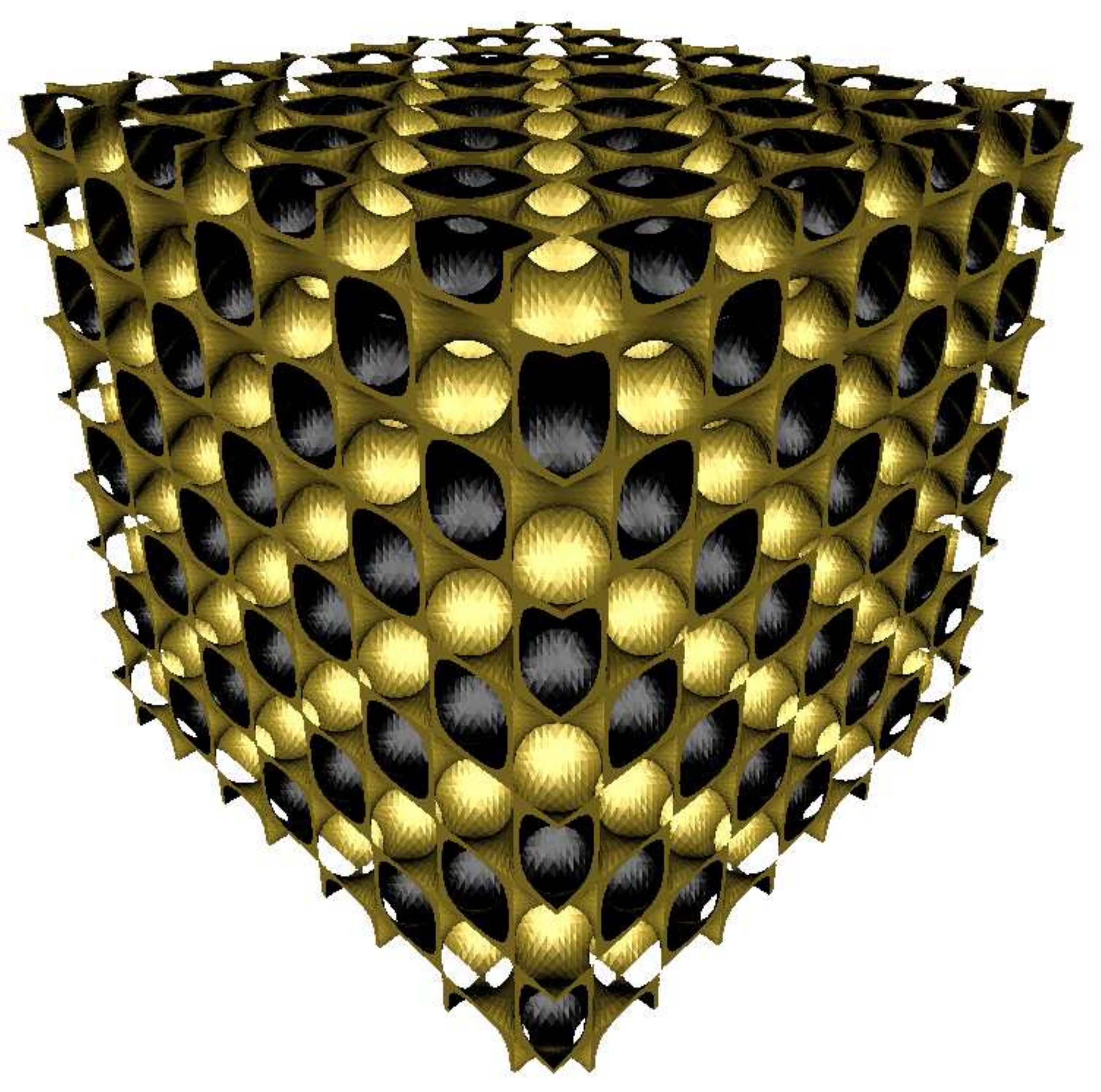}}
  \subfigure[]{
    \label{subfig:Gsheeet_units}
    \includegraphics[width=0.18\textwidth]{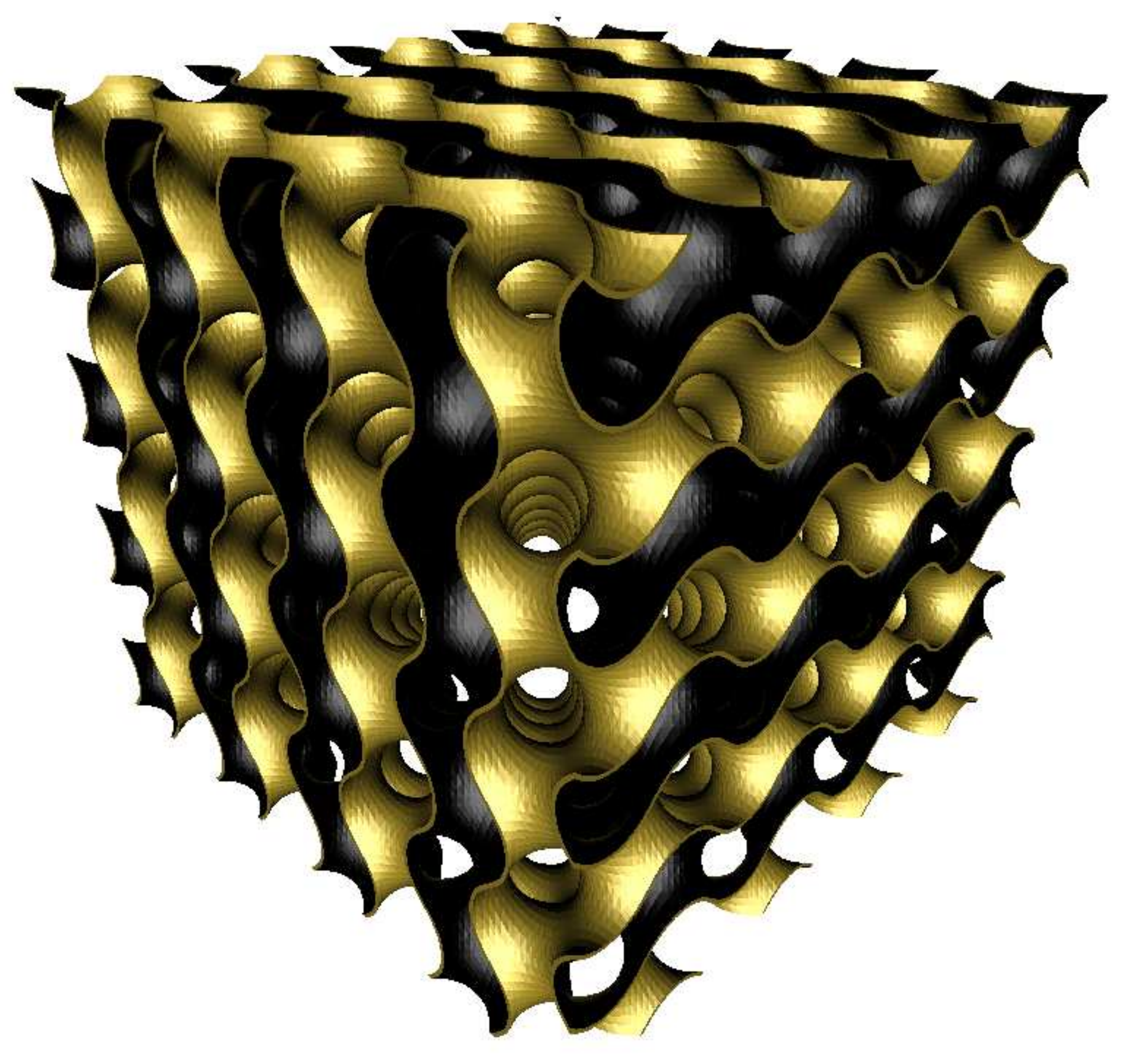}}
  \subfigure[]{
    \label{subfig:IWPsheet_units}
    \includegraphics[width=0.18\textwidth]{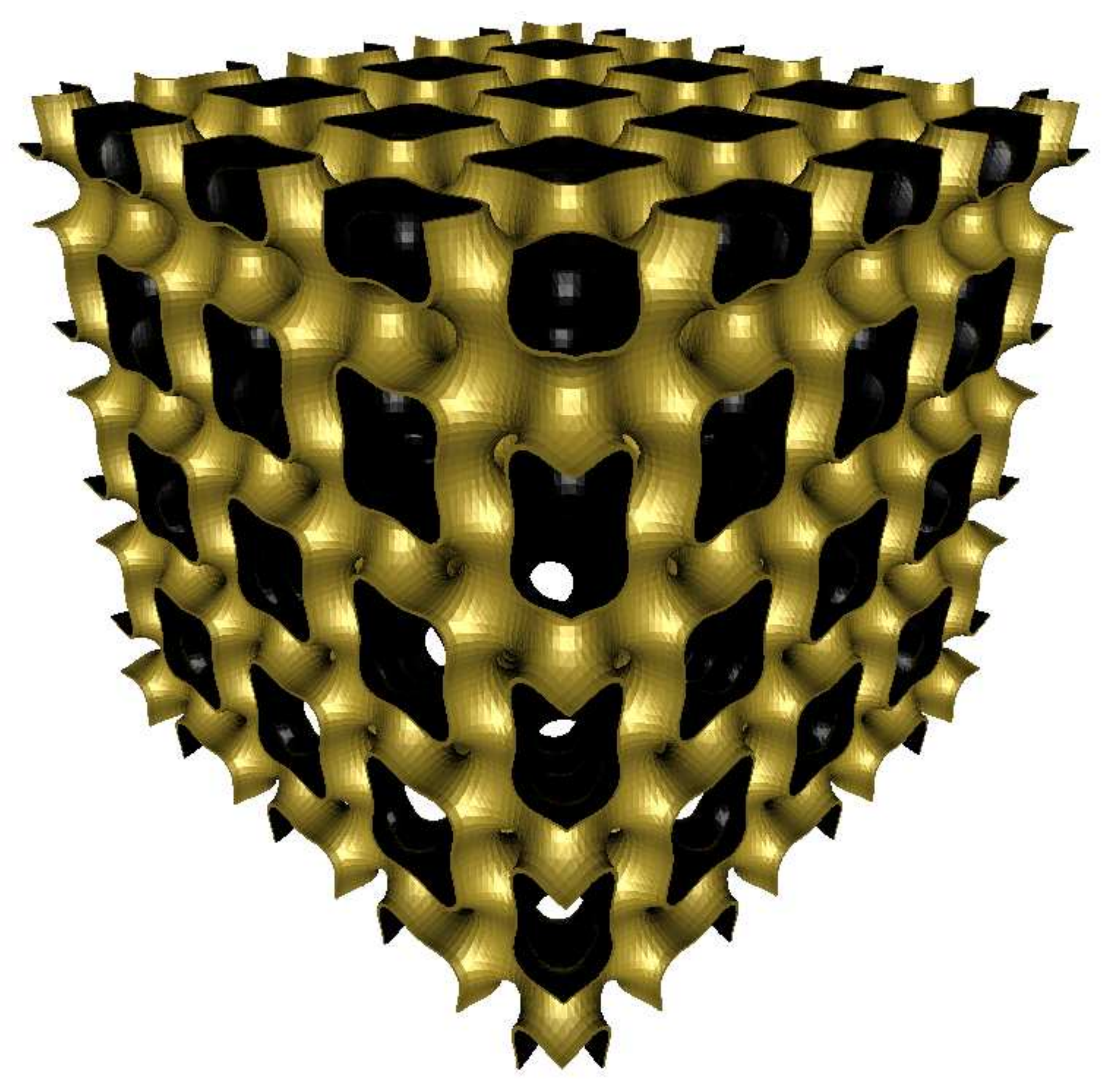}}
      \caption
      {\small
      Three types of volume TPMS structures for the four types of TPMS units.
      (a)(b)(c)(d) Pore structures for the P-type, D-type, G-type and  I-WP-type TPMSs.
      (e)(f)(g)(h) Rod structures for the P-type, D-type, G-type and  I-WP-type TPMSs.
      (i)(j)(k)(l) Sheet structures for the P-type, D-type, G-type and  I-WP-type TPMSs.
      }
   \label{fig:Pstructure}
  \end{center}
\end{figure*}

 Moreover, we define three types of \emph{volume TPMS structures}
    (refer to Fig.~\ref{fig:Pstructure}):
 \begin{itemize}
      \item {\it pore structure} represented by $\psi \geq C$,
      \item {\it rod structure} represented by $\psi \leq C$,
      \item {\it sheet structure} represented by $C - \epsilon \leq \psi \leq C$.
 \end{itemize}
 However, the triangular meshes of the three types of volume TPMS structures,
    generated by the polygonization,
    are open on the six boundary faces of the parameter domain,
    but they should be closed to form a solid.
 Take the pore structure ($\psi \geq C$) as an example.
 In the polygonization procedure of the TPMS $\psi=C$ by the MT algorithm,
    the triangles on the boundary faces of the parameter domain are categorized into two classes
    by the iso-value curve $\psi = C$ on the boundary faces:
    outside triangles, where the values of $\psi$ at the vertices of these triangles are larger than or equal to $C$,
    and inside triangles, where the values of $\psi$ at the vertices of these triangles are smaller than or equal to $C$,
 Therefore, the pore structure can be closed by adding the outside
    triangles into the triangular mesh generated by polygonizing $\psi=C$.
 The other two types of volume TPMS structures can be closed in
    similar ways.
 In Fig.~\ref{fig:Pstructure}, the three types of volume TPMS structures
    are illustrated.

 \subsection{TDF construction}
 \label{subsec:tdf_construction}

\begin{figure}[!htb]
  \centering
  \includegraphics[width=0.5\textwidth]{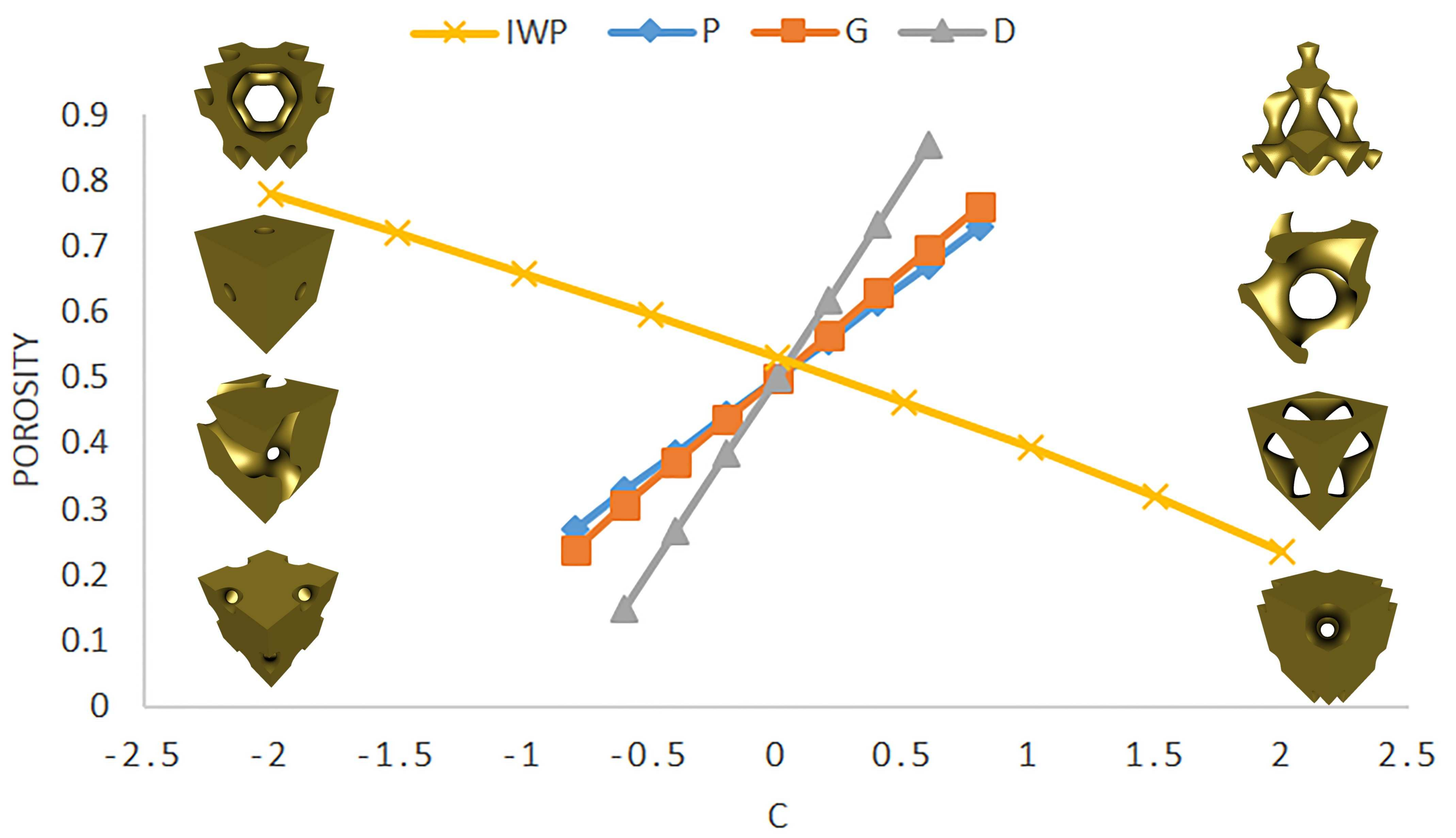}\\
  \caption{\small Relationship between the threshold $C$ and the porosity of the four types of TPMSs based on pore structures.}
  \label{fig:pore_relationship}
\end{figure}

\begin{figure}[!htb]
  \centering
  \includegraphics[width=0.5\textwidth]{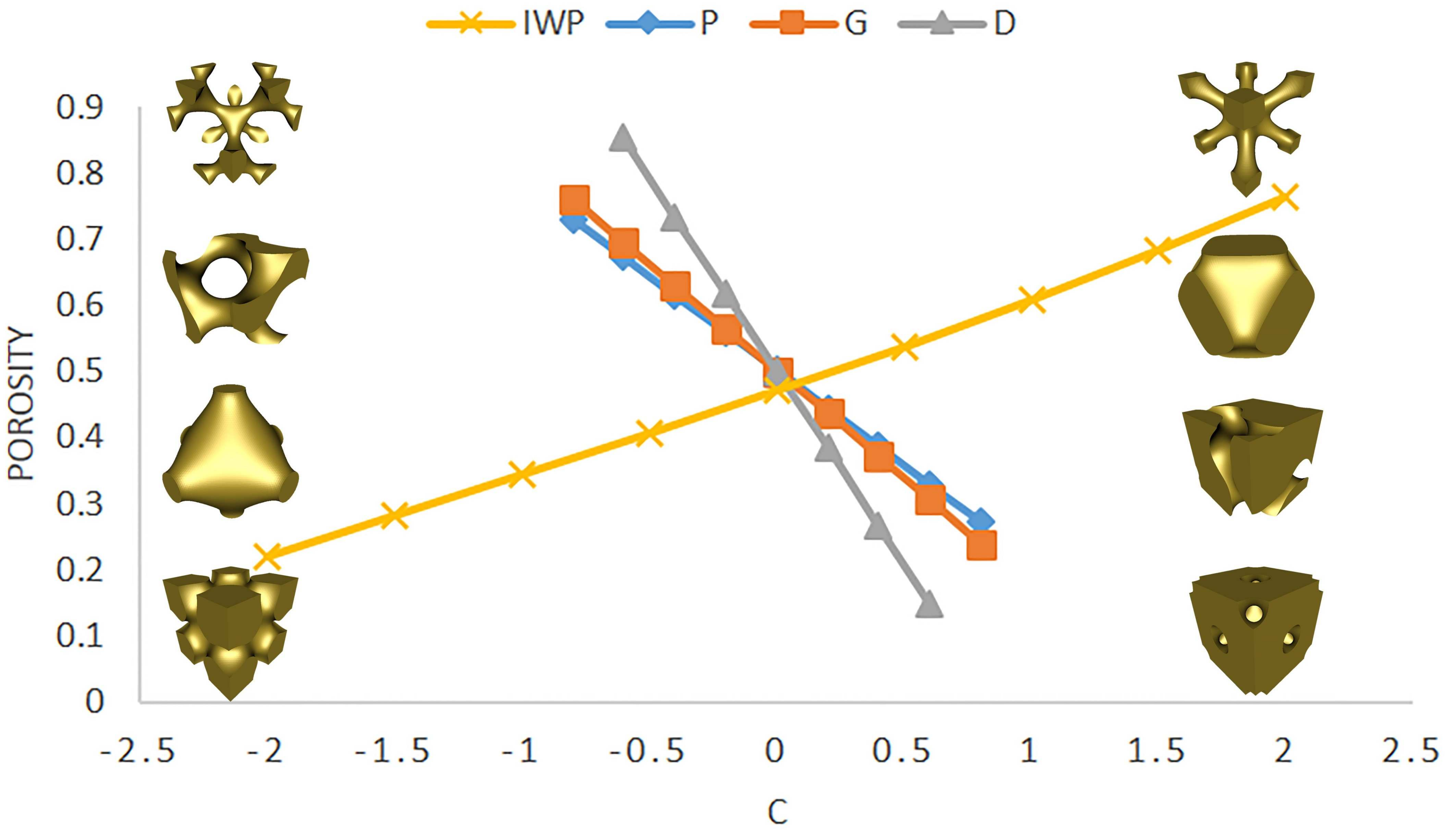}\\
  \caption{\small Relationship between the threshold $C$ and the porosity of the four types of TPMSs based on rod structures.}
  \label{fig:rod_relationship}
\end{figure}

\begin{figure}[!htb]
  \centering
  \includegraphics[width=0.5\textwidth]{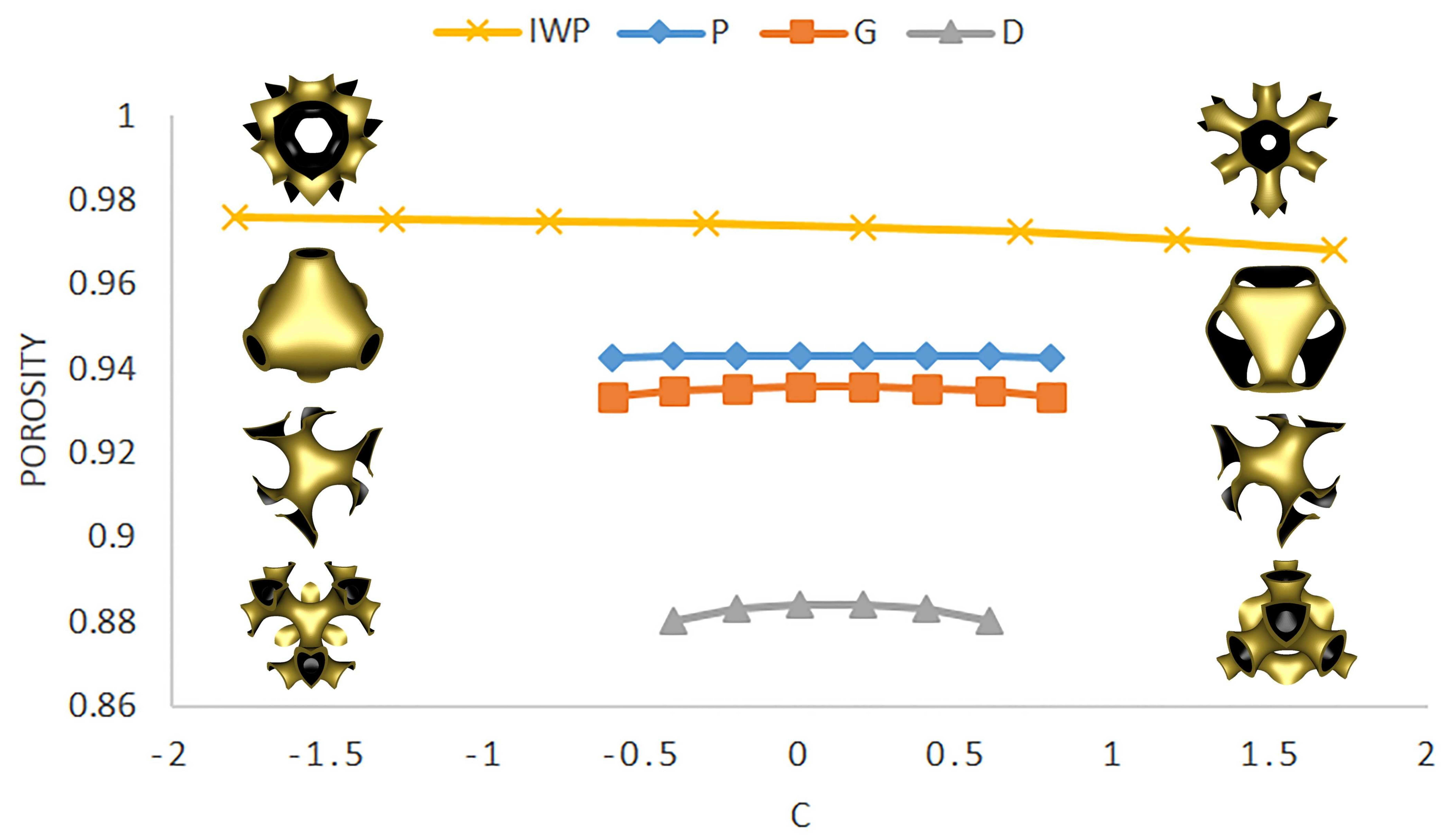}\\
  \caption{\small Relationship between the threshold $C$ and the porosity of the four types of TPMSs based on sheet structures.}
  \label{fig:sheet_relationship}
\end{figure}

 The porosity is an important parameter in porous scaffold design because it has direct influences on the transport of nutrition and waste.
 The porosity and pore size of porous scaffolds designed by TPMS units can be
    controlled by adjusting the threshold $C$ (see Table~\ref{tbl:nodal}).
 Moreover, the relationship between the porosity and the threshold $C$ is
    illustrated in Figs.\ref{fig:pore_relationship}-\ref{fig:sheet_relationship}.
 We can see that, for the three types of volume TPMS structures,
    i.e., pore, rod, and sheet,
    each has \emph{valid threshold range} and they are listed in Table~\ref{tbl:nodal}.


 To design heterogeneous porous scaffolds,
    we change the threshold $C$ to a trivariate function $C(u,v,w)$ defined on the parametric domain of a TBSS,
    which is called \emph{threshold distribution field} (TDF).
 Then, the \emph{TPMS in the parameter domain} is represented by the
    zero-level surface of,
 \begin{equation}
  \label{eq:combination}
  f(u,v,w)=\psi(u,v,w)-C(u,v,w)=0.
 \end{equation}
 Therefore, the TDF plays a critical role in the heterogeneous
    porous scaffold generation,
    and designing the TDF becomes a key problem in porous scaffold design.

 In this study, we developed some convenient techniques for generating a TDF
    in the cubic parameter domain of a TBSS.
 In brief, the cubic parameter domain of a TBSS is first discretized
    into a dense grid (in our implementation, it is discretized into a grid with a resolution of $50 \times 50 \times 50$),
    called a \emph{parametric grid}.
 Then, the threshold values at the grid vertices are assigned using the techniques presented later in this paper,
    which constitute a \emph{discrete TDF}.
 Finally, to save storage space,
    the discrete TDF is fitted by a trivariate B-spline function, i.e.,
 \begin{equation}
  \label{eq:distribution}
  C(u,v,w)=\sum_{i=0}^{n_u}\sum_{j=0}^{n_v}\sum_{k=0}^{n_w}
    N_{i,p}(u)N_{j,q}(v)N_{k,r}(w)C_{ijk},
 \end{equation}
 where the scales $C_{ijk}$ are the control points of the trivariate
    B-spline function.

 Now, we elucidate the techniques for generating the discrete TDF.

\begin{figure}[!htb]
  \begin{center}
  \subfigure[]{
    \label{subfig:curvature_model}
    \includegraphics[width=0.2\textwidth]{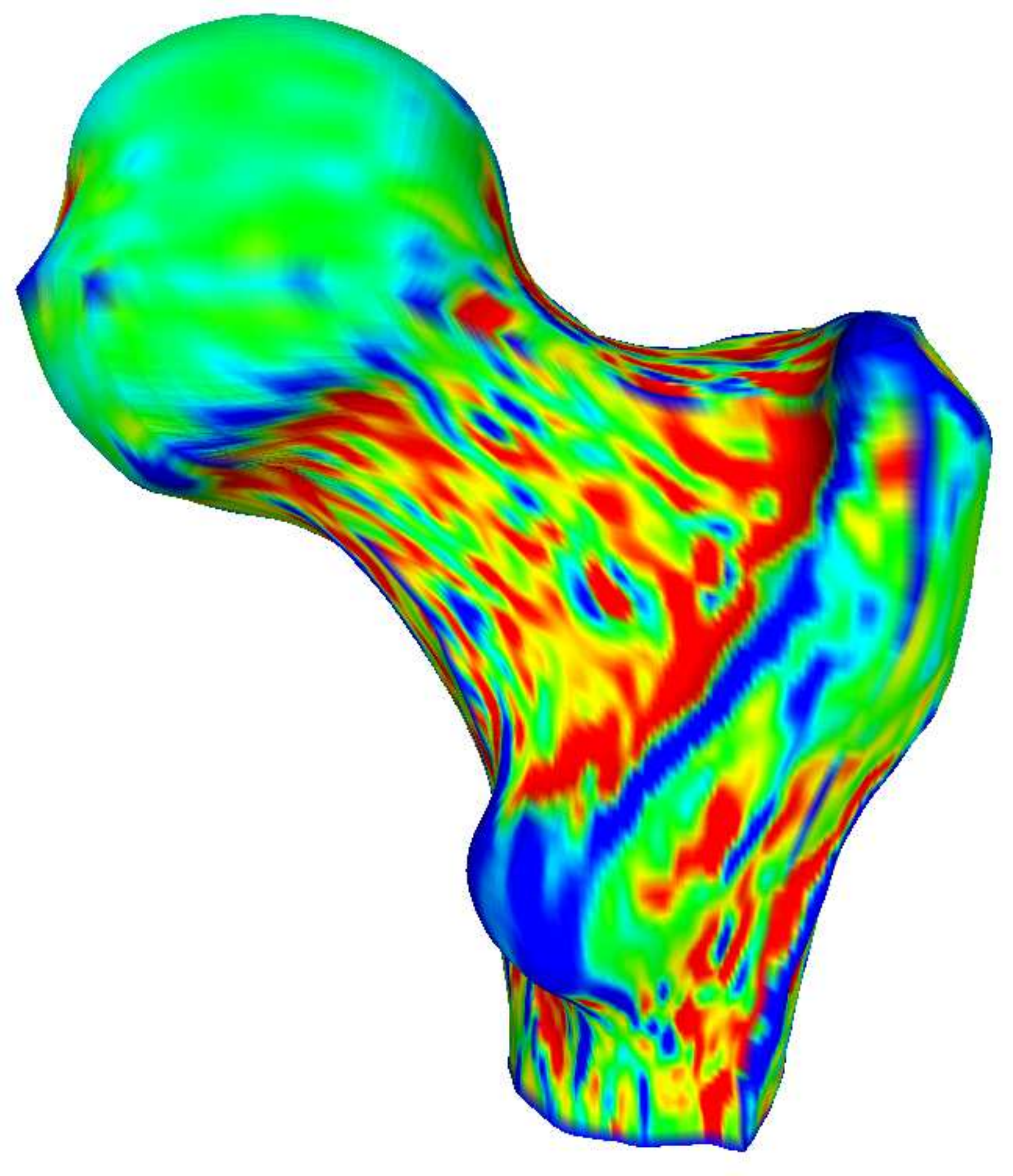}}
  \subfigure[]{
    \label{subfig:curvature_on_grid}
    \includegraphics[width=0.22\textwidth]{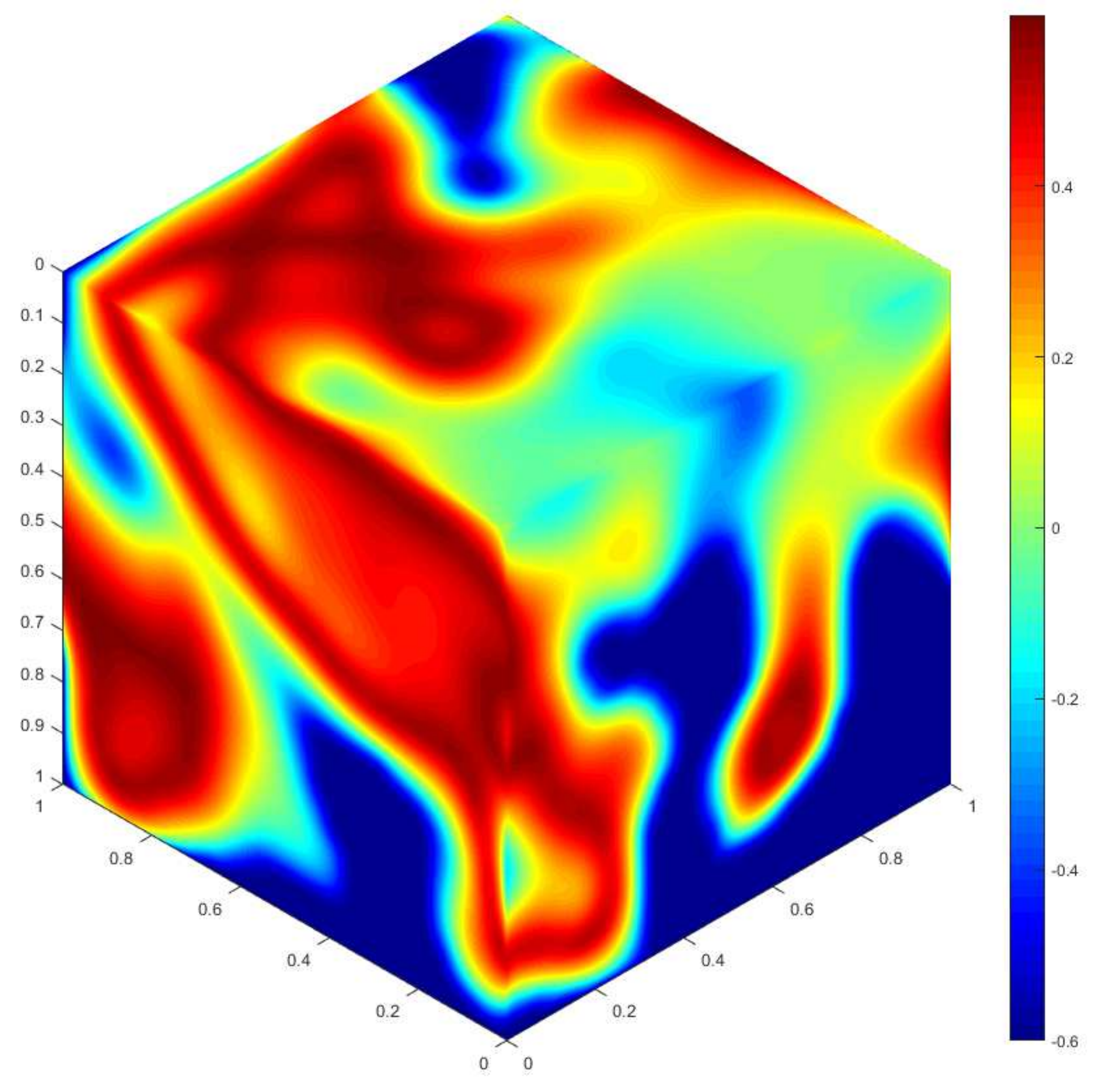}}
      \caption
      {\small
        Filling method.
        (a) Mean curvature distribution on the TBSS boundary surface.
        (b) Mean curvature distribution on the boundary vertices of the parametric grid.
      }
   \label{fig:curvature_distribution}
  \end{center}
\end{figure}

 \emph{Filling method.}
 Initially, all of the values at the parameter grid vertices are set to $0$.
 Then, the geometric quantities at points of the boundary surface of TBSS,
    such as mean curvature, Gauss curvature,
    are calculated and mapped to the boundary vertices of the parametric grid.
 Furthermore, the quantities at the boundary vertices are diffused into
    the inner parametric grid vertices by the Laplace smoothing operation~\cite{Field1988Laplacian}.
 Thus, the entire parametric grid is filled,
    and a discrete TDF is constructed.
 In Fig.~\ref{fig:curvature_distribution},
    the mean curvature distribution of the TBSS boundary surface is first calculated (Fig.~\ref{subfig:curvature_model}) and then is mapped to the boundary vertices of the parametric grid (Fig.~\ref{subfig:curvature_on_grid}).

\begin{figure}[!htb]
  \begin{center}
  \subfigure[]{
    \label{subfig:radial_distribution}
    \includegraphics[width=0.22\textwidth]{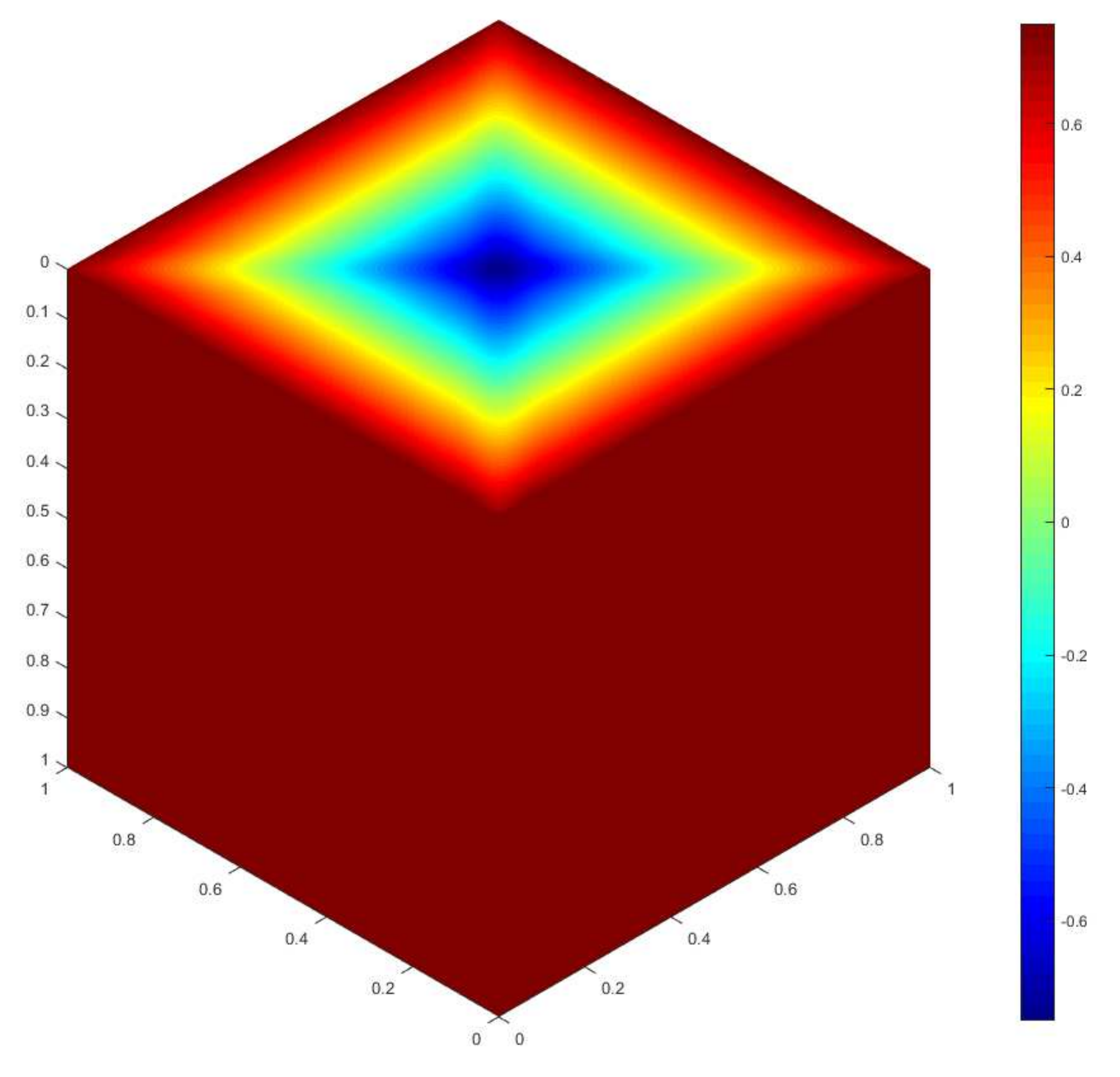}}
  \subfigure[]{
    \label{subfig:radial_distribution_modify}
    \includegraphics[width=0.22\textwidth]{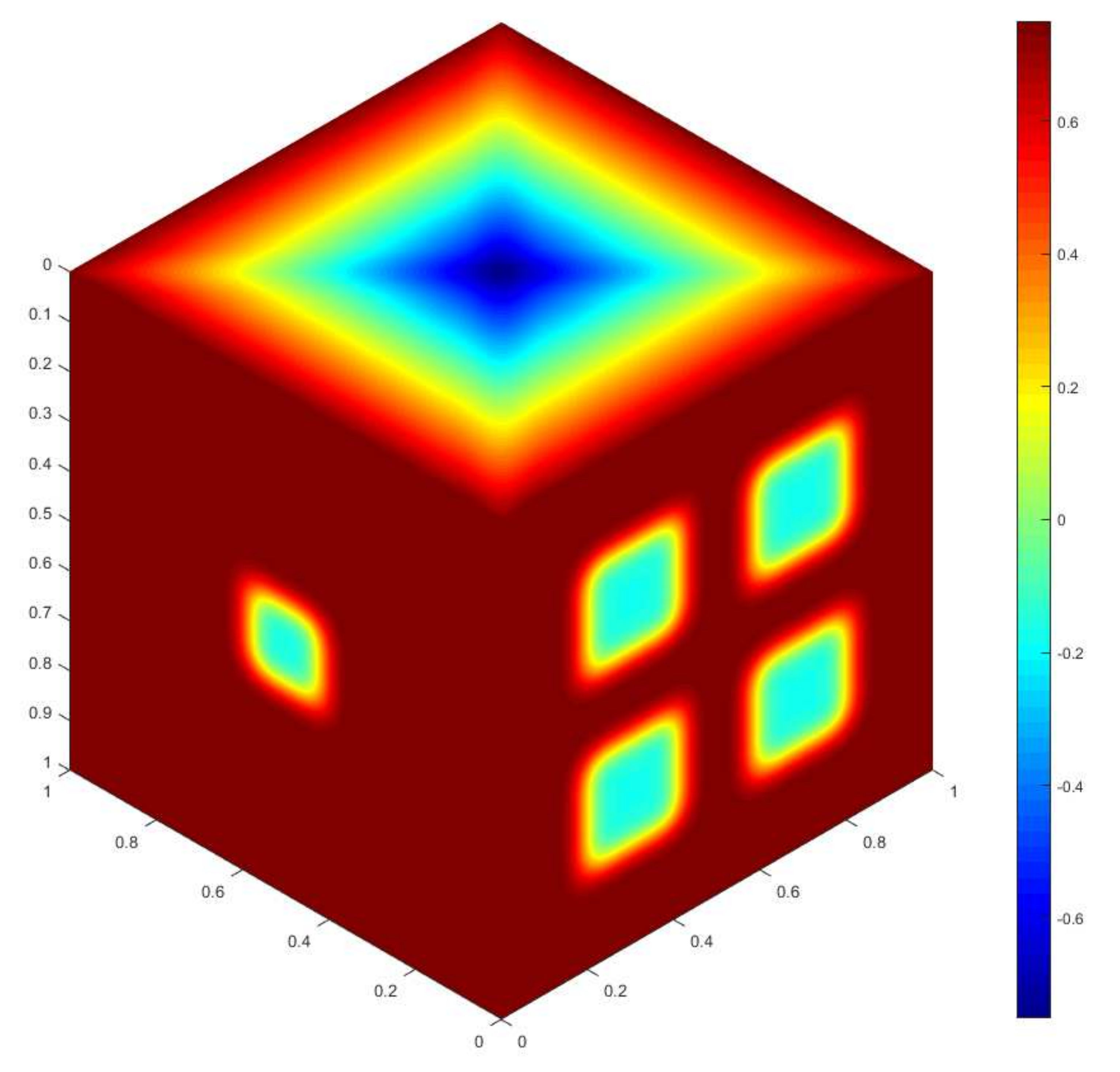}}
      \caption
      {\small
       Local modification of TDF.
      (a) TDF generated by the layer method.
      (b) TDF after local modification.
      }
   \label{fig:porosity_distribution}
  \end{center}
\end{figure}

 \emph{Layer method.} The parametric grid vertices are classified into layers
    according to their coordinates $(u_i,v_j,w_k)$.
 For example, the vertices with the same $w_k$ coordinates can be classified
    into the same layer.
 Vertices of the same layer are assigned the same threshold values.
 The TDFs in Figs.~\ref{subfig:radial_distribution} and ~\ref{subfig:layer_distribution} were generated using the layer method.
 In Fig.~\ref{subfig:radial_distribution},
    the vertices of the four side surfaces are taken as the first layer,
    the vertices adjacent to the first layer are taken being in the second layer,
    $\cdots$, and so on.
 In Fig.~\ref{subfig:layer_distribution},
    the vertices with the same $w_k$ coordinates are taken as the same layer.

 \emph{Prescribed function method.}
 The threshold values at the parametric grid vertices can be assigned by a
    function prescribed by users.
 For example, in Fig.~\ref{subfig:porosity_distribution_isis},
    the threshold value at the vertex with coordinates $(u_i,v_j,w_k)$ is assigned by the function,
    \begin{equation*}
        f(u_i,v_j,w_k) = \abs{u_i-v_j} + \abs{v_j-w_k} + \abs{u_i-w_k}.
    \end{equation*}

 After the discrete TDF is generated,
    the values at the grid vertices are linearly transformed into the valid threshold range according to the type of volume TPMS structure being produced.
 Then, we fit the discrete TDF with a trivariate
    B-spline function~\pref{eq:distribution},
    using the least squares progressive-iteration approximation (LSPIA) method~\cite{Deng2014Progressive}.
 The subscripts of the grid vertices, i.e., $(i,j,k)$,
    are the natural parametrization of the vertices.
 For the purpose of B-spline fitting,
    they are normalized into the interval $[0,1] \times [0,1] \times [0,1]$.
 The knot vectors of the B-spline function~\pref{eq:distribution} is
    uniformly defined under the B\'{e}zier end condition.
 In our implementation, the resolution of the control grid of the trivariate
    B-spline function is taken as $20 \times 20 \times 20$.
 Additionally, the initial values for LSPIA iteration at the control grid
    of the B-spline function~\pref{eq:distribution} are produced by linear interpolation of the discrete TDF.

 Suppose the LSPIA iteration has been performed for $l$ steps,
     and the $l^{th}$ B-spline function $C^{(l)}(u,v,w)$ is constructed:
 \begin{equation}
  \label{eq:k_iteration}
  C^{(l)}(u,v,w)=\sum_{i=0}^{n_u}\sum_{j=0}^{n_v}\sum_{k=0}^{n_w}
    N_{i,p}(u)N_{j,q}(v)N_{k,r}(w)C^{(l)}_{ijk}.
 \end{equation}
 to generate the $(l+1)^{th}$ B-spline function $C^{(l+1)}(u,v,w)$,
    the difference vector for each parametric grid vertex is calculated,
 \begin{equation}
  \label{eq:diff_data}
  \delta^{(l)}_{\alpha,\beta,\gamma} = T_{\alpha,\beta,\gamma}-C^{(l)}(u_{\alpha},v_{\beta},w_{\gamma}),
 \end{equation}
 where $T_{\alpha,\beta,\gamma}$ is the threshold value at the vertex
    $(\alpha,\beta,\gamma)$,
    and $(u_{\alpha},v_{\beta},w_{\gamma})$ are its parameters.
 Each difference vector $\delta^{(l)}_{\alpha,\beta,\gamma}$ is
    distributed to the control points $C^{(k)}_{i,j,k}$
    if the corresponding basis functions $N_{i,p}(u_{\alpha})N_{j,q}(v_{\beta})N_{k,r}(w_{\gamma})$ are non-zero.
 Moreover, a weighted average of all the difference vectors distributed to a control point is taken,
    leading to the \emph{difference vector for the control point},
  \begin{equation}
  \label{eq:diff_cont}
  \Delta^{(l)}_{ijk}=\frac{\sum_{I\in I_{\alpha\beta\gamma}}{N_{i,p}(u_I)N_{j,q}(v_I)N_{k,r}(w_I)\delta^{(l)}_l}}{\sum_{I\in I_{\alpha\beta\gamma}}{N_{i,p}(u_I)N_{j,q}(v_I)N_{k,r}(w_I)}},
 \end{equation}
 where $I_{\alpha\beta\gamma}$ is the set of indices such that $$N_{i,p}(u_{\alpha})N_{j,q}(v_{\beta})N_{k,r}(w_{\gamma}) \neq 0.$$

 Next, the $(l+1)^{th}$ control points $C^{(l+1)}_{ijk}$ are formed by adding the difference vectors $\Delta^{(l)}_{ijk}$ to the $l^{th}$ control points,
 \begin{equation}
  \label{eq:controlpoint}
  C^{(l+1)}_{ijk}=C^{(l)}_{ijk}+\Delta^{(l)}_{ijk}.
 \end{equation}
 Thus, the $(l+1)^{th}$ B-spline function $C^{(l+1)}(u,v,w)$ is produced:
  \begin{equation}
  \label{eq:t_next_iteration}
  C^{(l+1)}(u,v,w)=\sum_{i=0}^{n_u}\sum_{j=0}^{n_v}\sum_{k=0}^{n_w}
    N_{i,p}(u)N_{j,q}(v)N_{k,r}(w)C^{(l+1)}_{ijk}.
 \end{equation}
 The convergence of LSPIA iteration has been proved in~\cite{Deng2014Progressive}.
 After the LSPIA iterations stop,
    the iteration result $C(u,v,w)$ is taken as the TDF in the parametric domain.

 \emph{Local modification.}
 With the TDF $C(u,v,w)$ in the parametric domain,
  the TPMS in the parametric domain~\pref{eq:combination} can be generated.
 By mapping the TPMS into the TBSS,
    a porous scaffold is produced in the TBSS.
 However, if the generated porous scaffold does not satisfy the practical
    engineering requirements,
    the TDF $C(u,v,w)$ can be locally modified in the parametric domain,
    and then the porous scaffold can be rebuilt to meet the practical requirements.

 To locally modify the TDF,
    users first choose some vertices of the parameter grid,
    and change their threshold values to their desirable values.
 Then, a \emph{local} LSPIA iteration is invoked to fit the changed values at
    the chosen vertices.
 In the local LSPIA iteration,
    the difference vector $\delta$~\pref{eq:diff_data} is calculated only at the chosen vertices,
    and just the control points to which the difference vectors $\delta$ are distributed are adjusted.
 The other control points without distributed difference vectors remain
    unchanged.
 By locally modifying the TDF in Fig.~\ref{subfig:radial_distribution}
    using the method presented above,
    the TDF is changed as illustrated in Fig.~\ref{subfig:radial_distribution_modify}.

 \subsection{Generation of heterogeneous porous scaffold in TBSS}
 \label{subsec:porous_scaffold_generation}

 Until now, there was only one unknown in the heterogeneous porous scaffold generation:
    the period coefficients $\omega_x, \omega_y, \omega_z$ (Table~\ref{tbl:nodal}).
 It is worth noting that the internal connectivity of the scaffold
    is crucial to the transferring performance of the scaffold.
 For a TPMS unit,
    the smaller the unit volume,
    the worse the internal connectivity of the micro-holes,
    and the larger the unit volume,
    the better the internal connectivity of the micro-holes.
 The period coefficients $\omega_x, \omega_y, \omega_z$ (Table~\ref{tbl:nodal}) can
    be employed to adjust the number of TPMS units,
    as well as the size of TPMS units in the three parametric directions.
 The values of the period coefficients $\omega_x, \omega_y, \omega_z$ used in the examples
    in this paper are listed in Table~\ref{tbl:stat}.

\begin{figure*}[!htb]
  \begin{center}
    \subfigure[]{
    \label{subfig:axial_P_sheet_solid}
    \includegraphics[width=0.22\textwidth]{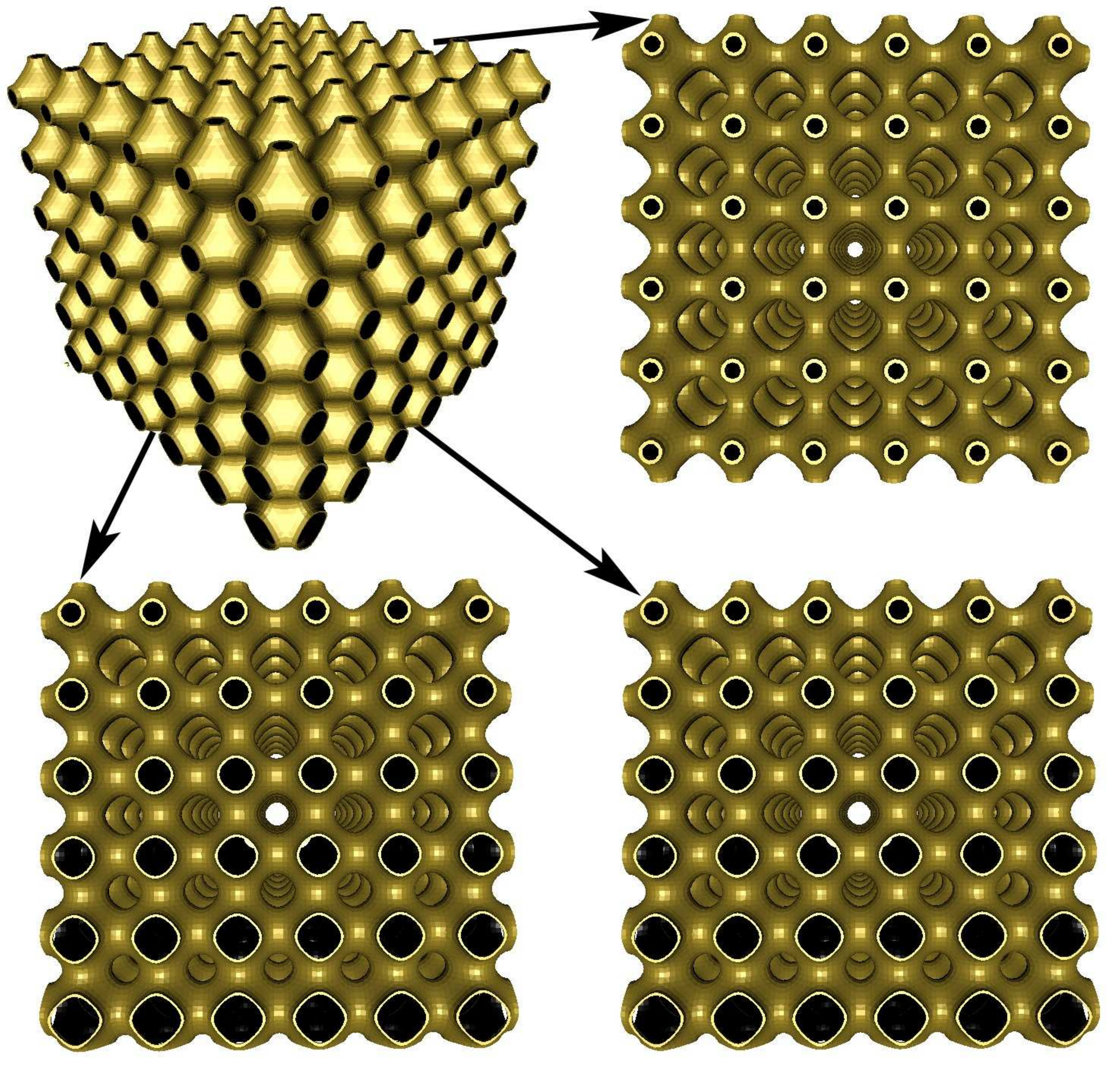}}
  \subfigure[]{
    \label{subfig:radial_P_sheet_solid}
    \includegraphics[width=0.22\textwidth]{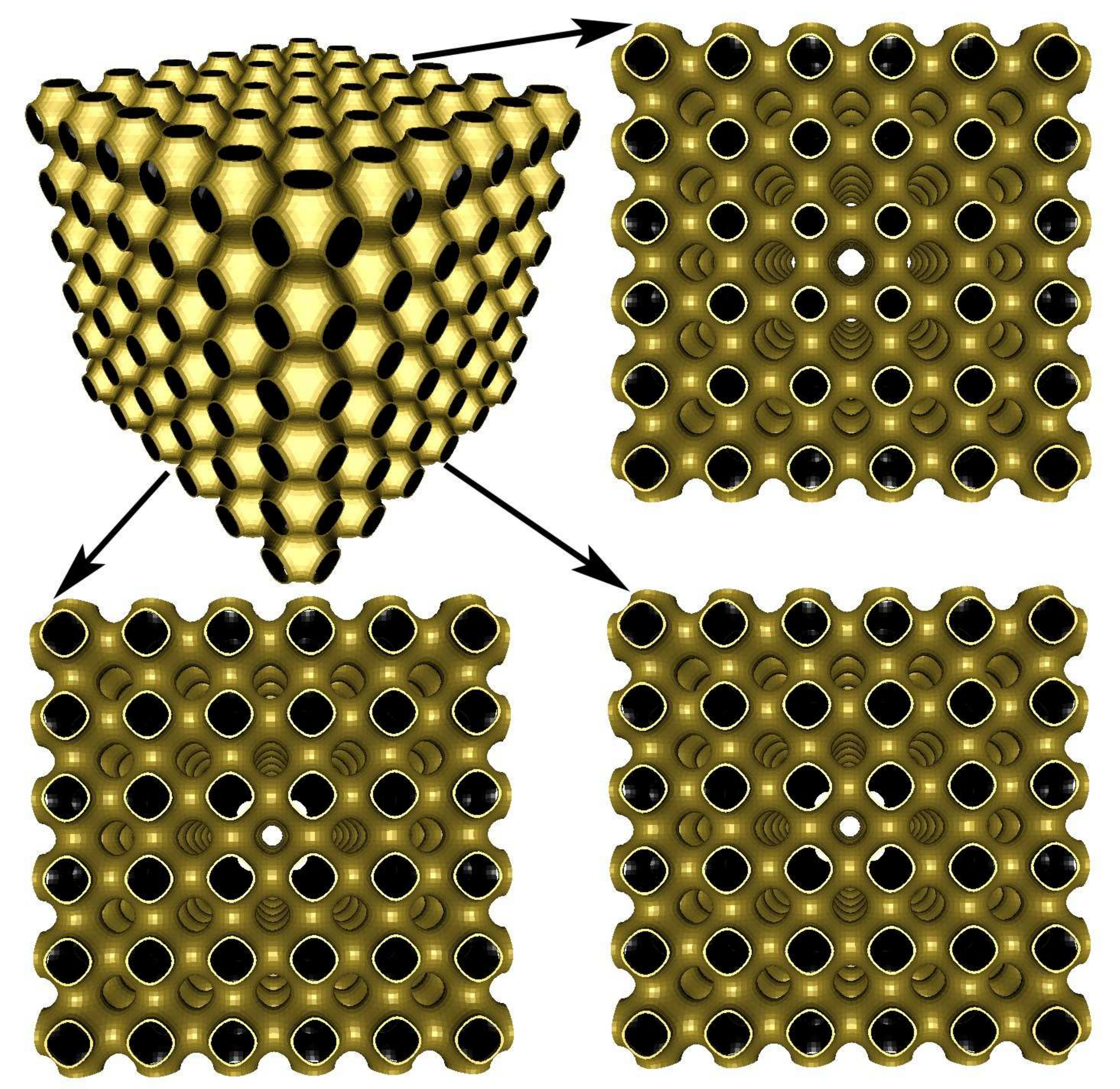}}
  \subfigure[]{
    \label{subfig:radial_modify_P_sheet_solid}
    \includegraphics[width=0.22\textwidth]{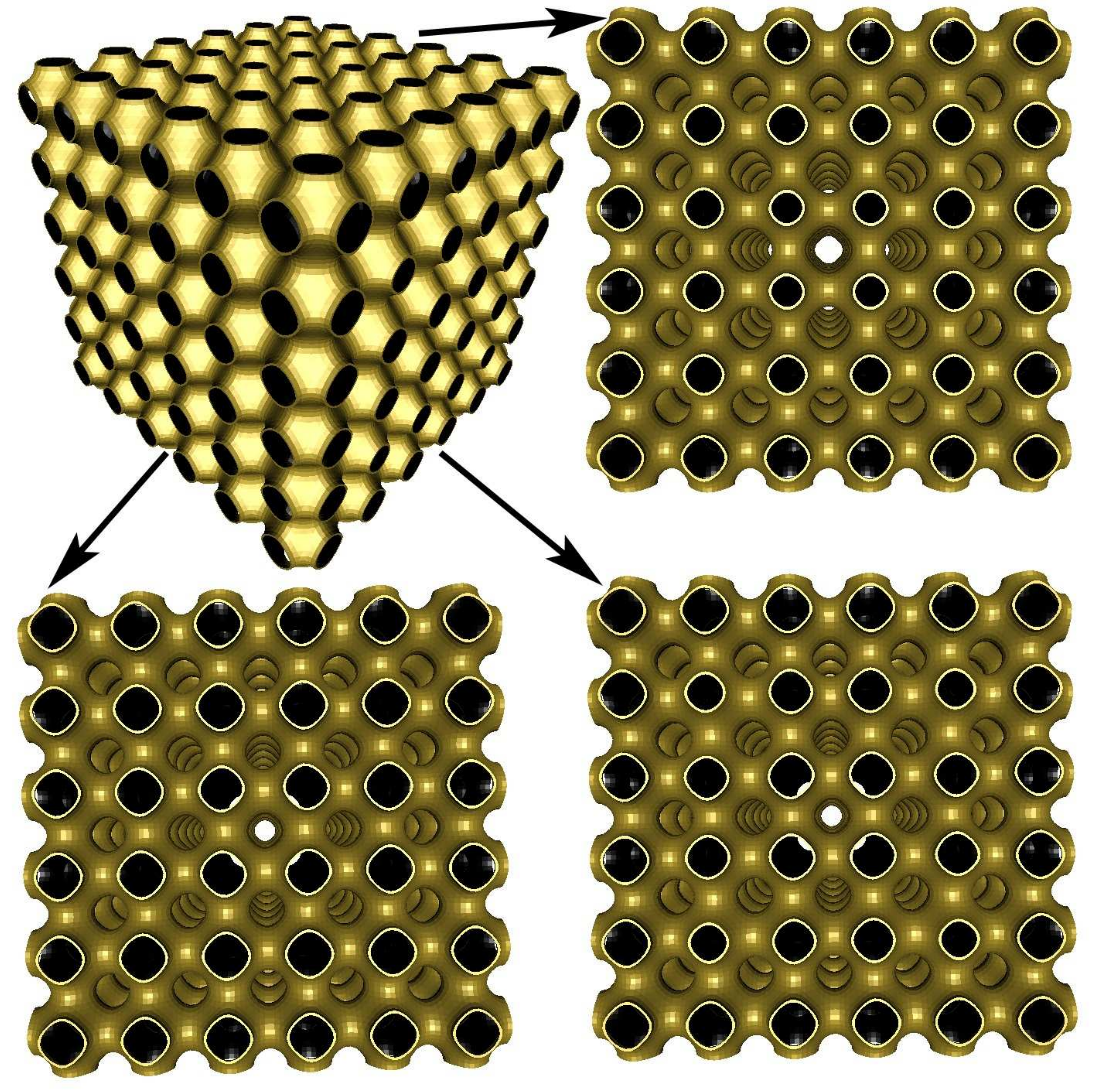}}
  \subfigure[]{
    \label{subfig:curvature_P_sheet_solid}
    \includegraphics[width=0.22\textwidth]{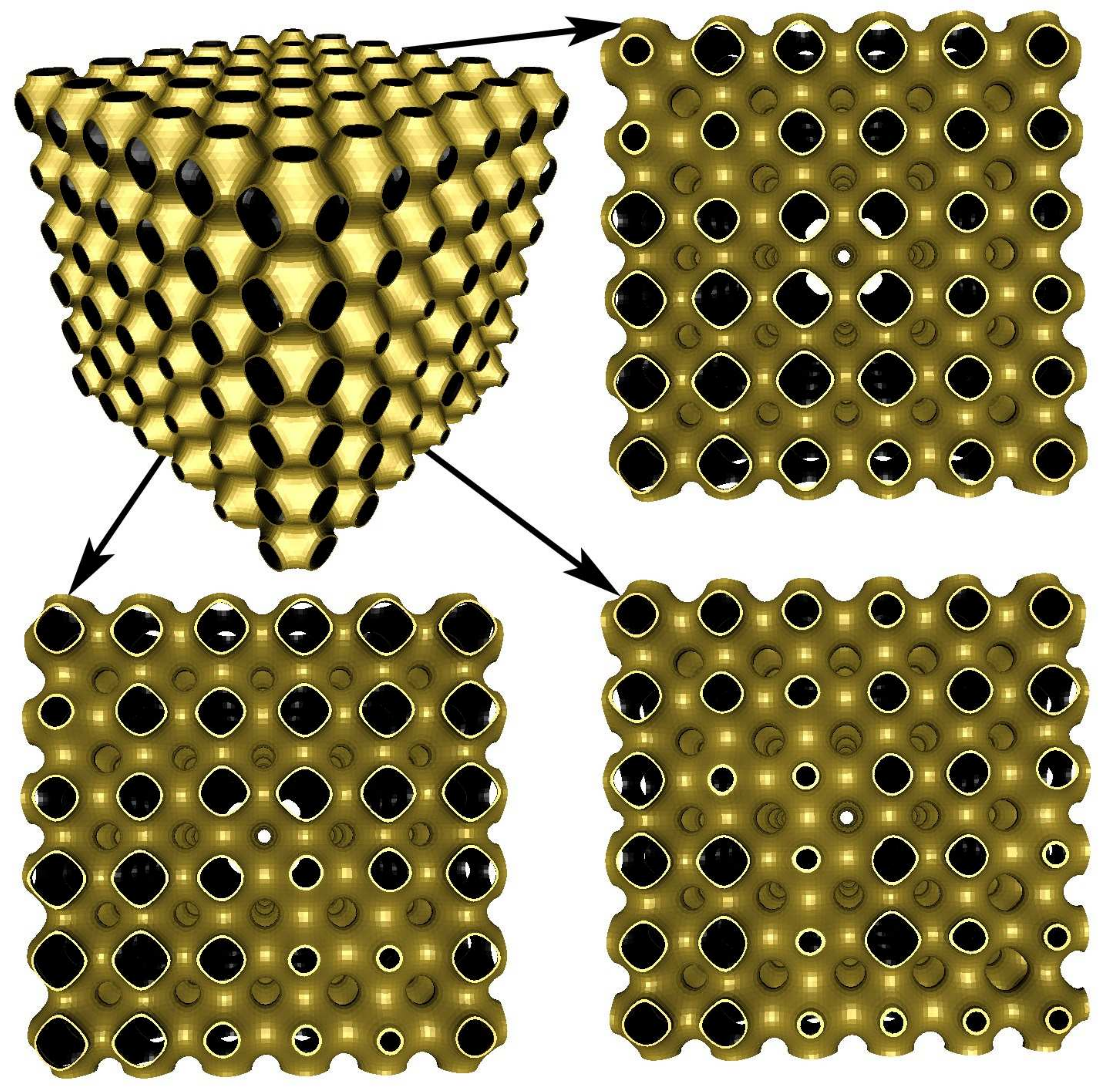}}
    \\
  \subfigure[]{
    \label{subfig:layer_distribution}
    \includegraphics[width=0.10\textwidth]{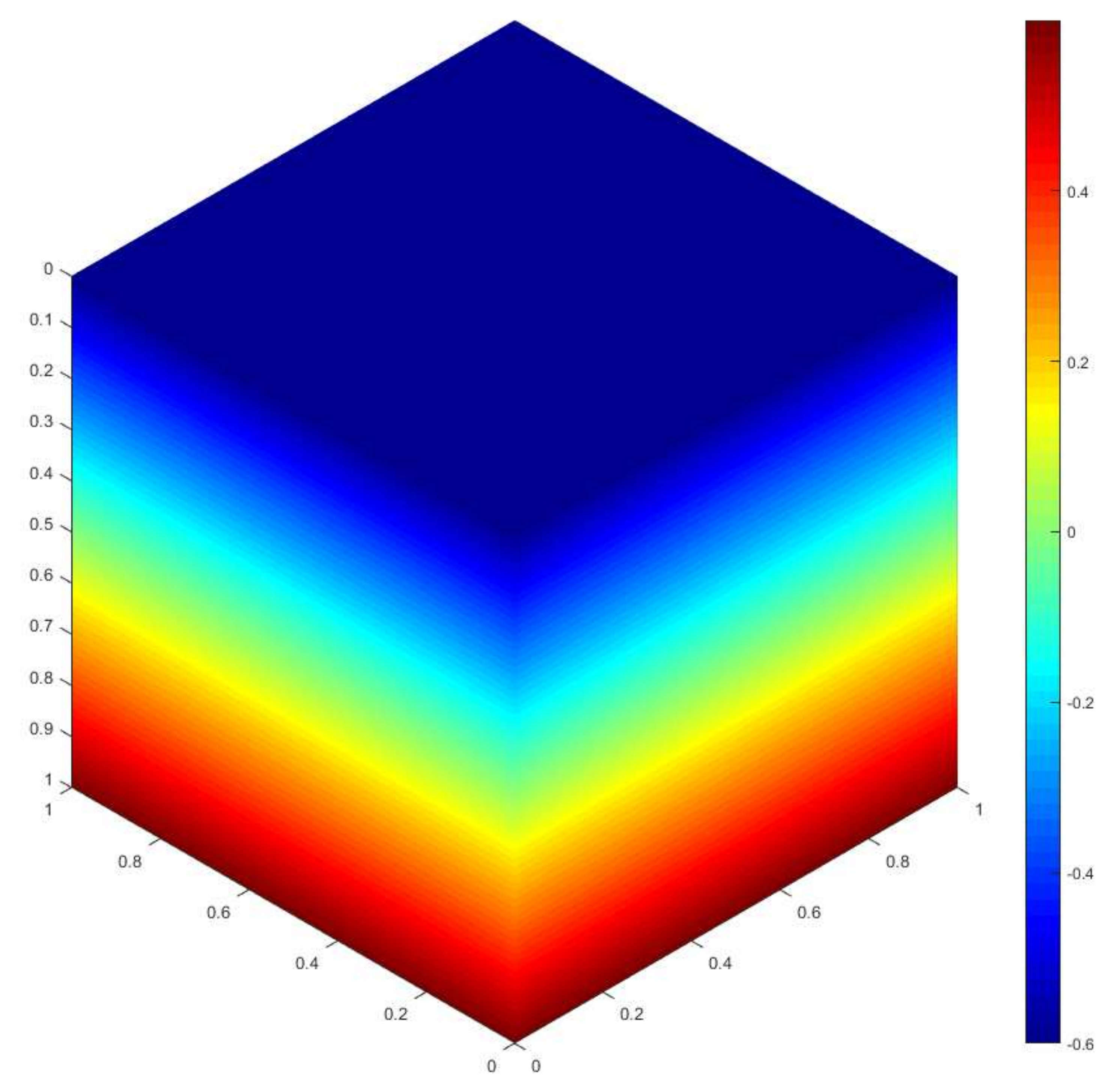}}
    \hspace{0.1\textwidth}
  \subfigure[]{
    \label{subfig:distribution_radial}
    \includegraphics[width=0.10\textwidth]{radial_distribution-eps-converted-to.pdf}}
  \hspace{0.1\textwidth}
  \subfigure[]{
    \label{subfig:distribution_radial_modify}
    \includegraphics[width=0.10\textwidth]{radial_distribution_modify-eps-converted-to.pdf}}
  \hspace{0.1\textwidth}
  \subfigure[]{
    \label{subfig:distribution_curvature}
    \includegraphics[width=0.10\textwidth]{porosity_distribution_balljoint-eps-converted-to.pdf}}
      \caption
      {\small
        Generation of the TPMSs (with their three-view drawing) (a,b,c,d) based on the corresponding TDF in the parametric domain (e,f,g,h).
        Note that the porosity of the TPMS is controlled by the TDF.
      }
   \label{fig:distribution_sheetsolid}
  \end{center}
\end{figure*}

 After the TDF in the parametric domain and the period coefficients are
    both determined (refer to Table~\ref{tbl:nodal}),
    the TPMS in the parametric domain (Eq.~\pref{eq:combination}), i.e.,
 \begin{equation*}
  f(u,v,w)=\psi(u,v,w)-C(u,v,w)=0,
 \end{equation*}
 can be calculated.
 For examples,
     the TPMSs (sheet structure) calculated based on the TDFs in Figs.~\ref{subfig:layer_distribution}-
    \ref{subfig:distribution_curvature} are illustrated in Figs.~\ref{subfig:axial_P_sheet_solid}-
    ~\ref{subfig:curvature_P_sheet_solid}.
 It can be seen clearly from Figs.~\ref{subfig:axial_P_sheet_solid}-
    ~\ref{subfig:curvature_P_sheet_solid} that,
    the porosity is controlled by the TDF in Figs.~\ref{subfig:layer_distribution}-
    \ref{subfig:distribution_curvature}.

\begin{figure}[!htb]
  \begin{center}
    \subfigure[]{
    \label{subfig:pore_volume_tpms}
    \includegraphics[width=0.23\textwidth]{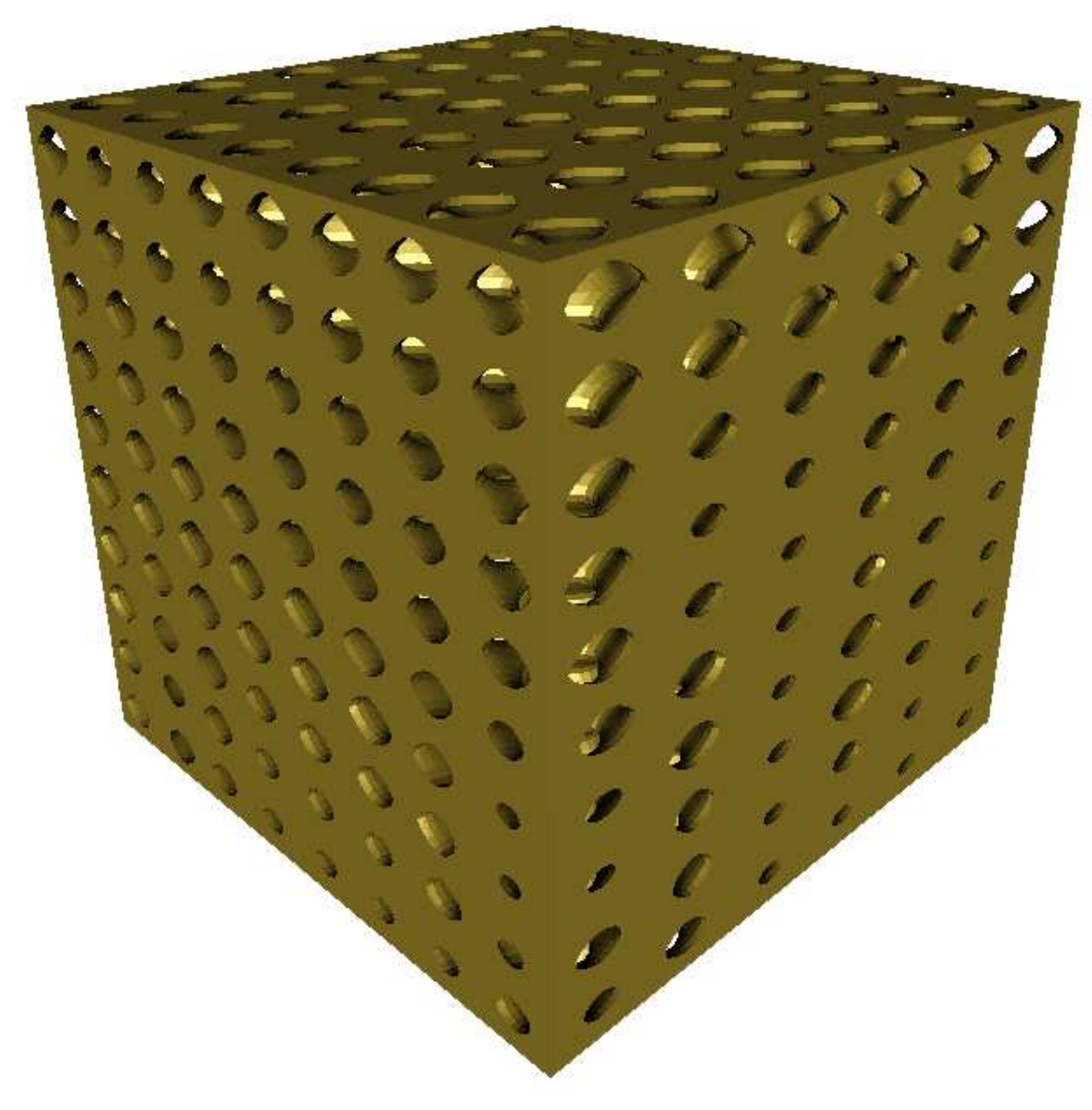}}
  \subfigure[]{
    \label{subfig:pore_balljoint_scaffold}
    \includegraphics[width=0.2\textwidth]{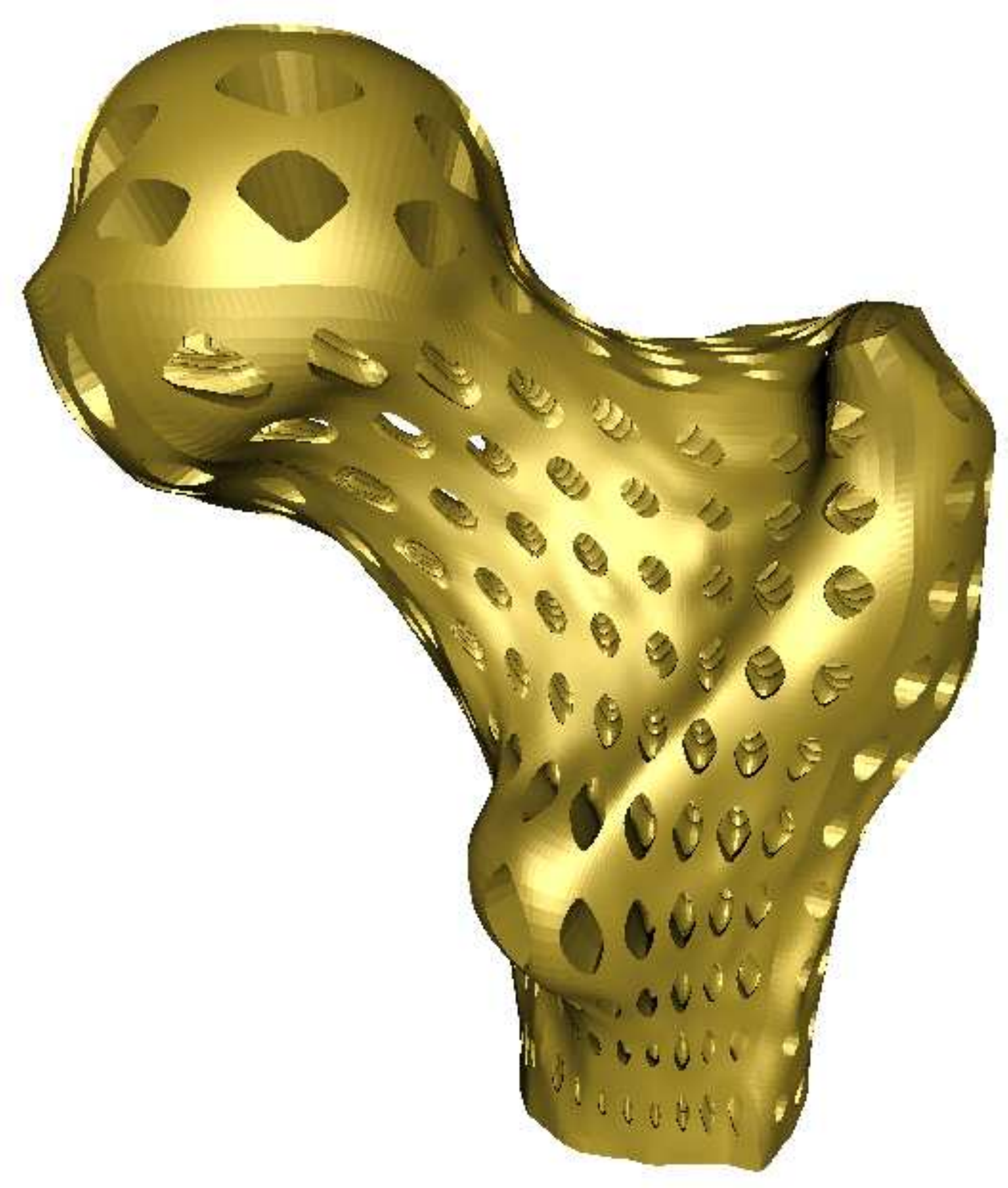}}
      \caption
      {\small
        Generation of heterogeneous porous scaffold.
        (a) TPMS in the parametric domain.
        (b) Heterogeneous porous scaffold in the TBSS.
      }
   \label{fig:porous_scaffold}
  \end{center}
\end{figure}

 Finally, the heterogeneous porous scaffold
     (Fig.~\ref{subfig:pore_balljoint_scaffold}) in the TBSS can be generated by mapping the TPMS in the parametric domain (volume TPMS structures) (Fig.~\ref{subfig:pore_volume_tpms}) into the TBSS, using the TBSS function.
 It should be noted that,
    to avoid fold-up, the Jacobian value of the TBSS should be positive.
 Because the TPMS in the parametric domain is unitary
    (Fig.~\ref{subfig:pore_volume_tpms}),
    it has completeness and continuity between adjacent TPMS units.
 Therefore, the heterogeneous porous scaffold in the TBSS is ensured to
    be complete and continuous
    (Fig.\ref{subfig:pore_balljoint_scaffold}).

 \subsection{Storage format based on TDF}
 \label{subsec:storage_format}

 Owing to their complicated geometric and topological structure,
    the storage costs for porous scaffolds are very large,
    usually requiring hundreds of megabytes (MB) (refer to Table~\ref{tbl:stat}).
 Therefore, the large storage cost becomes the bottleneck in porous scaffold
    generation and processing.
 In this study, we developed a new porous scaffold
    storage format that reduces the storage cost of porous scaffolds significantly.
 Using the new storage format,
    the space required to store the porous scaffold models presented in this paper ranges from $0.567$ MB to $1.355$ MB,
    while the storage space using the traditional STL file format ranges from $394.402$ MB to $1449.71$ MB.
 Thus, the new storage format reduces the storage requirement by at least $98\%$ compared with the traditional STL file format.
 Moreover, the generation of heterogenous porous scaffolds from the
    new file format costs a few seconds to a dozen seconds (Table~\ref{tbl:stat}).

 Specifically, because the TDF in the parametric domain and the
    period coefficients (Table~\ref{tbl:nodal}) entirely determine the heterogenous porous scaffold in a TBSS,
    the new storage format must only store the following information:
    \begin{itemize}
        \item period coefficients $\omega_x, \omega_y, \omega_z$,
        \item control points of the TDF $C(u,v,w)$,
        \item knot vectors of the TDF $C(u,v,w)$,
        \item control points of the TBSS $P(u,v,w)$,
        \item knot vectors of the TBSS $P(u,v,w)$.
    \end{itemize}
 Therefore, the new storage format is called the \emph{TDF format},
    and is summarized in Appendix for clarity.

\section{Implementation, results and discussions}
\label{sec:implementations}

 The developed heterogeneous porous scaffold generation method is implemented
    in the C++ programming language and tested on a PC with a 3.60 GHz i7-4790 CPU and 16 GB RAM.
 In this section, some examples are presented,
    and some implementation details are discussed.
 Moreover, the developed method is compared with classical
    scaffold generation methods.

 \subsection{Influence of threshold on heterogeneity of scaffold}
 \label{subsec:threshold_influence}

\begin{figure}[!htb]
  \begin{center}
  \subfigure[]{
    \label{subfig:influence_distribution}
    \includegraphics[width=0.18\textwidth]{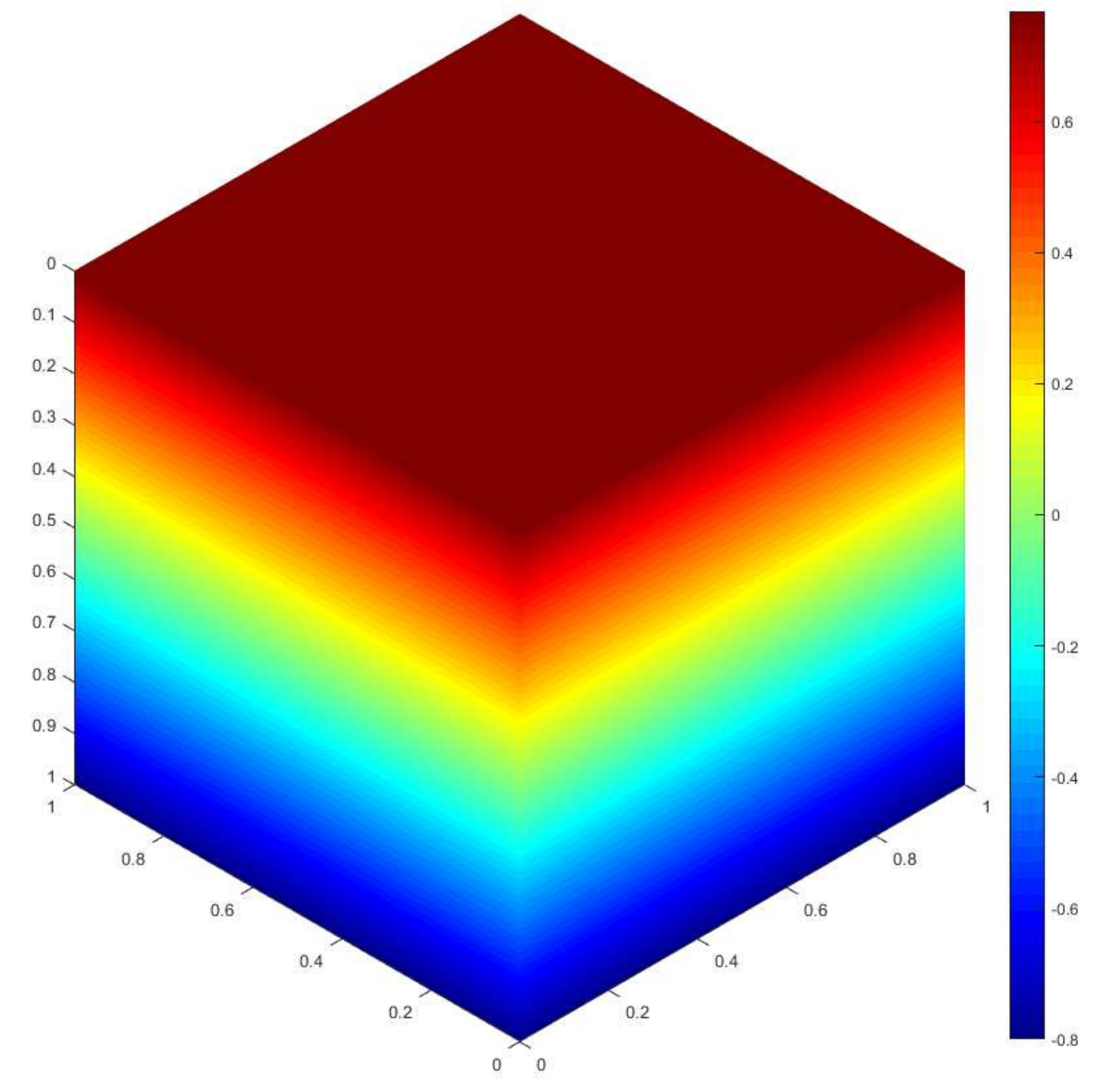}}
  \subfigure[]{
    \label{subfig:influence_porous_scaffold}
    \includegraphics[width=0.18\textwidth]{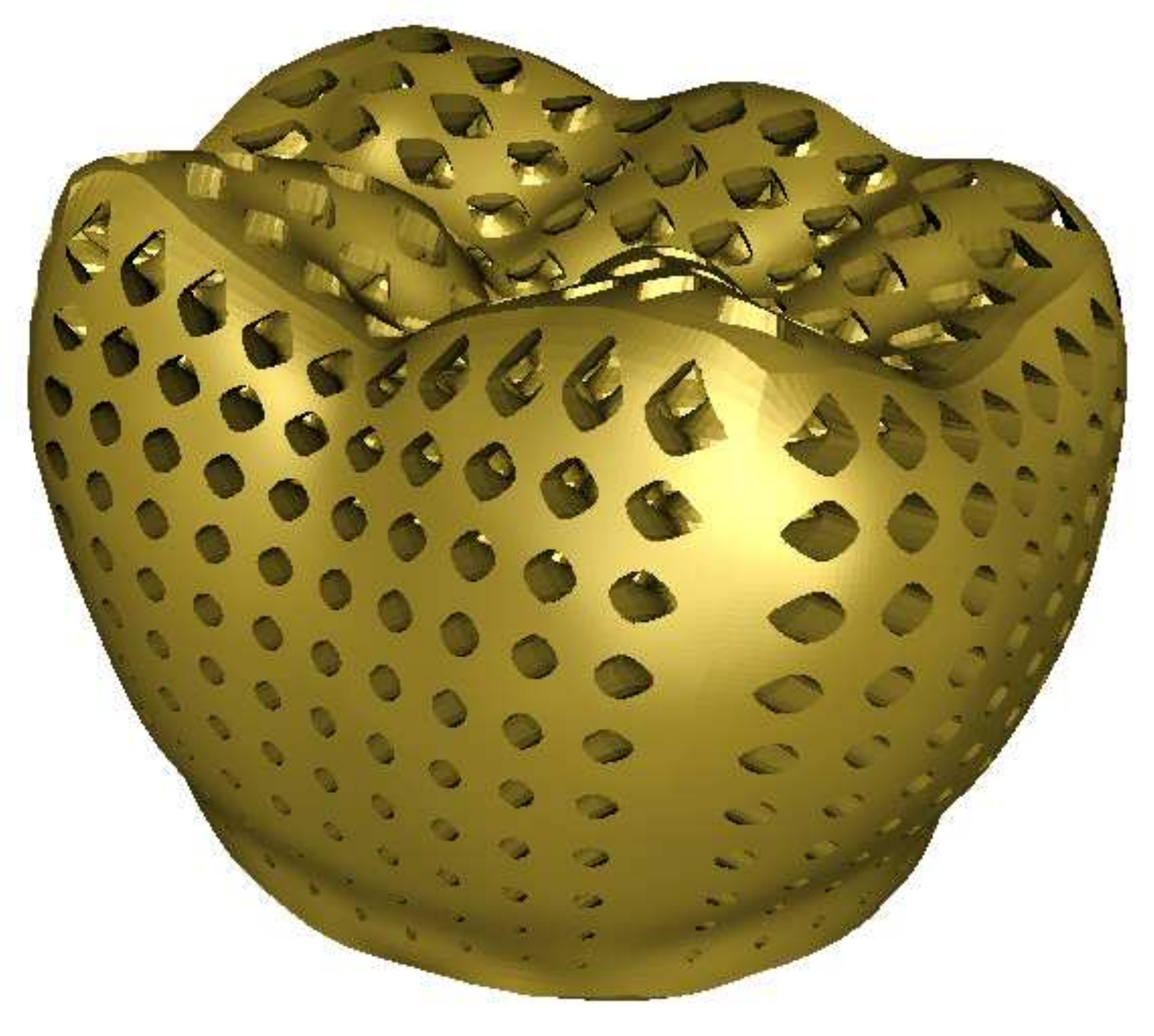}}
          \caption
          { \small
            Influence of threshold on the heterogeneity of scaffold.
            (a) TDF generated by the layer method.
            (b) Heterogenous porous scaffold (P-type, pore structure),
                produced based on the TDF in (a).
          }
   \label{fig:influence_threshold}
  \end{center}
\end{figure}

 The heterogeneity of a porous scaffold,
    i.e., its pore size and shape,
    is controlled by the TDF $C(u,v,w)$~\pref{eq:combination}.
 The larger the value of $C(u,v,w)$,
    the larger the pore size.
 In Fig.~\ref{subfig:influence_distribution},
    the TDF is generated by the layer method,
    which takes the parametric grid vertices with the same $w$ coordinates as the same layer.
 The small to large TDF values are visualized by blue to red colors.
 Fig.~\ref{subfig:influence_porous_scaffold} illustrates the heterogeneous porous
    scaffold (P-type, pore structure) generated based on the TDF in Fig.~\ref{subfig:influence_distribution}.
 We can see that, with the TDF values varying from large to small
    (Fig.~\ref{subfig:influence_distribution}),
    the pore size of the porous scaffold also changes from large to small
    (Fig.~\ref{subfig:influence_porous_scaffold}).

\subsection{Comparison with classical porous scaffold generation methods}
\label{subsec:comparison}

\begin{figure}[!htb]
  \begin{center}
    \subfigure[]{
    \label{subfig:porous_using_eightnodes_method}
    \includegraphics[width=0.45\textwidth]{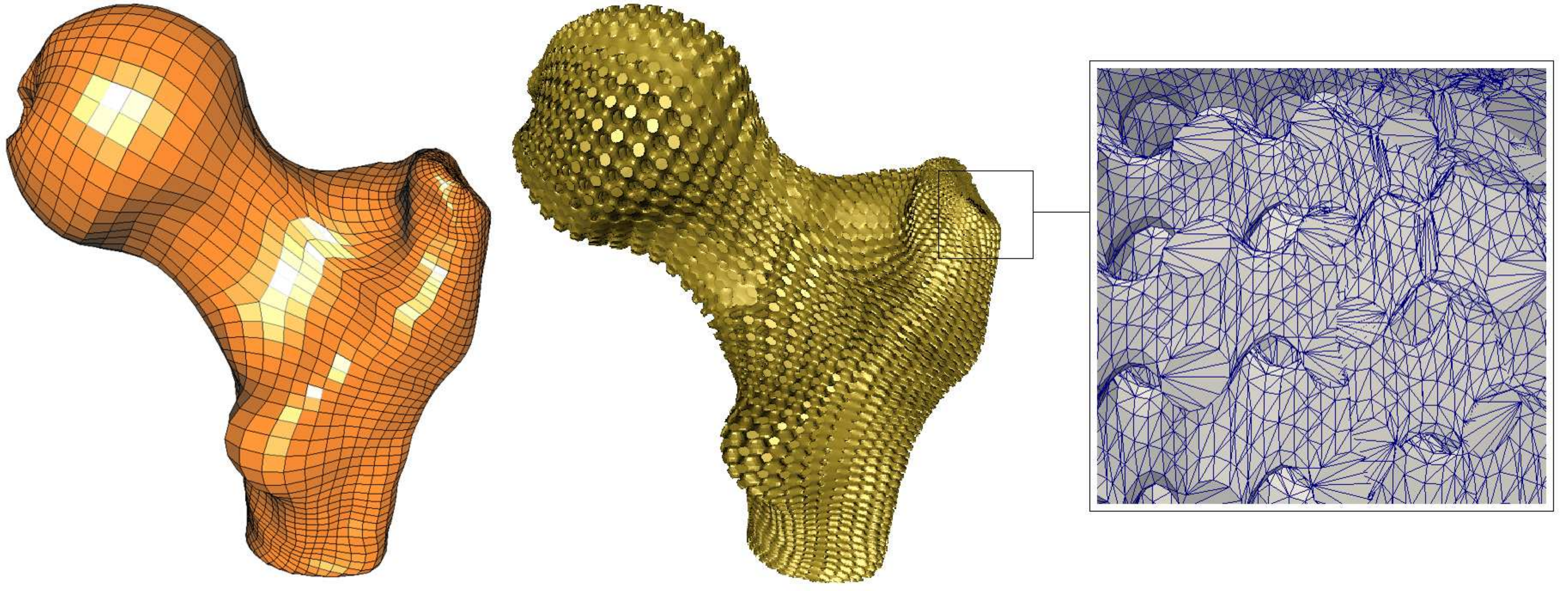}}
  \subfigure[]{
    \label{subfig:porous_using_tspline_method}
    \includegraphics[width=0.45\textwidth]{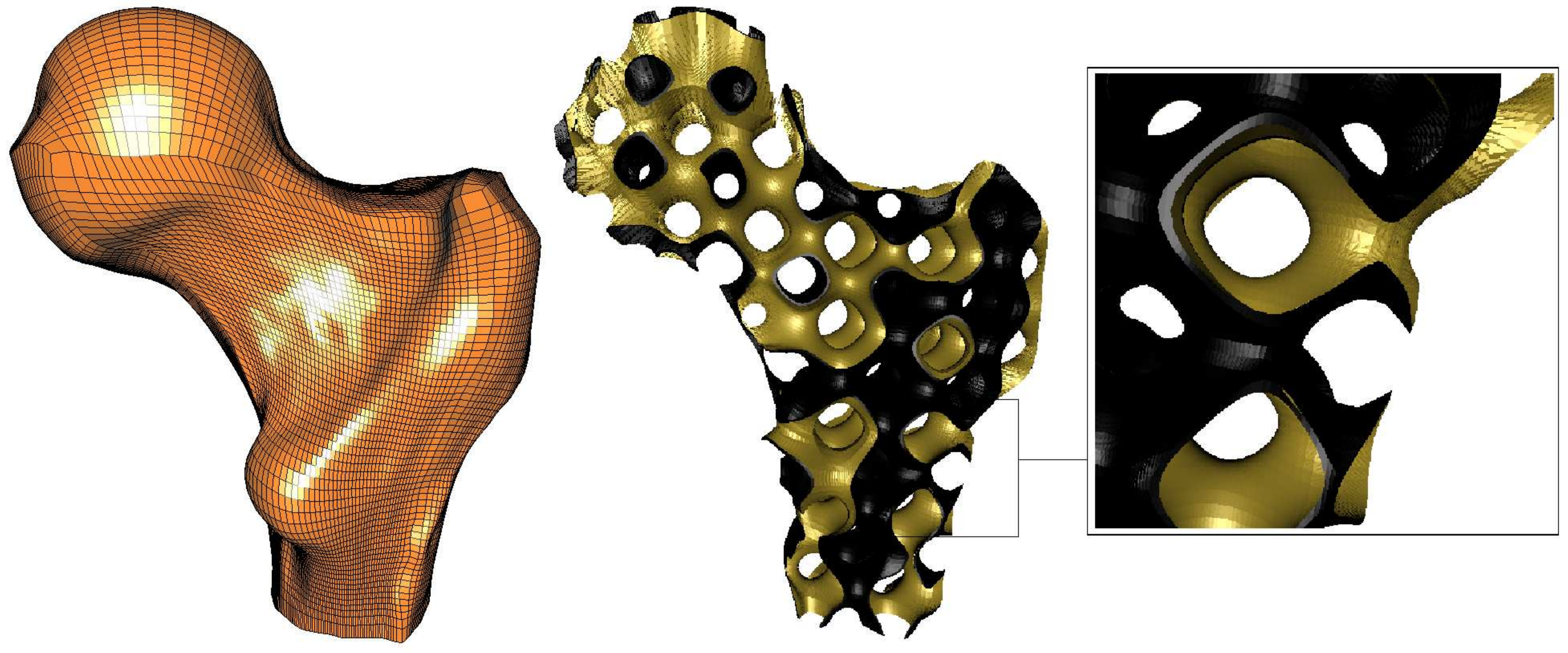}}
  \subfigure[]{
    \label{subfig:porous_using_our_method}
    \includegraphics[width=0.45\textwidth]{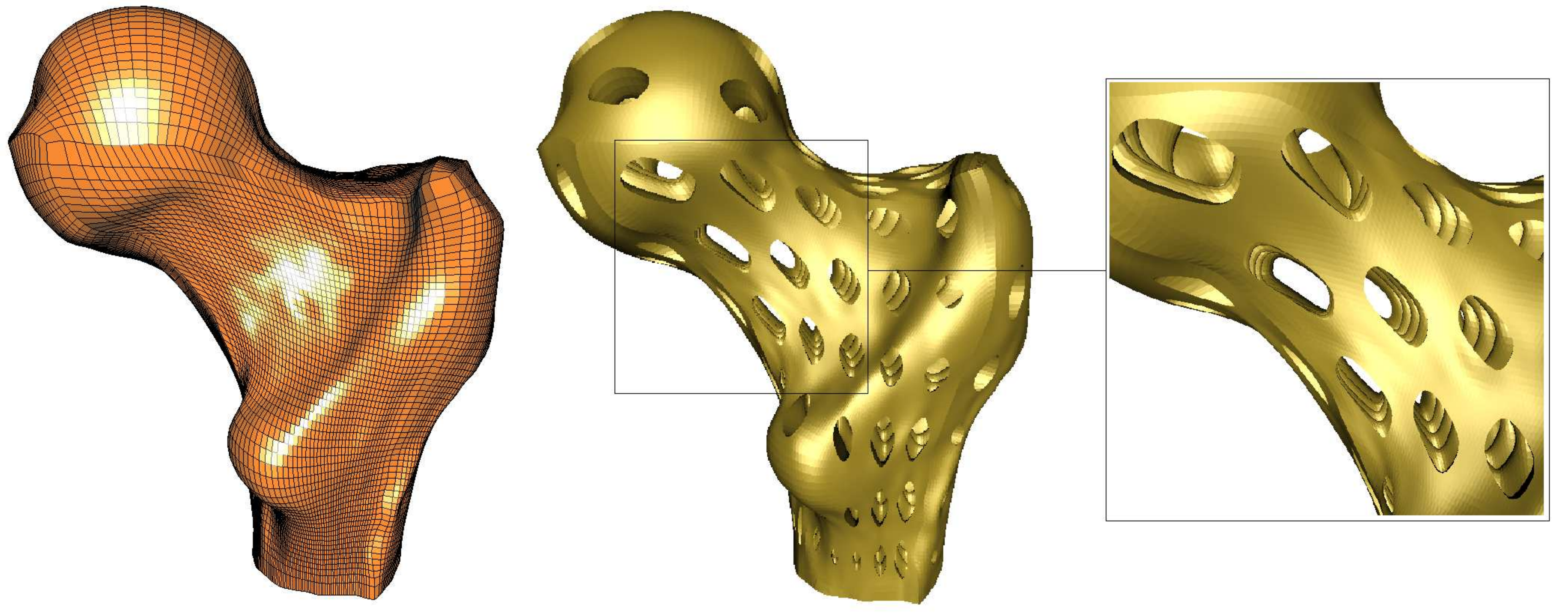}}
      \caption
      {\small
      Comparison with classical porous scaffold generating methods.
      The left-most column is the models input to the corresponding methods.
      (a)Method in~\cite{Yoo2011Computer} with hexahedral mesh as input.
      (b)Method in~\cite{Feng2018Porous} with T-spline solid as input.
      (c)Our method with TBSS as input.
      }
   \label{fig:comaprison}
  \end{center}
\end{figure}

 The heterogenous porous scaffold generation method developed in this study
    is compared here with two classical methods presented in Refs.~\cite{Yoo2011Computer,Feng2018Porous}.
 First, because the method proposed in~\cite{Yoo2011Computer}
    generates a porous scaffold by mapping a regular TPMS unit to each hexahedron of a hexahedral mesh model,
    it has the following shortcomings:
 (1) Continuity cannot be ensured between two adjacent TPMS units in the porous scaffold.
 (2) The threshold values $C$ of all TPMS units are the same.
 (3) The geometric quality of the porous scaffold is greatly influenced by
    the mesh quality of the hexahedral mesh model.
 As illustrated in Fig.~\ref{subfig:porous_using_eightnodes_method},
    the boundary mesh quality is poor,
    with many slender triangles.

 Secondly, the method presented in~\cite{Feng2018Porous} produces
    a porous scaffold by first immersing a trivariate T-spline model in an ambient TPMS
    and then taking the intersection of them as the porous scaffold.
 Therefore, this method cannot guarantee completeness of the TPMS units.
 As demonstrated in Fig.~\ref{subfig:porous_using_tspline_method},
    many boundary TPMS units are broken.

 However, because our method generates a heterogeneous porous scaffold by
    mapping a unitary TPMS in the parametric domain to a TBSS, it avoids the shortcomings of the other two methods.
 The heterogeneous porous scaffold generated by our method has the following
    properties (Fig.~\ref{subfig:porous_using_our_method}):
    \begin{itemize}
        \item Completeness of TPMS units and continuity between adjacent TPMS units are guaranteed.
        \item The TDF can be designed by users to easily control the porosity.
        \item Because the degree of a TBSS is relatively high,
            and a TBSS has high smoothness,
            the heterogenous porous scaffold generated by TBSS mapping is highly smooth.
        \item The TDF file format saves significant storage space.
    \end{itemize}

\begin{figure*}[!htb]
  \begin{center}
  \subfigure[]{
    \label{subfig:balljoint_bspline_solid}
    \includegraphics[width=0.18\textwidth]{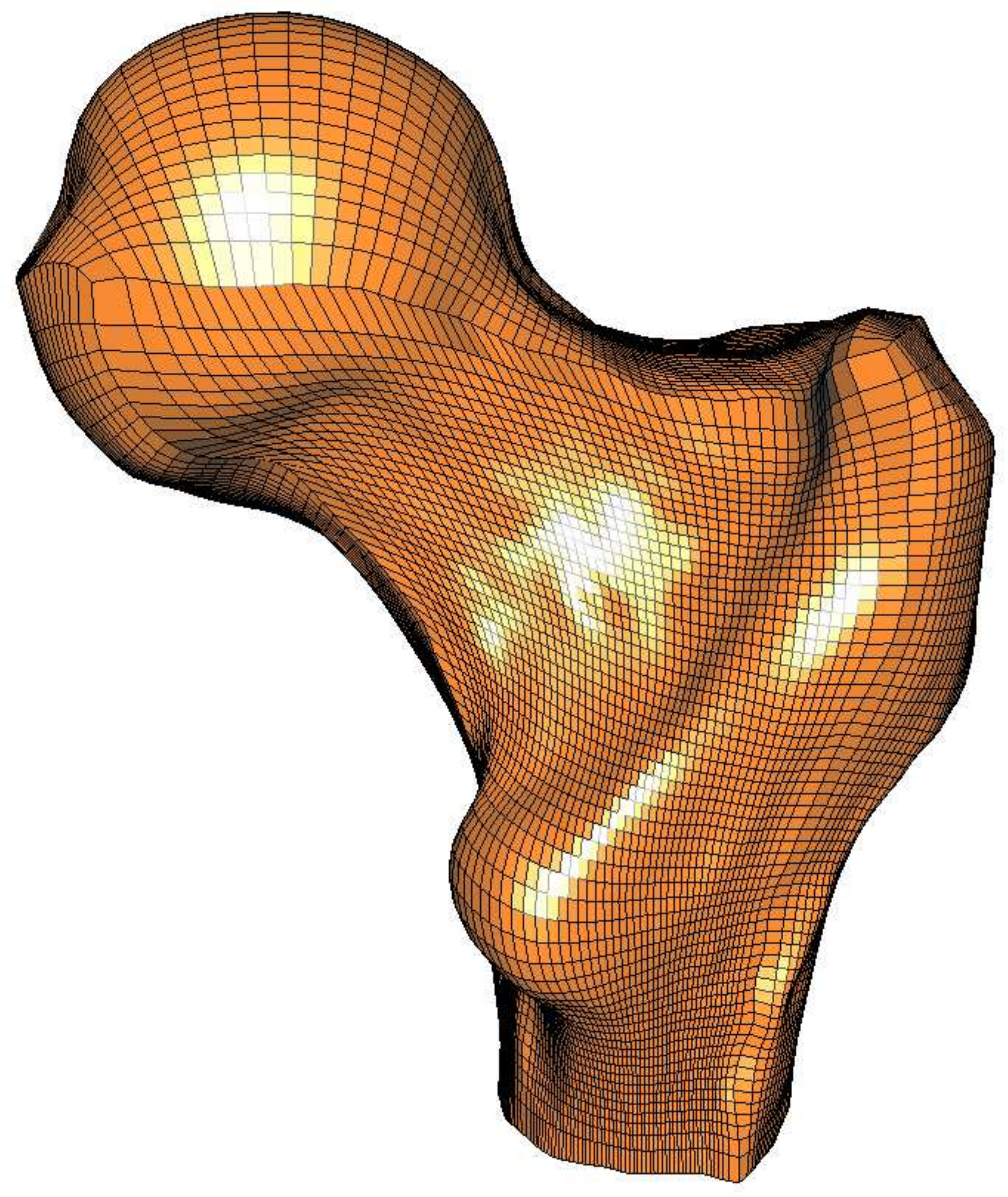}}
  \subfigure[]{
    \label{subfig:porosity_distribution_balljoint}
    \includegraphics[width=0.2\textwidth]{porosity_distribution_balljoint-eps-converted-to.pdf}}
  \subfigure[]{
    \label{subfig:balljoint_p_pore_scaffold}
    \includegraphics[width=0.18\textwidth]{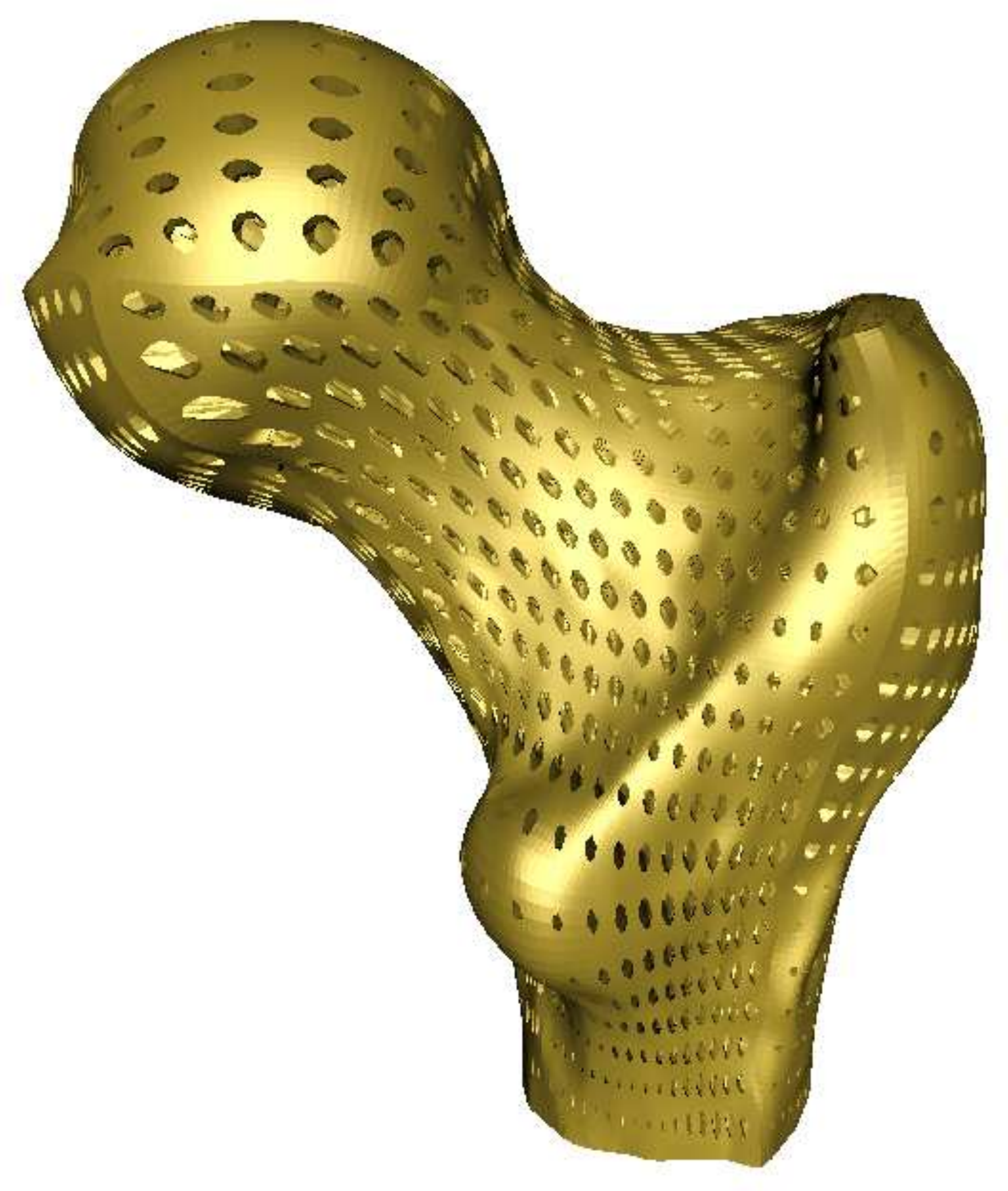}}
  \subfigure[]{
    \label{subfig:balljoint_p_rod_scaffold}
    \includegraphics[width=0.18\textwidth]{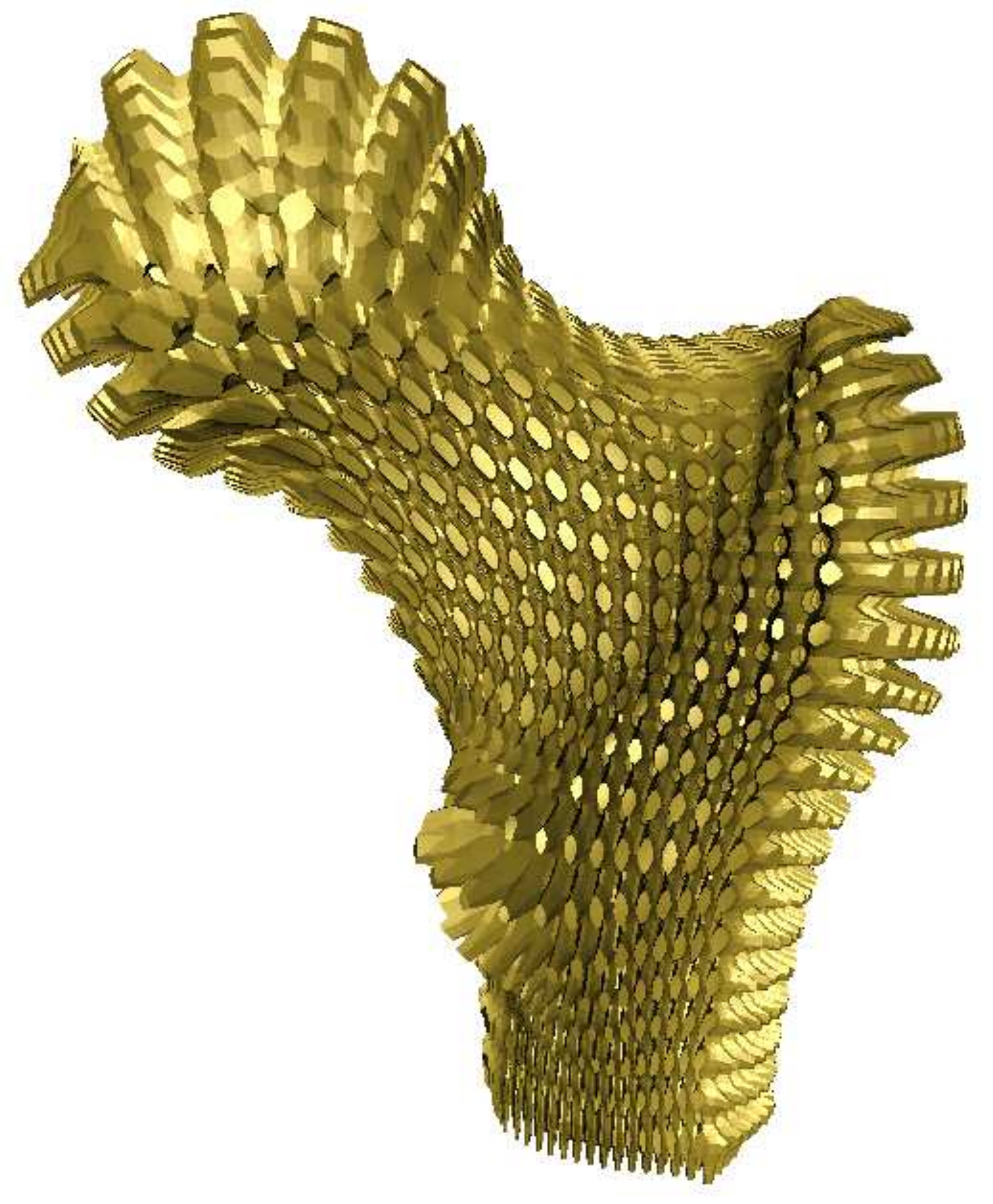}}
  \subfigure[]{
    \label{subfig:balljoint_p_sheet_scaffold}
    \includegraphics[width=0.18\textwidth]{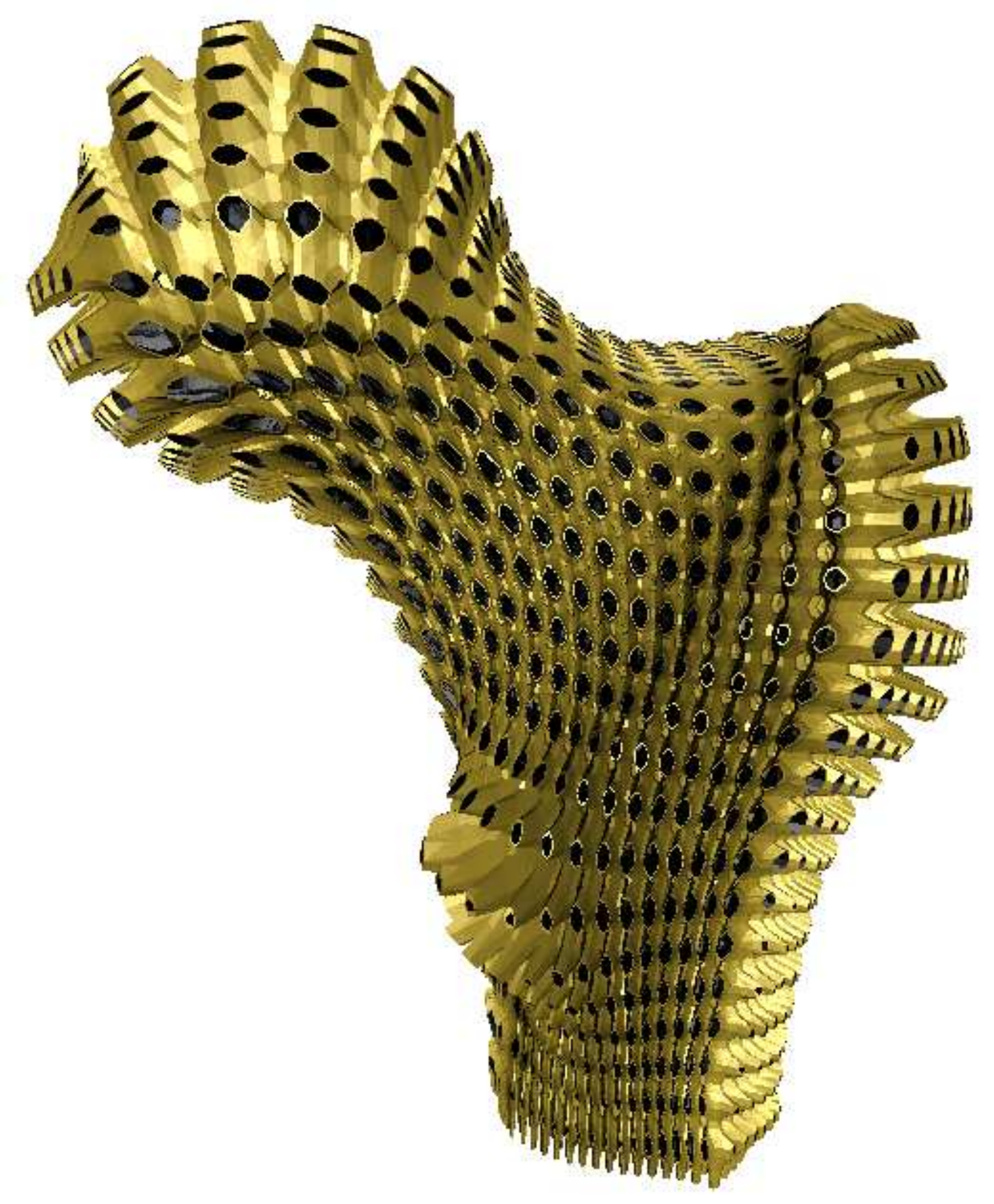}}
          \caption
          {\small
            Heterogeneous porous scaffold of \emph{Ball-joint}.
            (a) TBSS.
            (b) TDF in the parametric domain.
            (c) P-type pore structure.
            (d) P-type rod structure.
            (e) P-type sheet structure.
          }
   \label{fig:balljoint_porous_scaffold}
  \end{center}
\end{figure*}

\begin{figure*}[!htb]
  \begin{center}
  \subfigure[]{
    \label{subfig:venus_bspline_solid}
    \includegraphics[width=0.18\textwidth]{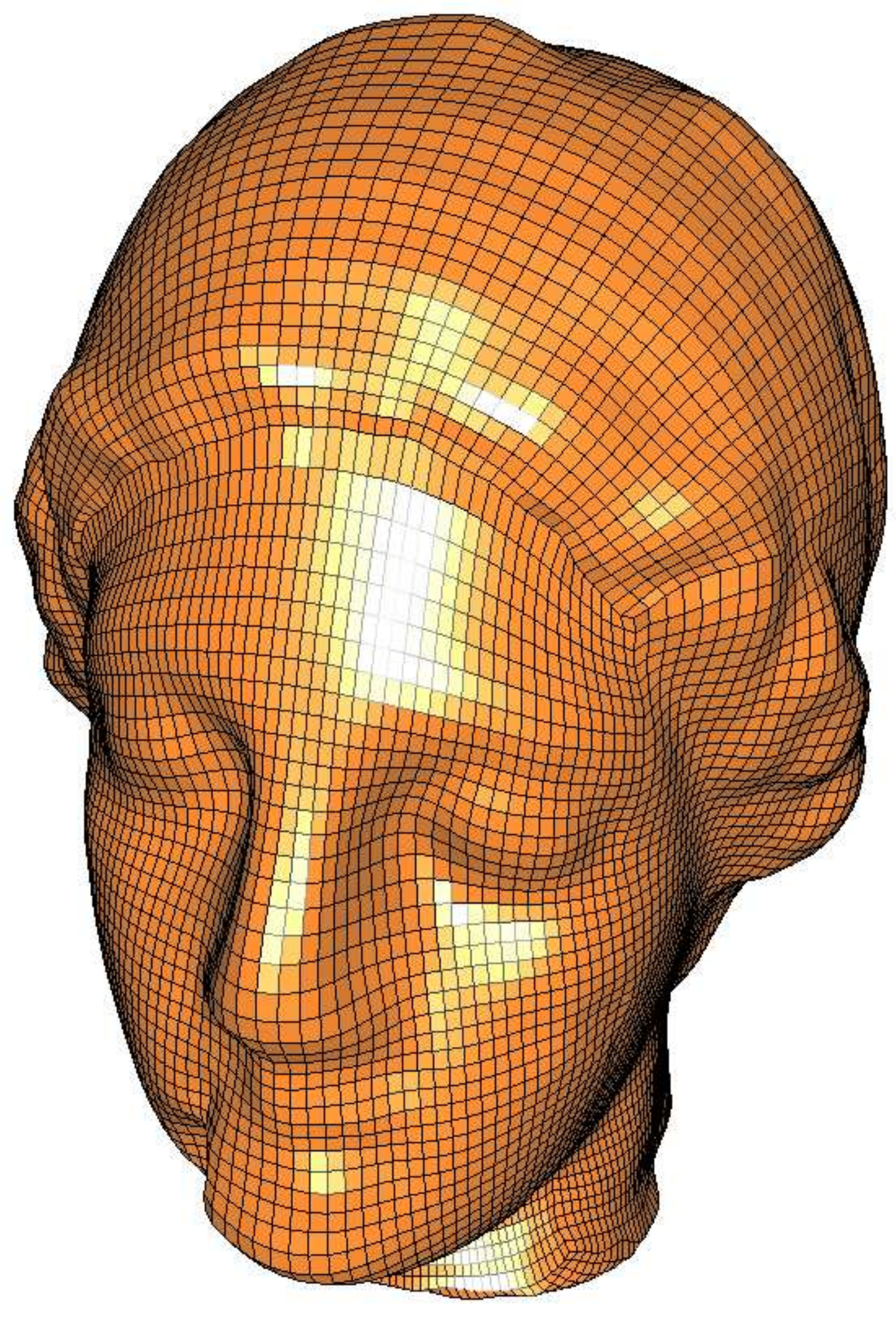}}
  \subfigure[]{
    \label{subfig:porosity_distribution_venus}
    \includegraphics[width=0.2\textwidth]{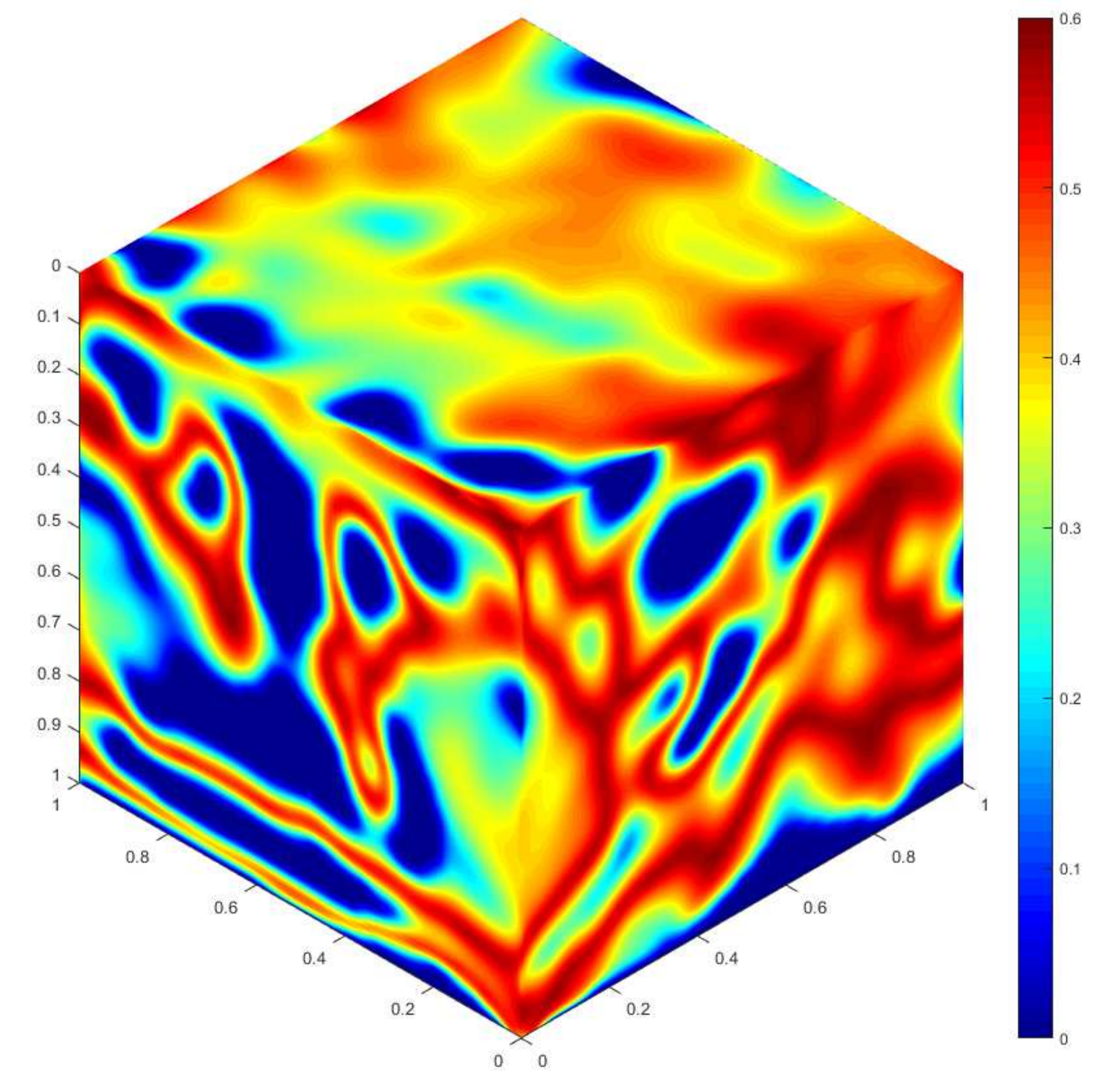}}
  \subfigure[]{
    \label{subfig:venus_d_pore_scaffold}
    \includegraphics[width=0.18\textwidth]{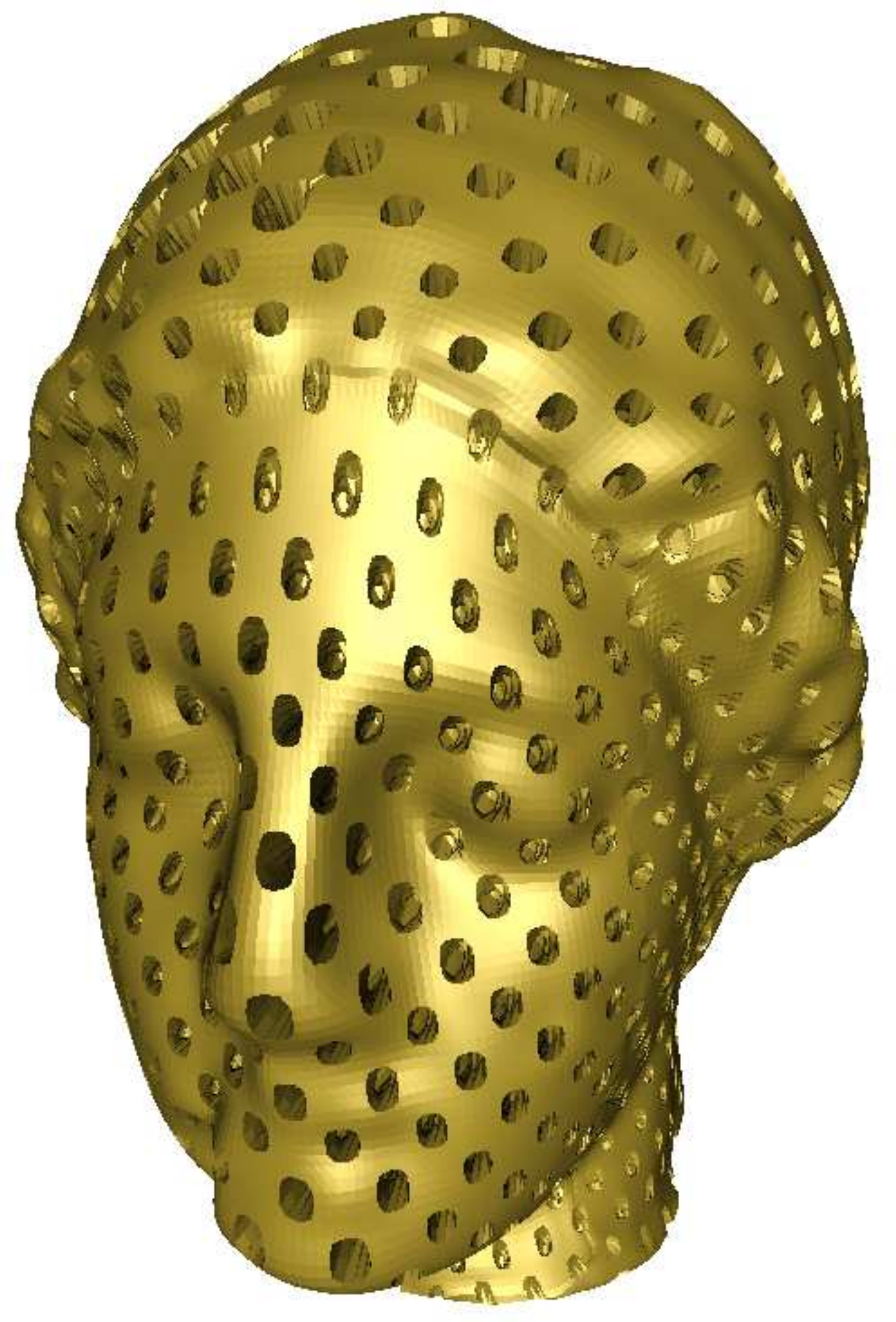}}
  \subfigure[]{
    \label{subfig:venus_d_rod_scaffold}
    \includegraphics[width=0.18\textwidth]{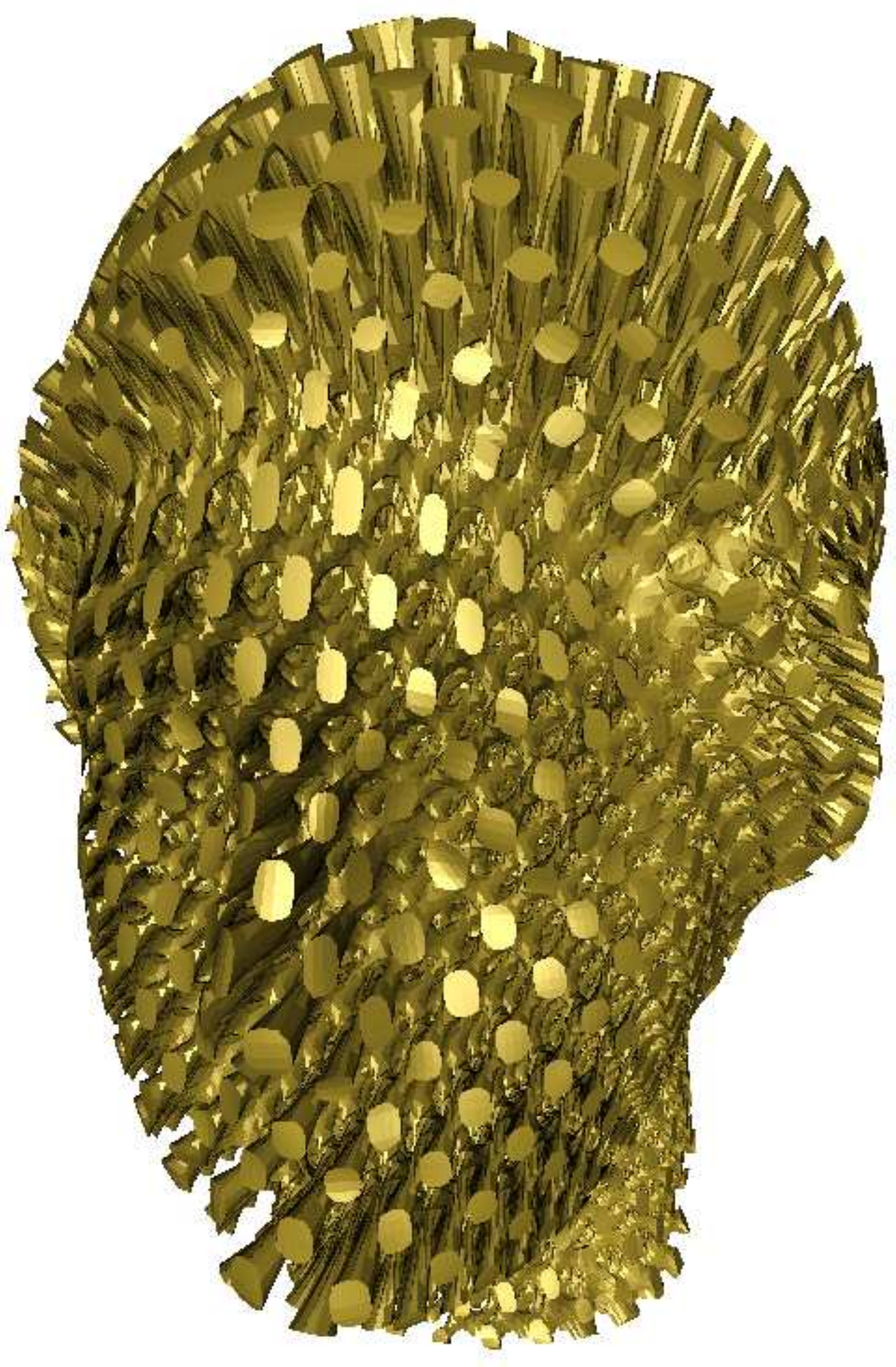}}
  \subfigure[]{
    \label{subfig:venus_d_sheet_scaffold}
    \includegraphics[width=0.18\textwidth]{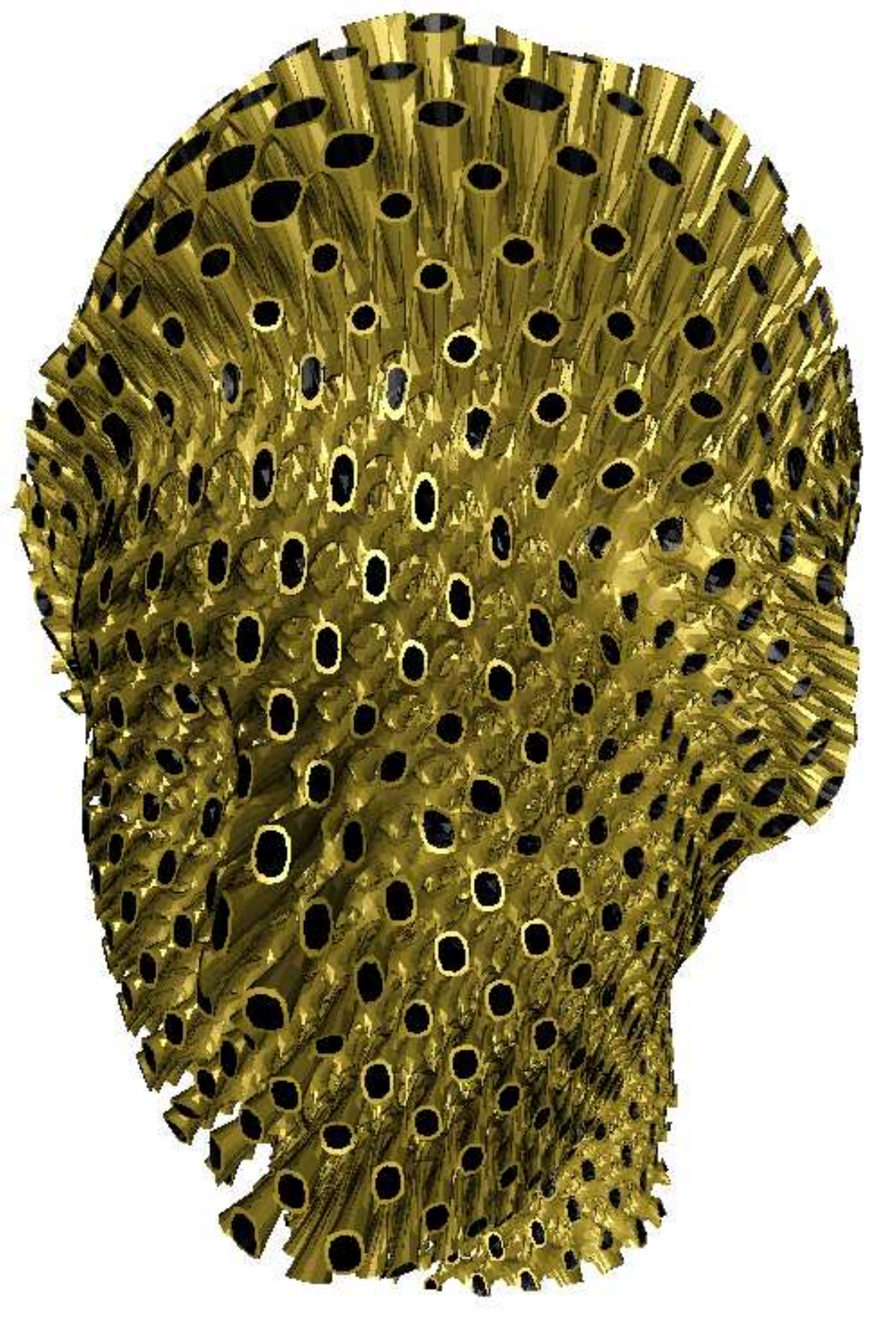}}
          \caption
          {\small
            Heterogeneous porous scaffold of \emph{Venus}.
            (a) TBSS.
            (b) TDF in the parametric domain.
            (c) D-type pore structure.
            (d) D-type rod structure.
            (e) D-type sheet structure.
          }
   \label{fig:venus_porous_scaffold}
  \end{center}
\end{figure*}

\begin{figure*}[!htb]
  \begin{center}
  \subfigure[]{
    \label{subfig:moai_bspline_solid}
    \includegraphics[width=0.12\textwidth]{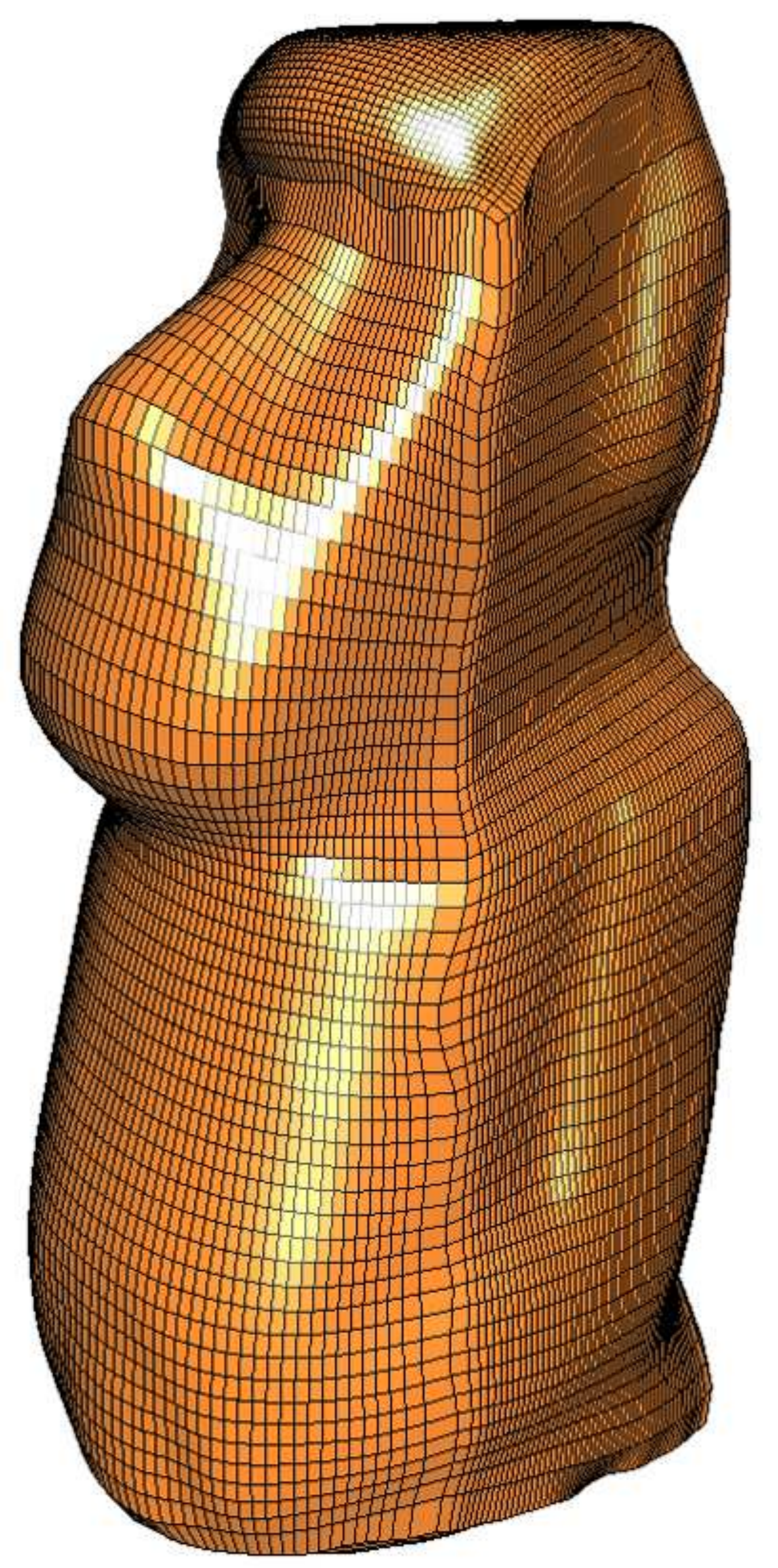}}
  \hspace{0.055\textwidth}
  \subfigure[]{
    \label{subfig:porosity_distribution_moai}
    \includegraphics[width=0.2\textwidth]{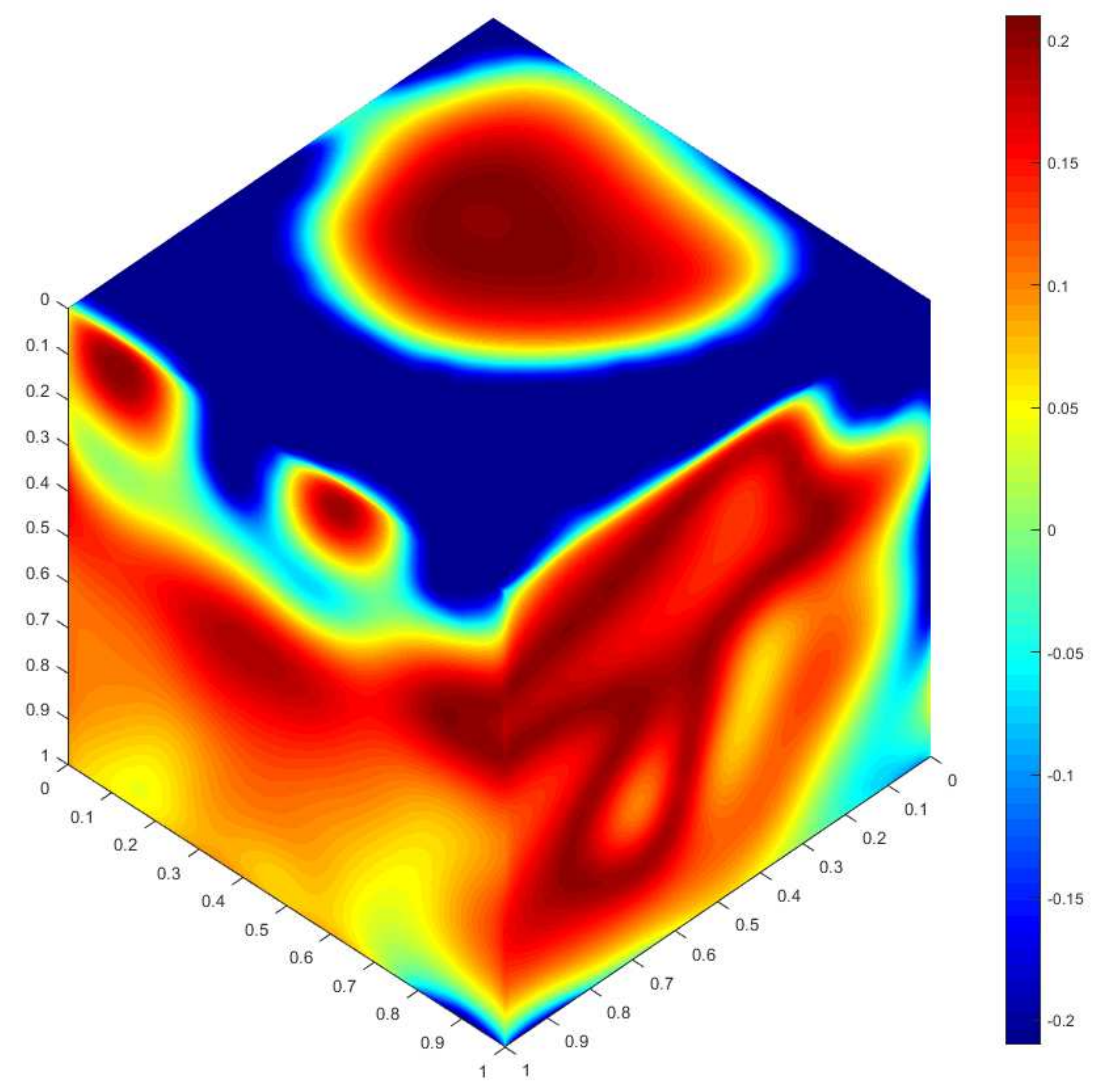}}
  \hspace{0.055\textwidth}
  \subfigure[]{
    \label{subfig:moai_iwp_pore_scaffold}
    \includegraphics[width=0.12\textwidth]{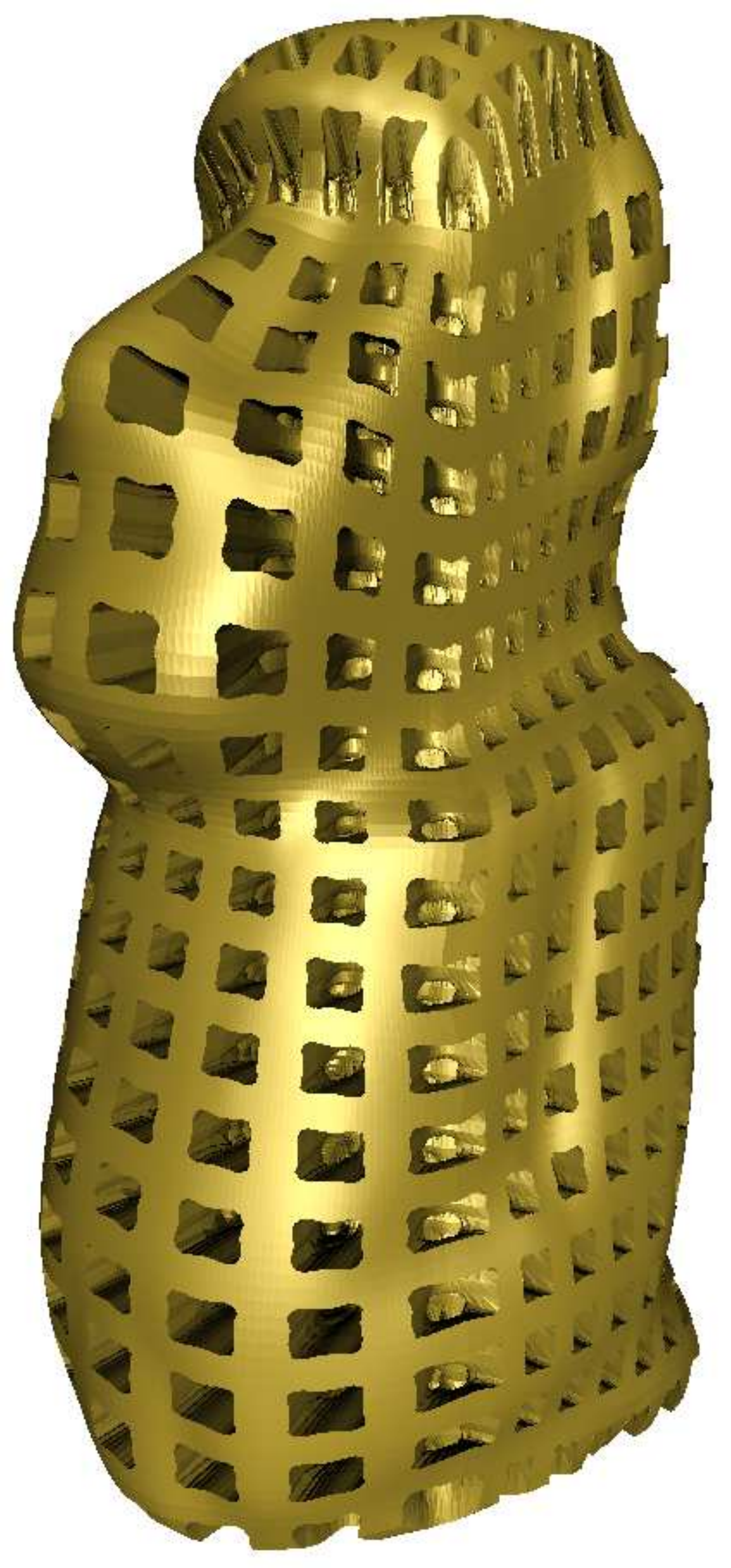}}
  \hspace{0.055\textwidth}
  \subfigure[]{
    \label{subfig:moai_iwp_rod_scaffold}
    \includegraphics[width=0.12\textwidth]{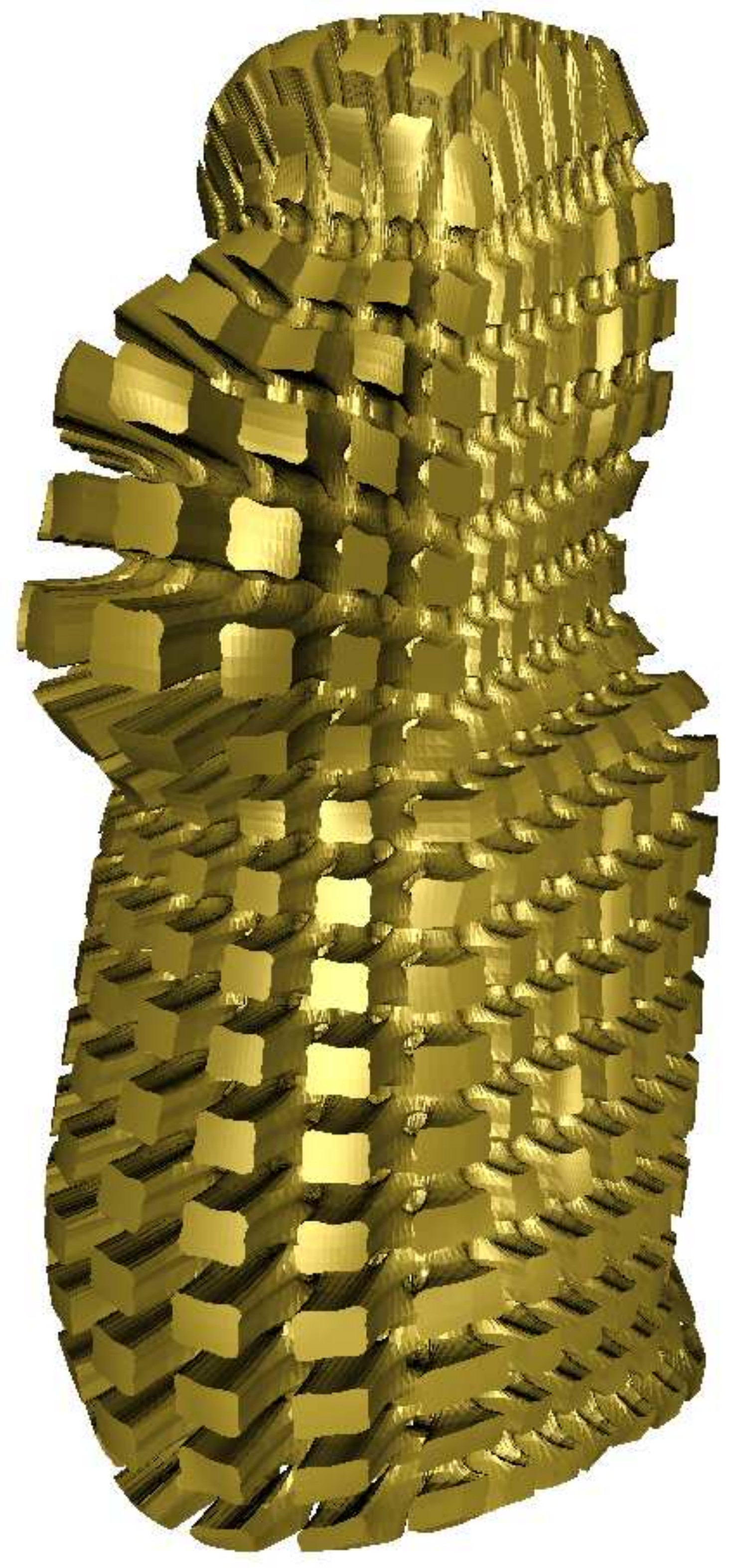}}
  \hspace{0.055\textwidth}
  \subfigure[]{
    \label{subfig:moai_iwp_sheet_scaffold}
    \includegraphics[width=0.12\textwidth]{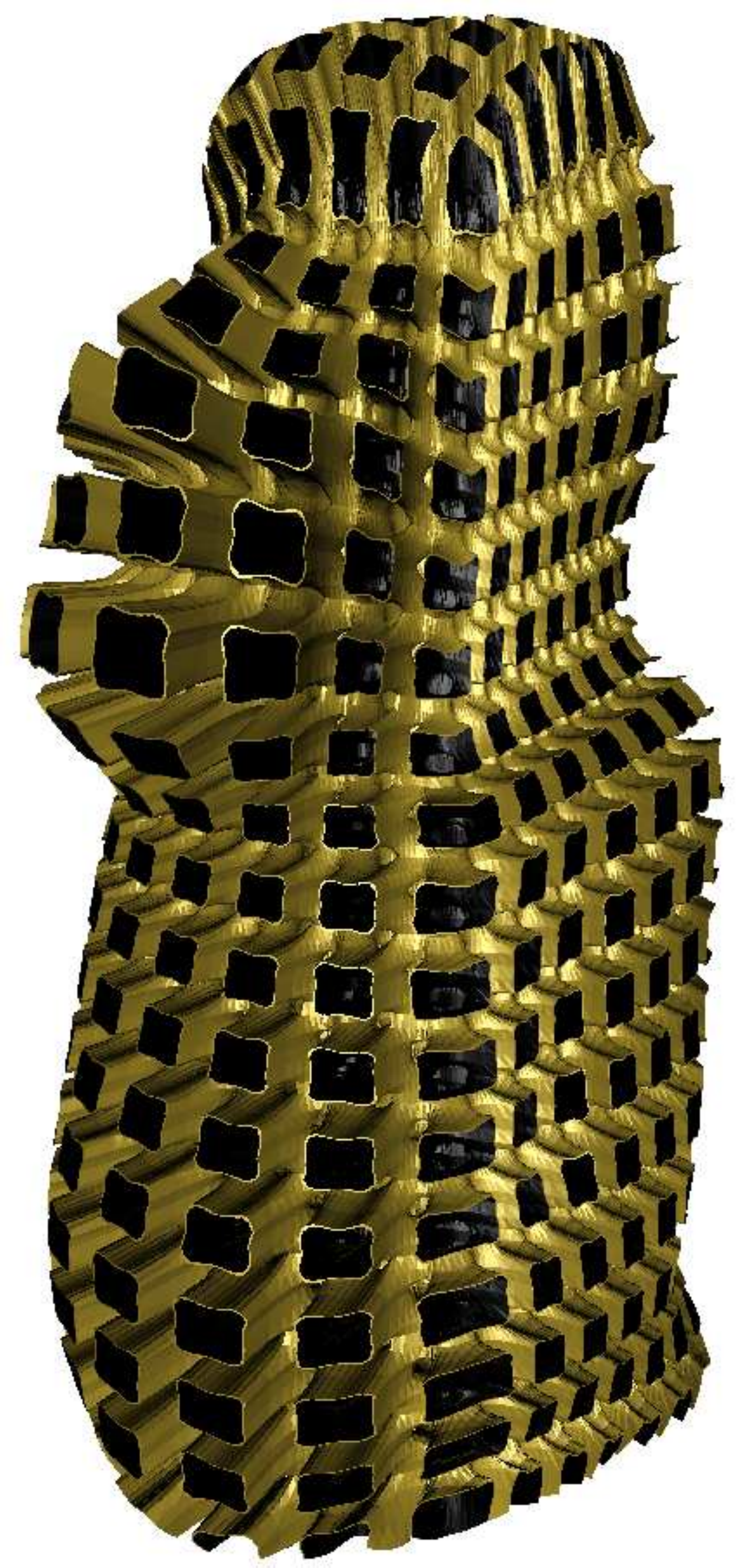}}
    \caption
    {
        \small
        Heterogeneous porous scaffold of \emph{Moai}.
        (a) TBSS.
        (b) TDF in the parametric domain.
        (c) I-WP-type pore structure.
        (d) I-WP-type rod structure.
        (e) I-WP-type sheet structure.
    }
   \label{fig:moai_porous_scaffold}
  \end{center}
\end{figure*}

\begin{figure*}[!htb]
  \begin{center}
  \subfigure[]{
    \label{subfig:tooth_bspline_solid}
    \includegraphics[width=0.18\textwidth]{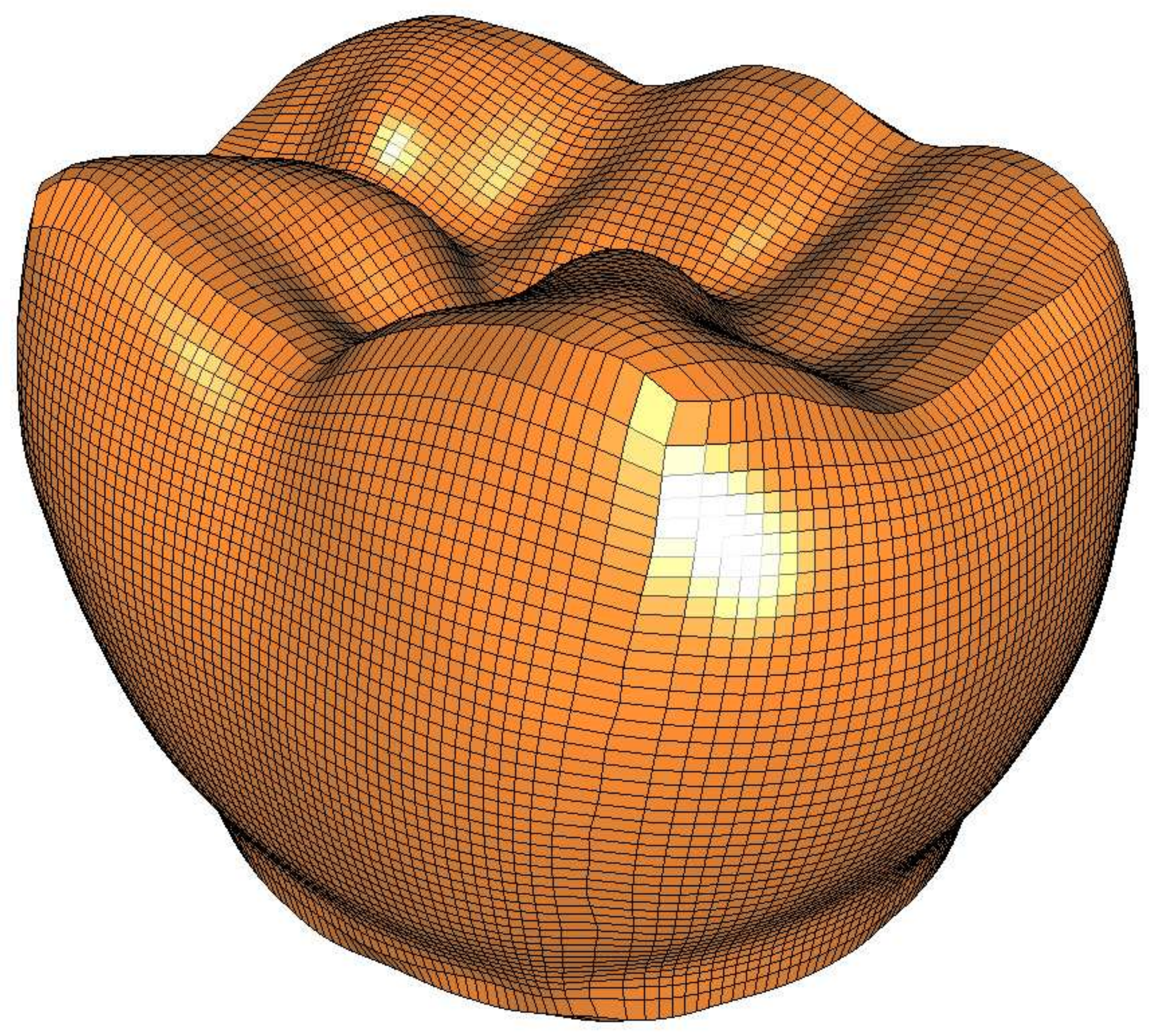}}
  \subfigure[]{
    \label{subfig:porosity_distribution_tooth}
    \includegraphics[width=0.2\textwidth]{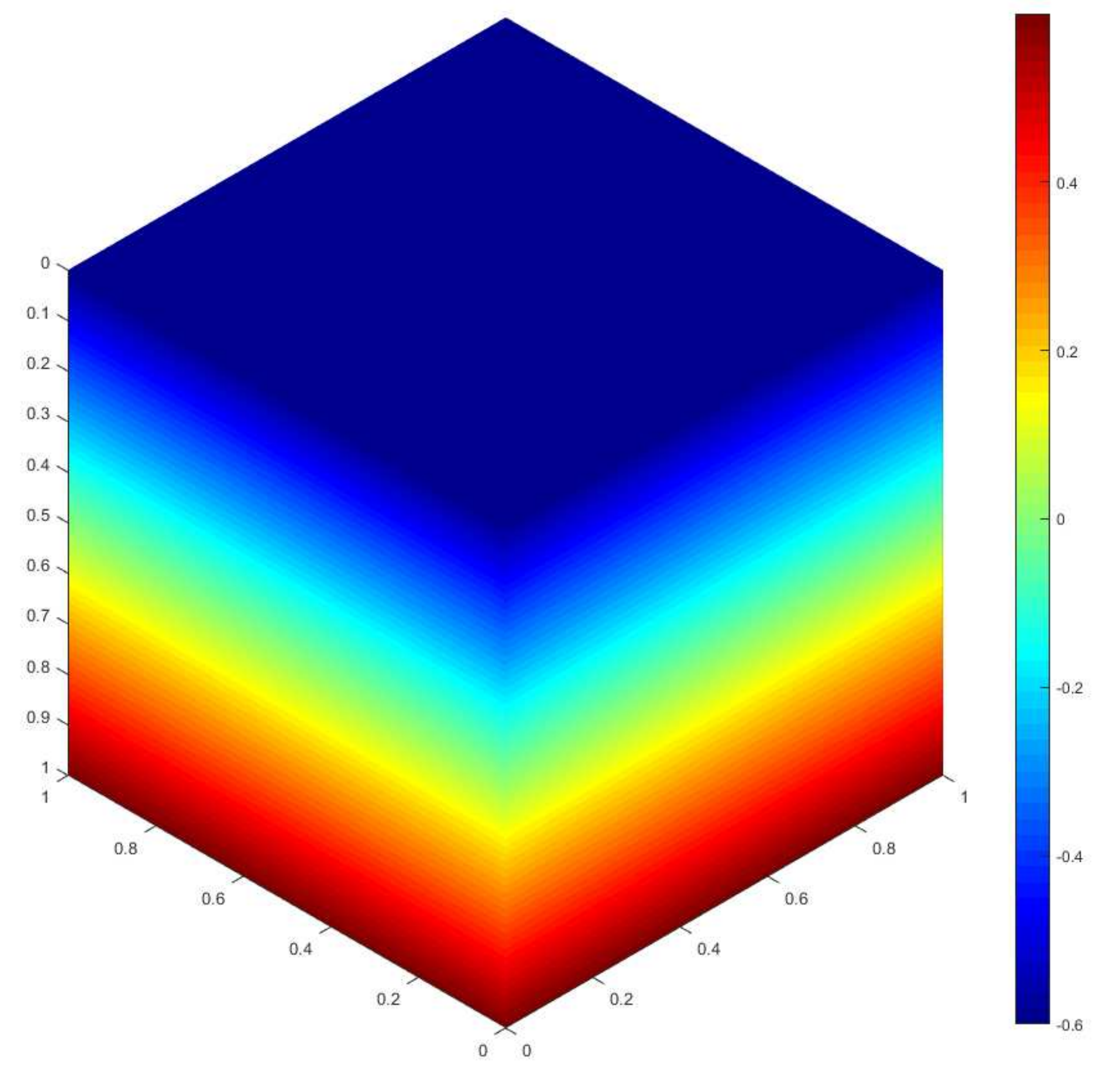}}
  \subfigure[]{
    \label{subfig:tooth_g_pore_scaffold}
    \includegraphics[width=0.18\textwidth]{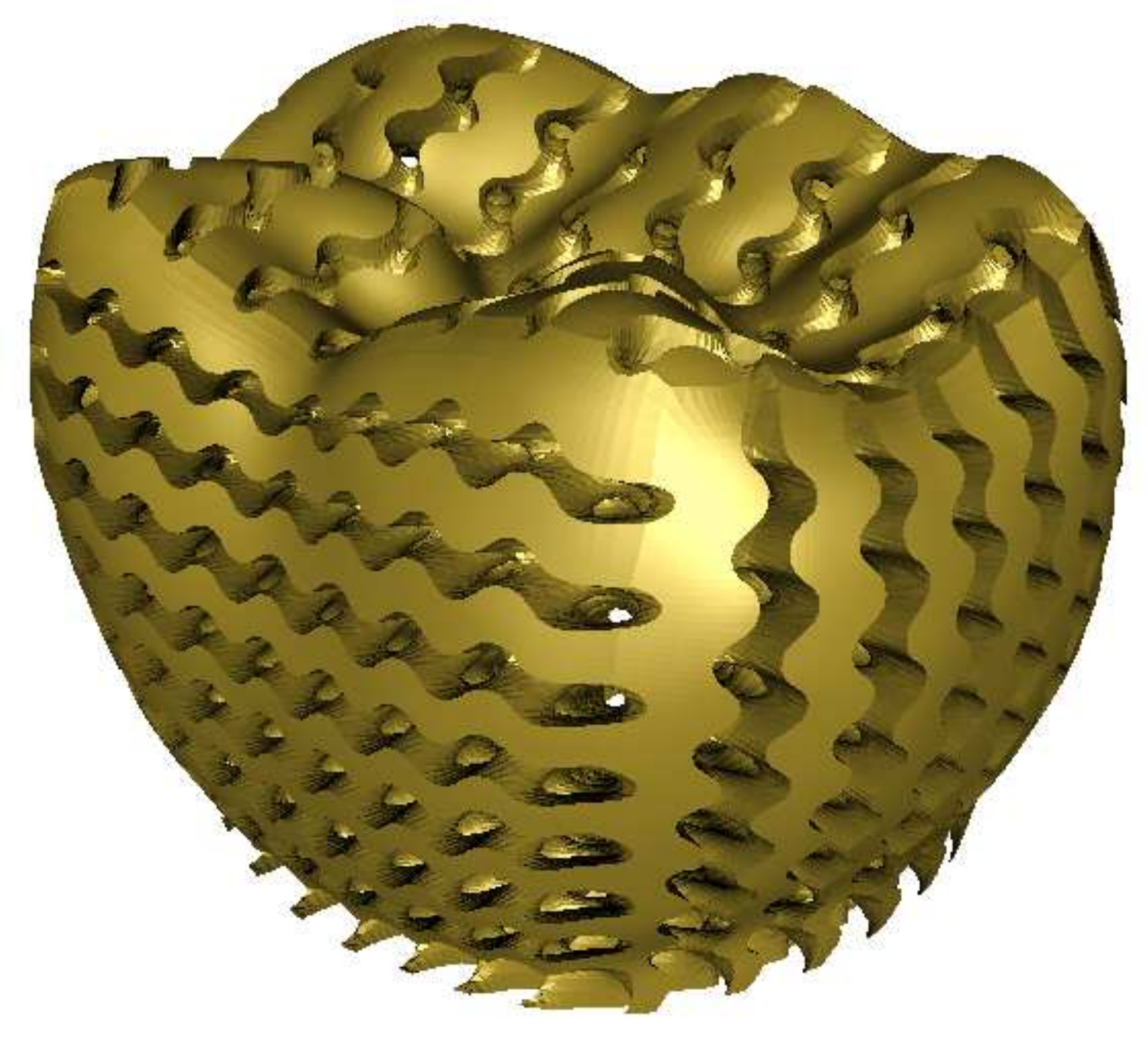}}
  \subfigure[]{
    \label{subfig:tooth_g_rod_scaffold}
    \includegraphics[width=0.18\textwidth]{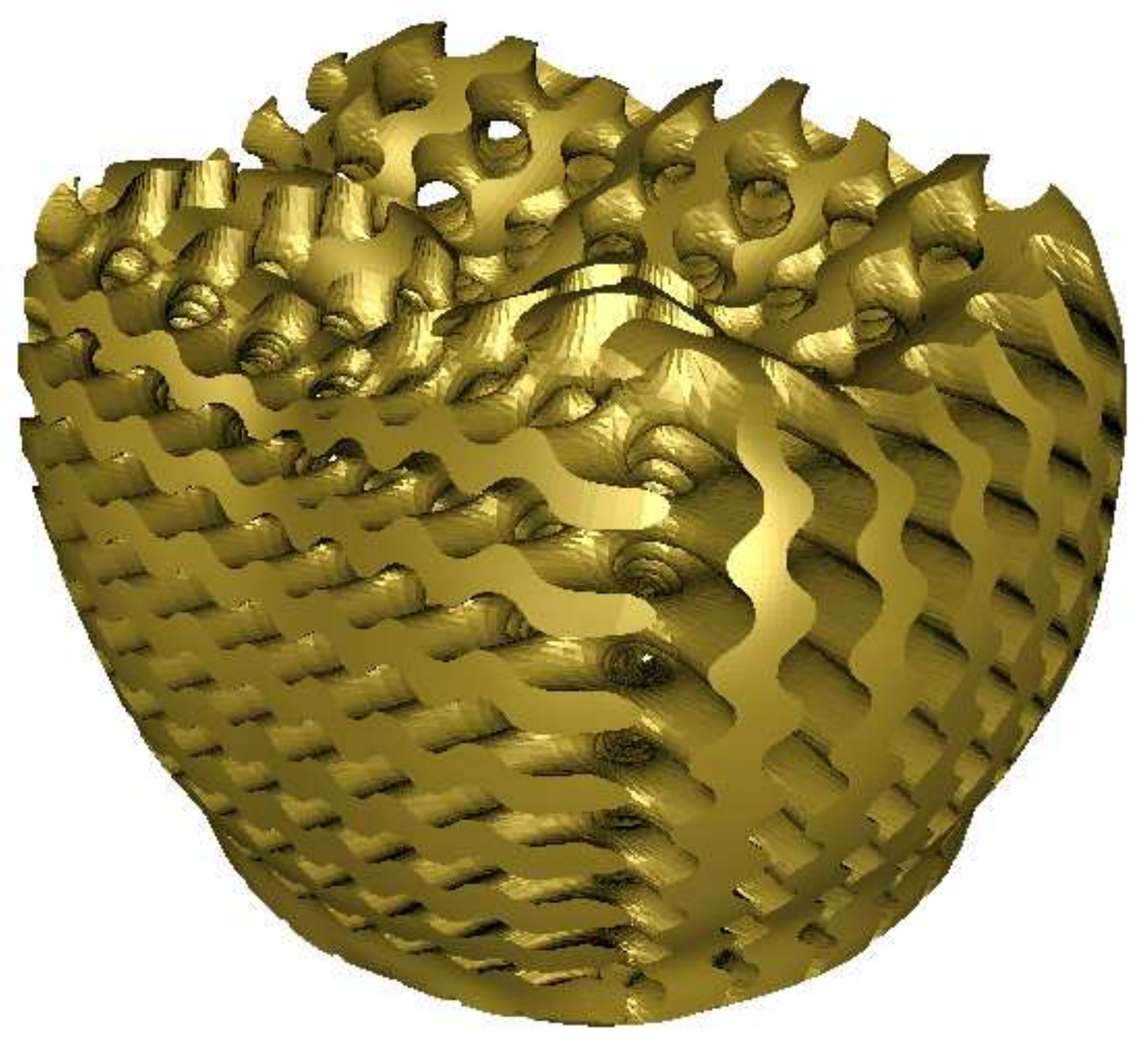}}
  \subfigure[]{
    \label{subfig:tooth_g_sheet_scaffold}
    \includegraphics[width=0.18\textwidth]{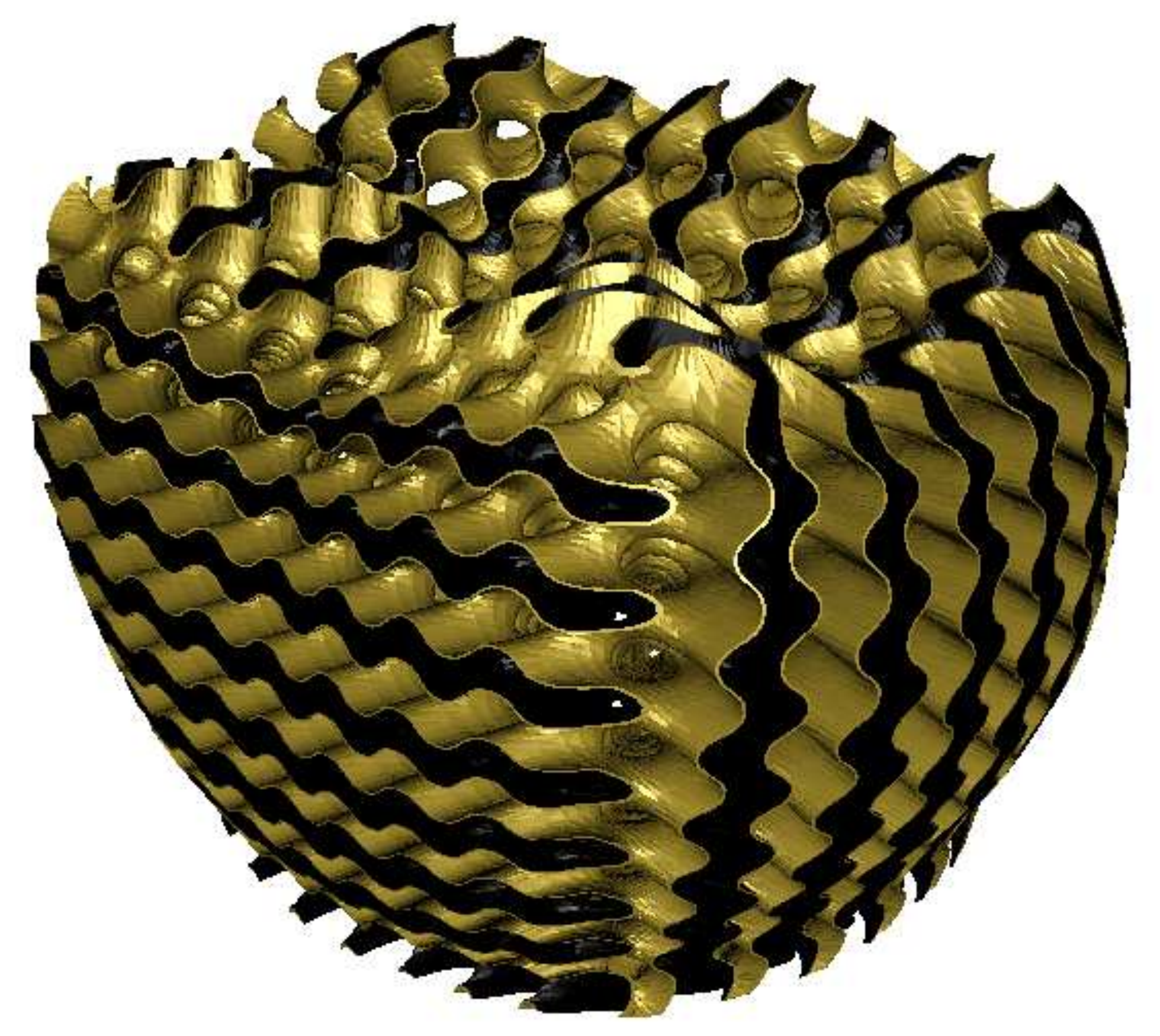}}
    \caption
    {
        \small
        Heterogeneous porous scaffold of \emph{Tooth}.
        (a) TBSS.
        (b) TDF in the parametric domain.
        (c) G-type pore structure.
        (d) G-type rod structure.
        (e) G-type sheet structure.
    }
   \label{fig:tooth_porous_scaffold}
  \end{center}
\end{figure*}

\begin{figure*}[!htb]
  \begin{center}
  \subfigure[]{
    \label{subfig:isis_bspline_solid}
    \includegraphics[width=0.1\textwidth]{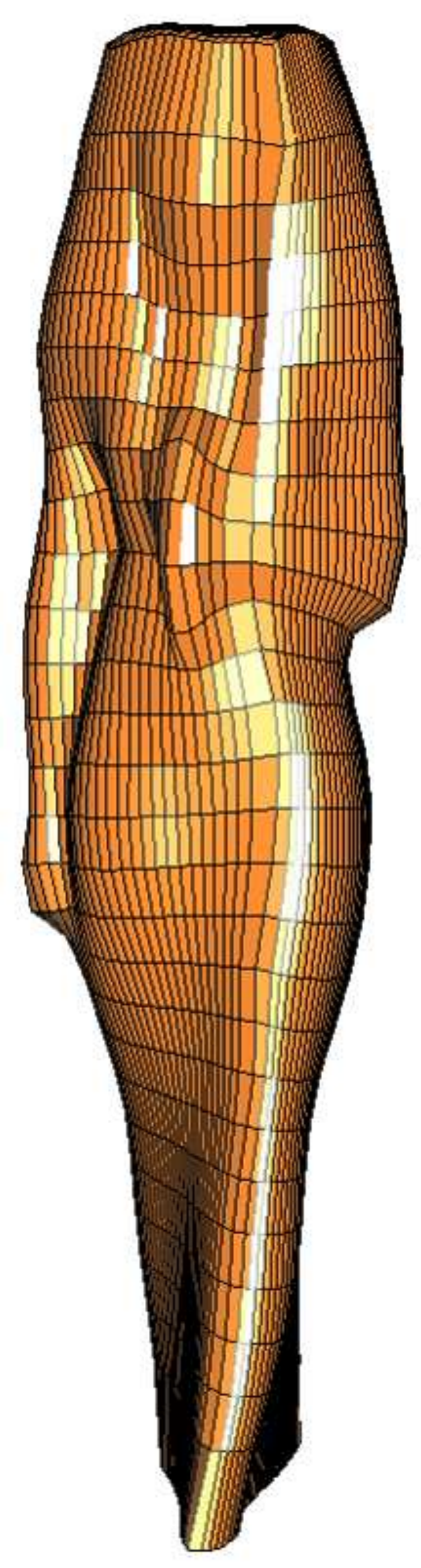}}
  \hspace{0.06\textwidth}
  \subfigure[]{
    \label{subfig:porosity_distribution_isis}
    \includegraphics[width=0.2\textwidth]{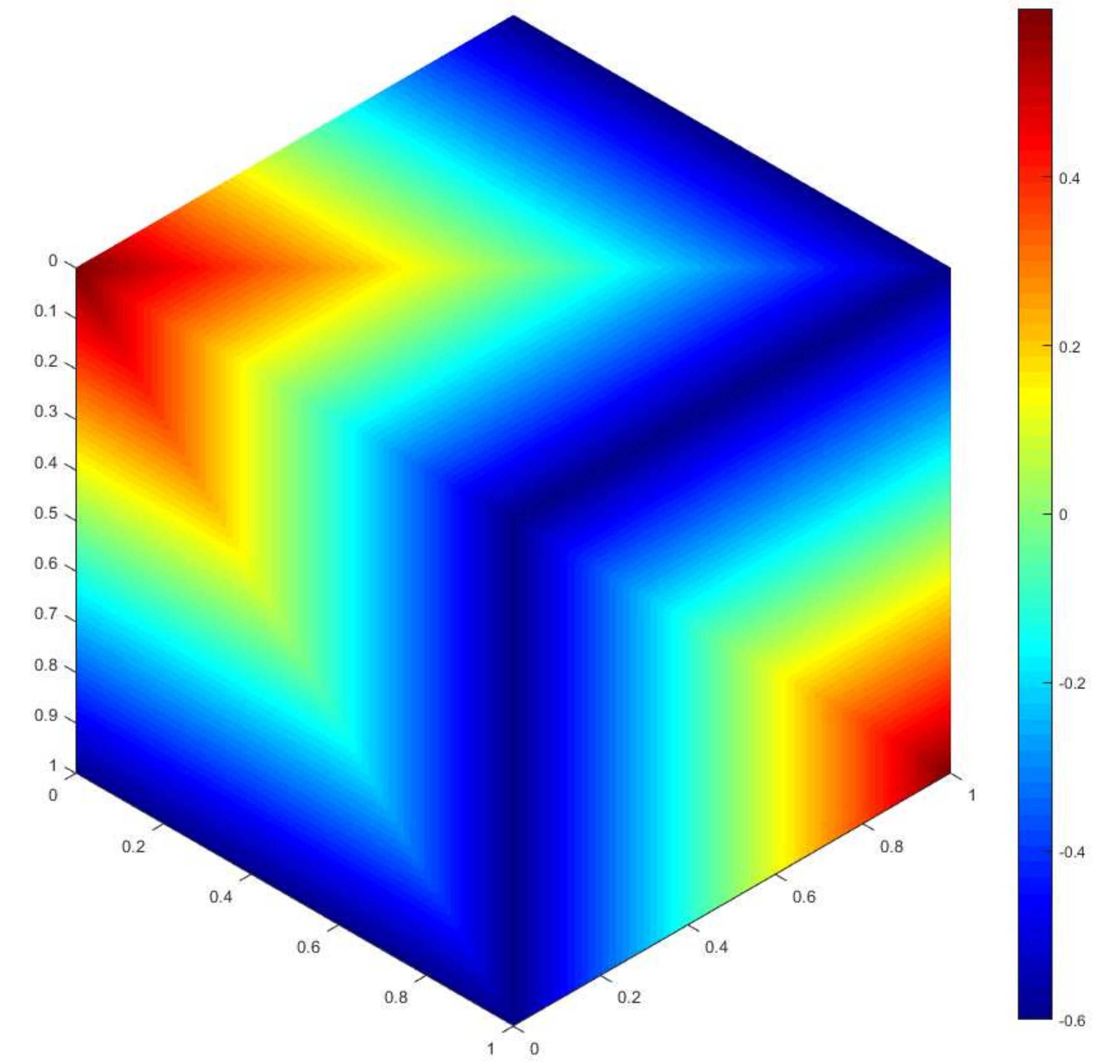}}
  \hspace{0.06\textwidth}
  \subfigure[]{
    \label{subfig:isis_p_pore_scaffold}
    \includegraphics[width=0.1\textwidth]{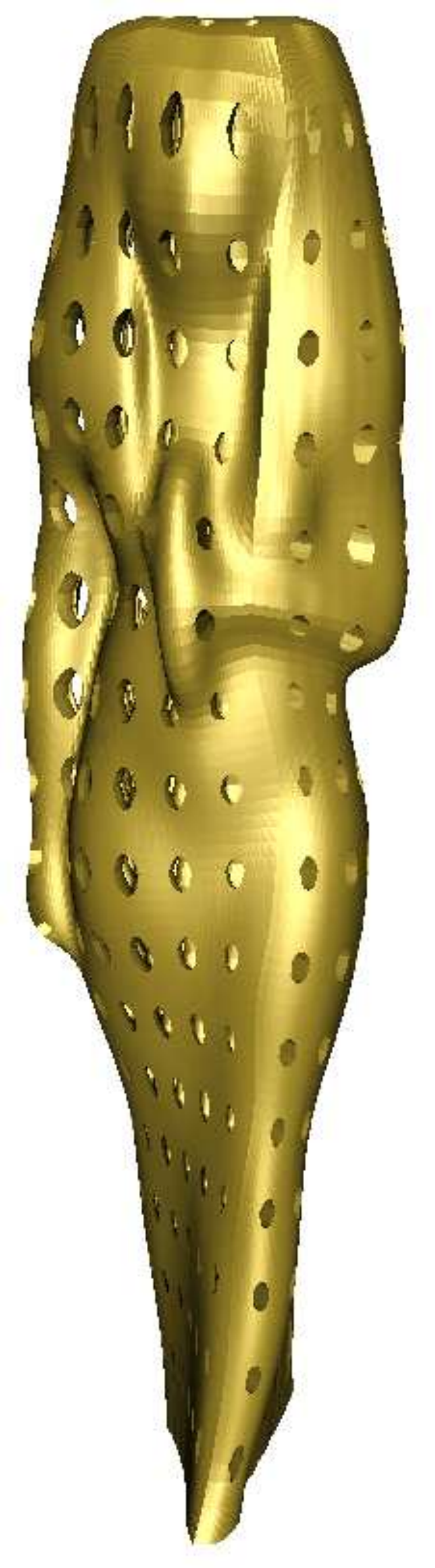}}
  \hspace{0.06\textwidth}
  \subfigure[]{
    \label{subfig:isis_p_rod_scaffold}
    \includegraphics[width=0.1\textwidth]{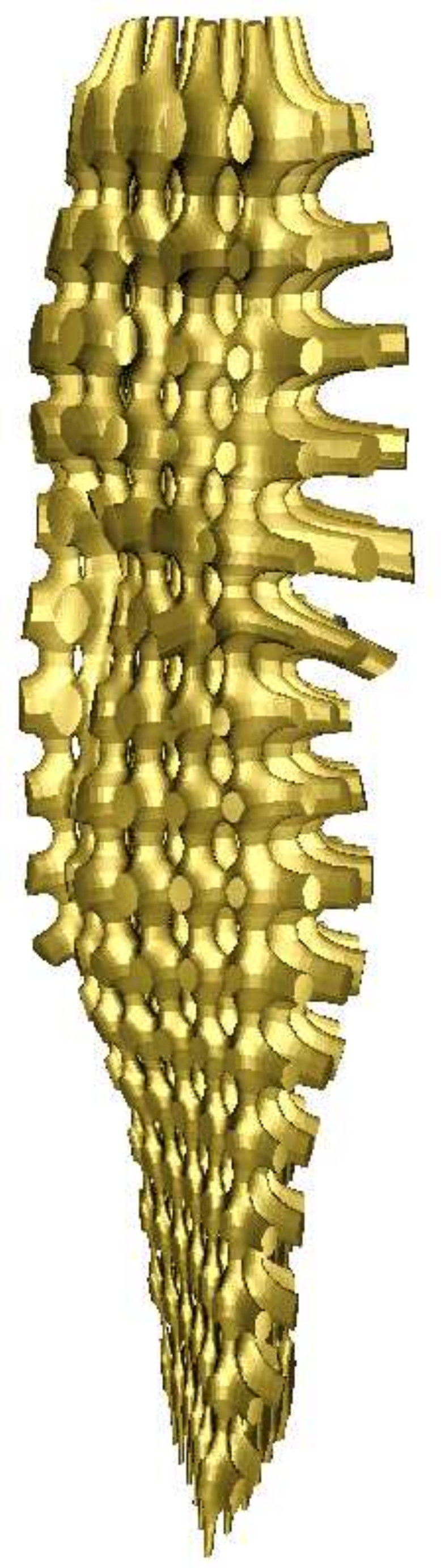}}
  \hspace{0.06\textwidth}
  \subfigure[]{
    \label{subfig:isis_p_sheet_scaffold}
    \includegraphics[width=0.1\textwidth]{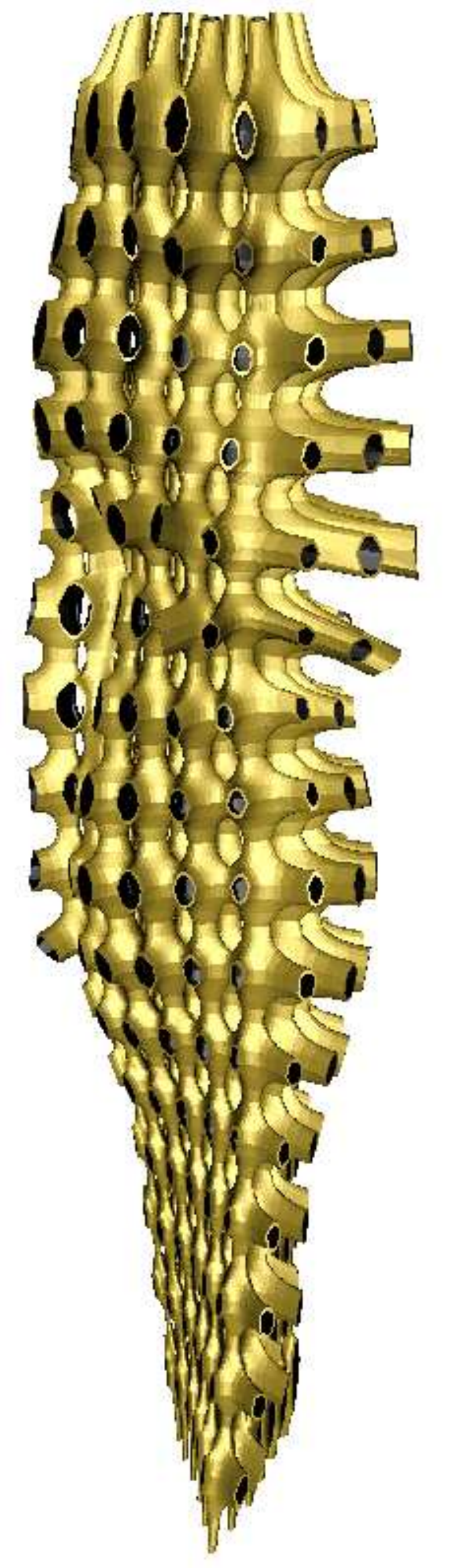}}
    \caption
    {
        \small
        Heterogeneous porous scaffold of \emph{Isis}.
        (a) TBSS.
        (b) TDF in the parametric domain.
        (c) P-type pore structure.
        (d) P-type rod structure.
        (e) P-type sheet structure.
    }
   \label{fig:isis_porous_scaffold}
  \end{center}
\end{figure*}


  \begin{table*}[!htb]
  \centering
  \footnotesize
  \caption{Statistical data of the heterogenous porous scaffold generation method developed in this paper.}
  \label{tbl:stat}
  \begin{threeparttable}
  \begin{tabular}{| c | c | c | c | c | c | c | c | c |}
  \hline
    \multirow{2}{*}{Model} & \multirow{2}{*}{Type} & \multirow{2}{*}{Structure}& \multirow{2}{*}{Period coefficients} & \multicolumn{3}{|c|}{Run time(s)\tnote{1}} & \multicolumn{2}{|c|}{Storage space(MB)\tnote{2}}\\
    \cline{5-7} \cline{8-9}
    & & &  & {TDF} & {TPMS} &{Porous scaffold} & {STL format} & {TDF format}\\
  \hline
    \multirow{3}{*}{Ball joint} & \multirow{3}{*}{P} & pore  & \multirow{3}{*}{$ (16,14,18)$}& \multirow{3}{*}{2.745} &0.326 & 3.194 & 741.367 & \multirow{3}{*}{0.810}\\
    \cline{3-3} \cline{6-8}
     &  & rod &  &  &0.319 & 3.109 & 721.599 & \\
    \cline{3-3} \cline{6-8}
     & & sheet &  &  &0.658 & 6.651 & 1449.71 & \\
  \hline
    \multirow{3}{*}{Venus} & \multirow{3}{*}{D}& pore & \multirow{3}{*}{$ (10,10,10)$}  & \multirow{3}{*}{2.728}& 0.293 & 3.024 & 701.682 & \multirow{3}{*}{0.947}\\
    \cline{3-3} \cline{6-8}
     & & rod &  & &0.286 & 3.007 & 695.564 & \\
    \cline{3-3} \cline{6-8}
     & & sheet &  & &0.528 & 6.581 & 1399.55 & \\
  \hline
    \multirow{3}{*}{Moai} & \multirow{3}{*}{I-WP}& pore & \multirow{3}{*}{$ (6,6,16)$} & \multirow{3}{*}{2.736} & 0.290 & 2.554 &  557.324& \multirow{3}{*}{0.824}\\
    \cline{3-3} \cline{6-8}
     & & rod &  & &0.284 & 2.555 & 554.906 & \\
    \cline{3-3} \cline{6-8}
     & & sheet &  &  &0.583 & 5.530 & 1105.41 & \\
  \hline
    \multirow{3}{*}{Tooth} & \multirow{3}{*}{G}& pore & \multirow{3}{*}{$ (8,6,8)$} & \multirow{3}{*}{2.732}  &0.238 & 1.974 & 394.402 & \multirow{3}{*}{0.567}\\
    \cline{3-3} \cline{6-8}
     & & rod &  &  &0.237 & 1.956 & 394.422 & \\
    \cline{3-3} \cline{6-8}
     & & sheet &  &  &0.499 & 4.156 & 773.741 & \\
  \hline
    \multirow{3}{*}{Isis} & \multirow{3}{*}{P}& pore  & \multirow{3}{*}{$(6,6,16)$} & \multirow{3}{*}{2.754} &0.238 & 2.089 & 473.401 & \multirow{3}{*}{1.355}\\
    \cline{3-3} \cline{6-8}
     & & rod &  &  &0.234 & 1.994 & 454.359 & \\
    \cline{3-3} \cline{6-8}
     & & sheet &  &  &0.521 & 4.236 & 908.205 & \\
  \hline
  \end{tabular}
 \begin{tablenotes}
    \item[1] Run time (in second) for TDF construction, generation of volume TPMS in parametric domain and generation of heterogenous porous scaffold.
    \item[2] Storage space (in megabyte) of heterogenous porous scaffolds using the traditional STL file format and the TDF file format developed in this paper.
 \end{tablenotes}
 \end{threeparttable}
 \end{table*}

 \subsection{Results}
 \label{subsec:results}

 In this section,
    some heterogenous porous scaffold results generated by our method are presented (Figs.~\ref{fig:balljoint_porous_scaffold}-\ref{fig:isis_porous_scaffold}) to demonstrate the effectiveness and efficiency of the
    method.
 In Figs.~\ref{fig:balljoint_porous_scaffold}-\ref{fig:isis_porous_scaffold},
    (a) is the input TBSS,
    (b) is the TDF in the parametric domain,
    and (c,d, and e) are the heterogeneous porous scaffolds of pore structure, rod structure, and sheet structure with different TPMS types.
 Moreover, statistical data are listed in Table~\ref{tbl:stat},
    including,
    period coefficients $(\omega_x, \omega_y, \omega_z)$,
    run times for generating the TDF,
    TPMS in parametric domain,
    and heterogenous porous scaffold in the TBSS,
    and storage costs of porous scaffolds with the traditional STL file format and new TDF file format.

\subsection{TDF file format}
\label{subsec:tdf_file_format}

 In Table~\ref{tbl:stat},
    the storage spaces required to store the porous scaffold using the traditional STL file format and the new TDF file format are listed.
 Using the TDF file format,
    storing the porous scaffolds costs $0.567$ to $1.355$ MB,
    while using the STL file format, it costs $394.402$ to $1449.71$ MB.
 Therefore, at least $98\%$ of storage space is saved by using the new
    TDF file format.
 Moreover, in Table~\ref{tbl:stat},
    the time cost for generating the heterogeneous porous scaffold from the TDF file format is also listed,
    including the run time for generating volume TPMS structures and porous scaffolds.
 We can see that the time costs range from $2$ to $7$ seconds,
    which is acceptable for user interaction.

 Finally, the TDF file format brings some extra benefits.
 Traditionally, heterogeneous porous scaffolds have been stored as linear
    mesh models.
 However, the TDF file format stores a trivariate B-spline function.
 Therefore, in theory, a porous scaffold can be generated to any
    prescribed precision using the TDF file format.
 In addition, the period coefficients and control points of the
    trivariate B-spline function,
    stored in the TDF file format,
    can be taken as some types of \emph{parameters}.
 Therefore, a heterogeneous porous scaffold can be changed by altering
    these parameters,
    just like in the parametric modeling technique.

\section{Conclusion}
\label{sec:conclusion}
 In this study,
    we developed a method for generating a heterogeneous porous scaffold in a TBSS by the TDF designed in the parametric domain of the TBSS.
 First, the TDF is easy to be designed in the cubic parameter domain,
    and is represented as a trivariate B-spline function.
 The TDF can be employed to control the porosity of the porous scaffold.
 Then, a TPMS can be generated in the parameter domain based on the TDF and
    the period coefficients.
 Finally, by mapping the TPMS into the TBSS,
    a heterogeneous porous scaffold is produced.
 Moreover, we presented a new file format (TDF) for storing the porous scaffold that saves significant storage space.
 By the method developed in this study,
    both completeness of the TPMS units and continuity between adjacent TPMS units can be guaranteed.
 Moreover, the porosity of the porous scaffold can be controlled easily by
    designing a suitable TDF.
 More importantly, the TDF file format not only saves significant storage space,
    but it can also be used to generate a porous scaffold to any prescribed precision.
 In terms of future work, determining how to change the shape of a porous scaffold using the
    parameters stored in the TDF file format is a promising research direction.

\section*{Acknowledgements}
This work is supported by the National Natural Science Foundation of China
    under Grant No.61872316.

\bibliographystyle{elsart-num}
\bibliography{PorousScaffold}

\newpage
\section*{Appendix: TDF file format}
 \noindent
 \#period coefficients($\omega_x,\omega_y,\omega_z$)\\
 $\omega_x$ $\omega_y$ $\omega_z$\\
 \#resolution of control grid of TDF\\
 $n_u+1$ $n_v+1$ $n_w+1$\\
 \#control points of TDF\\
 $C_{0,0,0}$\\
 $C_{0,0,1}$\\
 \indent \vdots\\
 $C_{n_u,n_v,n_w}$\\
 \#knot vector in $u$-direction of TDF\\
 $u_0$ $u_1$ $\cdots$ $u_{n_u+4}$\\
 \#knot vector in $v$-direction of TDF\\
 $v_0$ $v_1$ $\cdots$ $v_{n_v+4}$\\
 \#knot vector in $w$-direction of TDF\\
 $w_0$ $w_1$ $\cdots$ $w_{n_w+4}$\\
 \#resolution of control grid of TBSS\\
 $m+1$ $n+1$ $l+1$\\
 \#control points of TBSS\\
 $Px_{0,0,0}$ $Py_{0,0,0}$ $Pz_{0,0,0}$\\
 $Px_{0,0,1}$ $Py_{0,0,1}$ $Pz_{0,0,1}$\\
 \indent \vdots \qquad\quad \vdots \qquad\quad \vdots\\
 $Px_{m,n,l}$ $Py_{m,n,l}$ $Pz_{m,n,l}$\\
 \#knot vector in $u$-direction of TBSS\\
 $U_0$ $U_1$ $\cdots$ $U_{m+4}$\\
 \#knot vector in $v$-direction of TBSS\\
 $V_0$ $V_1$ $\cdots$ $V_{n+4}$\\
 \#knot vector in $w$-direction of TBSS\\
 $W_0$ $W_1$ $\cdots$ $W_{l+4}$\\

\end{document}